\documentclass[10pt, twoside]{article}
\usepackage[paperwidth=170mm, paperheight=238mm]{geometry}
\usepackage{graphicx}
\usepackage{amsmath, amsfonts, amssymb, amsthm, bbm}
\usepackage{bm}
\usepackage{mathtools}
\usepackage{fancyhdr}
\usepackage{subfig}
\usepackage{lettrine}
\usepackage[explicit]{titlesec}
\usepackage{watermark}
\usepackage{color}
\usepackage[final]{pdfpages}
\usepackage{setspace}
\usepackage{chngcntr}
\usepackage{nameref}
\usepackage{enumitem}
\usepackage[perpage]{footmisc}
\usepackage{physics}
\usepackage{float}
\usepackage{eqnarray}

\usepackage[font=small]{caption}
\usepackage[colorlinks, citecolor=blue, linkcolor=black, urlcolor=blue]{hyperref}

\usepackage[T1]{fontenc}
\usepackage{lmodern}

\numberwithin{equation}{section}
\numberwithin{figure}{section}
\numberwithin{table}{section}

\definecolor{grey}{rgb}{0.5,0.5,0.5}
\definecolor{l-grey}{rgb}{0.8,0.8,0.8}
\definecolor{white}{rgb}{1,1,1}
\definecolor{black}{rgb}{0,0,0}
\definecolor{myred}{rgb}{0.831, 0.165, 0.184}
\definecolor{myblue}{rgb}{0.149, 0.471, 0.698}
\definecolor{mypurple}{rgb}{0.698, 0.400, 1}

\newcommand{\beq}{\begin{equation}}
\newcommand{\eeq}{\end{equation}}
\newcommand{\beqa}{\begin{eqnarray}}
\newcommand{\eeqa}{\end{eqnarray}}
\renewcommand{\ket}[1]{|#1\rangle}
\renewcommand{\bra}[1]{\langle#1|}

\let\origdoublepage\cleardoublepage
\newcommand{\clearemptydoublepage}{%
  \clearpage
  {\pagestyle{empty}\origdoublepage}%
}

\let\cleardoublepage\clearemptydoublepage

\titleformat{\section}[display]{\vspace*{190pt} \bfseries\sffamily \Huge}
{\begin{picture}(0,0)\put(-60,-30){\textcolor{grey}{\thesection}}\end{picture}}
{0pt}
{#1}
[]
\titlespacing*{\section}{40pt}{10pt}{40pt}[40pt]

\titlespacing*{\subsection}{0pt}{30pt}{20pt}[0pt]
\titleformat{\subsection}[display]{\Large \sffamily}{}{0pt}{\thesubsection \ #1}[]

\begin{document}

\pagestyle{fancy}
\renewcommand{\headrulewidth}{0pt}
\fancyhead{}
\fancyfoot{}






\begin{center}

\hrule

\vspace{16pt}
{\huge Control sequences\\
for Nitrogen-Vacancy centers \\
in the high frequency regime\par}
\vspace{16pt}

\hrule

\vspace{35pt}

{\Large {\bf Carlos Munuera Javaloy} }

\vspace{35pt}

\emph{Supervised by} \\

\vspace{15pt}

{\large

Dr. Jorge Casanova

}

\vspace*{\fill}

\includegraphics[height=2.5cm]{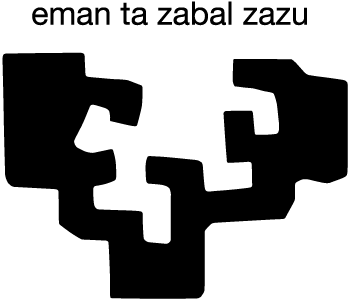}

\vspace{20pt}

Departamento de Qu\'imica F\'isica\\
Facultad de Ciencia y Tecnolog\'ia\\
\vspace{4pt}
Universidad del Pa\'is Vasco

\vspace{15pt}

{\large May 2024}

\end{center}

\pagebreak

This document is a PhD thesis developed during the period from October 2019 to May 2024 at 
the University of the Basque Country.

\vspace*{250pt}


\bigskip

\bigskip

\noindent Bilbao, May 2024

\vspace*{\fill}

{\setstretch{1.2}


\noindent This document was generated with the 2015  \LaTeX  \ distribution.

\noindent The \LaTeX \  template is adapted from a \href{https://gitlab.com/iagobaapellaniz/PhD-Thesis}{template} by Iagoba Apellaniz.

}

\cleardoublepage

\vspace*{60pt}
\vfill
\begin{center}
\emph{A mis padres, a mi hermana y a mi abuela.}
\end{center}
\vfill

\cleardoublepage

\vspace*{150pt}
\hspace*{\fill}\begin{minipage}{\textwidth-90pt}
\emph{Well, this little maneuver's gonna cost us 51 years!}
\end{minipage}
\\
\vspace*{10pt}
\begin{flushright}
{Joseph Cooper, film "Interstellar"}
\end{flushright}


\titleformat{\section}[display]
{\vspace*{160pt}
\bf\sffamily \Huge}
{{\textcolor{black}{\thesection}}. #1}
{0pt}
{#1}
[]
\titlespacing*{\section}{75pt}{10pt}{40pt}[40pt]

\textsf{\tableofcontents}


\section*{Abstract}
\pagenumbering{roman}
\fancyfoot[LE,RO]{\thepage}
\phantomsection
\addcontentsline{toc}{section}{Abstract}

In recent years, the field of quantum sensing has garnered increasing attention due to its potential to revolutionize various scientific and technological domains. Among the different quantum sensors, the nitrogen-vacancy (NV) color center in diamond stands out for its ease of use, ability to be read out and initialized with a laser, and long coherence times even at room temperature. Over the past years, numerous quantum control sequences have been developed to successfully deploy NV sensors in diverse situations, such as measuring nearby spin clusters, classical AC signals, and static magnetic fields. However, the NV center faces limitations when coupling to high frequency signals. More specifically, as the frequency of the target signal increases, stronger driving fields over NVs are needed, ultimately reaching the limits of current experimental capabilities.

In this thesis, we propose several protocols to address this high-frequency problem in different scenarios. 

We start by applying Shortcuts to Adiabaticity (STA) methods to improve previous designs, creating robust pulses that tailor the interaction between the NV center and AC signals, effectively replicating the dynamics of instantaneous pulses.

Next, we focus on the interaction between an NV center and other electron spins. This is a scenario where high frequency Larmor precession is ubiquitous due to the large gyromagnetic ratio of the electron. In this context, we utilize the natural ZZ interactions of the system, which commute with the Larmor precession terms, avoiding the need to compensate for fast rotations through external controls. Specifically, we incorporate a Double Electron-Electron Resonance (DEER) sequence to detect the coupling between two electron labels using a single NV center, providing substantial numerical evidence to support this possibility.

We then explore the application of surface electrons to hyperpolarize the nuclear spins of a sample placed on top of the diamond. We design a sequence that can simultaneously transfer polarization from the NV center to the surface electron and from the surface electron to the sample. We demonstrate that this mechanism increases the polarization transfer efficiency due to the strong dipolar interaction between the NV center and the electron an the proximity of the latter (note this is at the diamond surface) with target nuclei.

Next, we address the high-frequency problem in NV-based nuclear magnetic resonance (NMR) spectroscopy by proposing a sequence that induces a slow signal parallel to the external magnetic field using RF drivings resonant with the nuclear spin sample. The frequency of this signal depends only on the Rabi frequency of the radio frequency (RF) field. This is a tunable quantity, thus easily trackable by an NV center-based sensor when the Rabi is calibrated to tens of kHz. By generating the signal in discrete intervals and ensuring that the magnetization vector returns to its original position, we create distinct free evolution and measurement windows. This approach maps target quantities such as chemical shifts of the sample into the signal amplitude. Applying a discrete Fourier transform to the resulting measurements yields the sample's energy shifts.

Finally, we revisit this approach from a different perspective. This time, target parameters are encoded in the phase and amplitude of the induced signal. This can be achieved by applying a detuned RF, which generates a detectable signal that can be inspected via continuous measurements. In addition, the introduced RF driving takes the form of a Lee-Goldburg 4 sequence, that effectively removes dipole-dipole couplings. This driving generates a complex NMR signal due to its cyclic phase changes. Thus, we have developed a geometric frame that enables to design a pulsed protocol on the NV center to measure with optimal resolution.

\cleardoublepage


\section*{Resumen}
\fancyfoot[LE,RO]{\thepage}
\phantomsection
\addcontentsline{toc}{section}{Resumen}

\vfill
\lettrine[lines=2, findent=3pt,nindent=0pt]{L}{os} recientes avances tecnológicos, como el desarrollo de dispositivos capaces de controlar con exactitud los campos de radio y microondas, los láseres y los dispositivos criogénicos, nos han permitido manipular las propiedades cuánticas de la materia con una precisión sin precedentes. Este nivel de control se puede ejercer en diversas plataformas, incluyendo los circuitos superconductores, los puntos cuánticos semiconductores, los iones atrapados y los átomos de Rydberg. Todas estas plataformas pueden ser controladas sin destruir la coherencia cuántica, al menos durante cierto tiempo. Estos avances han desencadenado una carrera mundial con vistas al desarrollo de la nueva ola de tecnologías cuánticas. Entre ellas destacan la computación cuántica y las comunicaciones cuánticas que, a causa de su potencial impacto en el futuro, han acaparado la atención tanto de los medios de comunicación como de los inversores.

Sin embargo, otro ámbito de la tecnología cuántica que está adquiriendo un reconocimiento cada vez mayor son los sensores cuánticos. A pesar de ser menos conocidos que las otras dos áreas ya mencionadas, los sensores cuánticos están demostrando aplicaciones prácticas, aprovechándose de la sensibilidad intrínseca de los sistemas cuánticos a perturbaciones externas, como fuerzas, rotaciones o campos magnéticos. Este enfoque permite trazar un camino para el desarrollo de dispositivos que demuestran una sensibilidad inigualable, con frecuencia a escalas nanométricas. De este modo, estos dispositivos no sólo alcanzan, y a veces superan, la sensibilidad de los mejores sensores clásicos (como los SQUID para medir campos magnéticos), sino que también se espera que, en algunos casos, escalen con el límite de Heisenberg. Por lo tanto, se concibe la posibilidad de que su precisión mejore con el número de medidas en vez de con su raíz cuadrada. Todo ello supone una serie de avances que ofrecen, como consecuencia, la creación de herramientas que impulsan el progreso en un amplio espectro de campos, como la ciencia de los materiales, las imágenes biomédicas y la geología.

En lo que se refiere a los sensores cuánticos, éstos se valen de fenómenos cuánticos como la superposición y el entrelazamiento para detectar magnitudes físicas. Además, estos sensores ofrecen la ventaja de que las mediciones se basan en constantes físicas fundamentales, lo que implica que se reduzca significativamente la necesidad de calibrarlos con respecto a su contrapartida clásica. Asimismo, los sensores cuánticos tienen aplicaciones prometedoras en una amplia gama de campos como, por ejemplo, los magnetómetros de alta sensibilidad, los radares avanzados, los detectores de ondas gravitacionales y la neuroimagen.

En cuanto a las diversas plataformas que se pueden utilizar para la detección cuántica, se incluyen tales como sistemas de espines, circuitos superconductores, iones atrapados y vapores atómicos. Entre estas plataformas, el centro de nitrógeno vacante (NV) en el diamante es una opción particularmente popular, debido a su tamaño atómico y a que puede mantener su coherencia a temperatura ambiente. Esta tesis se centra en el  centro NV y su función como detector de campo magnético.

El centro NV es un defecto de color en la red del diamante, caracterizado por la sustitución de dos carbonos adyacentes por un átomo de nitrógeno y una vacante. El estado fundamental del centro NV puede manipularse con precisión usando campos de microondas. Además, la fluorescencia del NV es dependiente del estado de su espín, lo que pemite la lectura y la inicialización del estado cuántico del NV empleando un láser verde y registrando su fluorescencia. Este hecho nos permite un control coherente y preciso, además de una medición sencilla de su estado cuántico. Asimismo, otro rasgo a resaltar son sus excepcionales tiempos de coherencia, con $T_1$ del orden de milisegundos y $T_2$ de cientos de microsegundos en NVs individuales. No obstante, lo más destacable es que estos tiempos de coherencia se mantienen incluso a temperatura ambiente, haciendo del NV un dispositivo especialmente atractivo para aplicaciones relacionadas con la detección cuántica. A su vez, el diamante, conocido por sus propiedades tales como su resistencia, estabilidad y biocompatibilidad, ofrece un medio excelente para el sensor cuántico de NV dada sus características, aumentando de manera significativa su utilidad en diversas aplicaciones. Entre dichas aplicaciones nos encontramos, por ejemplo, ante los sensores de campo magnético en yunques de diamante, que son utilizados para experimentos de alta presión; en los agentes de contraste para aplicaciones de resonancia magnética y en los sensores cuánticos intracelulares.

Predominantemente utilizados para la detección de campos magnéticos, los centros NV también han demostrado otros usos como sensores de temperatura, presión y campos eléctricos. Por ello, los NVs pueden usarse en aplicaciones diversas: desde servir como sensores de campo magnético en puntas de microscopios de fuerza atómica y sondas para resonancia magnética nuclear (RMN), hasta actuar como medios para la hiperpolarización. Además, algunos dispositivos basados en NV han llegado a comercializarse. Ejemplos de ello lo constituyen las sondas de diamante basadas en NVs, que se utilizan para la obtención de imágenes de campos magnéticos con resolución nanométrica, como es el caso de Quantilever de Qnami y las puntas QS de QZabre.

La exploración de los centros NV como sistema cuántico único comenzó en 1997, cuando Gruber et al. demostraron por primera vez resonancia magnética detectada ópticamente (ODMR) en un único centro NV. Años más tarde, en 2008, Maze et al. y Balasubramanian et al., demostraron de forma independiente la detección de campos magnéticos haciendo uso de centros NV individuales, desplazando así el diálogo en torno a los NVs de qubits con vistas al procesamiento de información cuántica a sensores cuánticos prometedores.

Desde su surgimiento, la gama de aplicaciones de los centros NV se ha expandido rápidamente. En 2011, Dolde et al. se encargaron de introducir la detección de campos eléctricos con centros NV, ampliando aún más el conjunto de magnitudes físicas medibles. En 2013, Kucsko et al. lograron la detección de temperatura utilizando centros NV. En este trabajo, realizaron termometría dentro de una única célula viva empleando nanodiamantes. Tan solo un año después, en 2014, Doherty et al. demostraron el uso de la ODMR para medir altas presiones en yunques de diamante.

Recientemente, en 2015, Shi et al. consiguieron realizar la detección de una sola molécula marcada con una etiqueta de espín electrónico haciendo uso de un solo centro NV. Mientras tanto, Staudacher et al. llevaron a cabo las primeras mediciones de protones estadísticamente polarizados en la superficie del diamante, para lo que se sirvieron de un protocolo de espectroscopia de correlación. En 2018, Glenn et al. lograron un hito con la primera demostración de RMN de alta resolución utilizando centros NV a través de un protocolo de muestreo continuo. En años posteriores, mejoraron el experimento empleando técnicas clásicas de hiperpolarización para aumentar la sensibilidad del método. También en 2018, I. Schwartz et al. desarrollaron y demostraron la secuencia PulsePol, mejorando así el potencial de los centros NV como fuentes de hiperpolarización.

Los logros en el terreno experimental que han sido mencionados, fueron complementados con avances a nivel teórico, que han introducido nuevas secuencias y profundizado nuestra comprensión de la física de los centros NV. Sin embargo, todavía quedan pendientes grandes retos por superar para poder explotar plenamente el potencial de los centros NV como sensores cuánticos. En este sentido, un reto notable es la cota en las frecuencias de AC máximas que pueden medir los centros NV, normalmente restringidas a unos 40 MHz debido a la limitación en la potencia de los controles que pueden aplicarse sobre los NV. Este límite de frecuencia dificulta considerablemente aplicaciones como la RMN basada en NVs, donde el uso de campos magnéticos potentes es deseable para aumentar la polarización y mejorar la resolución de los desplazamientos químicos. Sin embargo, el uso de grandes campos magnéticos conlleva a la generación señales de Larmor de alta frecuencia, a las que el NV no puede acoplarse. Superar este límite de frecuencia, al que nos referiremos como el problema de la alta frecuencia, es el objetivo central de esta tesis.

Por lo tanto, en esta tesis investigamos esquemas de control en centros NV para poder abordar el problema de la alta frecuencia en aplicaciones de nanoescala y microescala. Esta exploración abarca múltiples protocolos y aplicaciones, mejorando potencialmente las capacidades de detección cuántica tanto en centros NV individuales como en conjuntos de NVs.

La tesis se estructura en siete capítulos, siendo el primero de ellos la introducción.

En lo que se refiere al capítulo 2, introducimos los conceptos fundamentales y las herramientas matemáticas relacionadas con el centro NV que suponen un factor esencial para comprender el resto de la presente tesis. De este modo, comenzamos examinando la estructura energética del Hamiltoniano del estado fundamental electrónico del centro NV. A continuación, discutimos los fundamentos del control de los centros NV mediante campos de microondas externos, centrándonos en técnicas como el eco de espín y la secuencia Carr-Purcell-Meiboom-Gill (CPMG). Dicho proceso se sigue mediante el esbozo de los retos técnicos relacionados con las secuencias de control en presencia de blancos (tanto señales clásicas como espines individuales) de alta frecuencia, detallando así sus implicaciones y sentando las bases para las soluciones desarrolladas en esta tesis. En la parte final del capítulo 2, introducimos los fundamentos de la espectroscopia de RMN basada en NV, a la que dedicaremos los últimos capítulos.

En cuanto al capítulo 3, desarrollamos una técnica que se sirve de la conformación de pulso para mitigar la pérdida de sensibilidad en escenarios en los que no se puede ignorar la anchura del pulso. Asimismo, presentamos un método basado en Shortcuts to Adiabaticity (STA) que compensa eficazmente el efecto de la anchura de pulso finita y, además, produce pulsos robustos.

A continuación, en el capítulo 4, abordamos el problema de las altas frecuencias centrándonos en las interacciones ZZ entre los centros NV y los espines objetivo, puesto que estas interacciones no se suprimen por las altas frecuencias de Larmor. En este capítulo, exploramos el escenario de medir el acoplamiento entre dos etiquetas electrónicas unidas a una molécula en la superficie de un diamante a través de un solo centro NV. Para ello, proponemos el uso de una secuencia tipo Double electron-electron resonance (DEER), e investigamos modelos analíticos, regímenes de parámetros y el uso de inferencia Bayesiana. Este enfoque podría ayudar a dilucidar la estructura y la dinámica molecular, proporcionando un método para investigar comportamientos bioquímicos complejos.

En el capítulo 5, continuamos explorando la interacción ZZ y desarrollamos un protocolo para transferir la polarización desde el centro NV a un espín de electrón en la superficie, que posteriormente puede utilizarse para polarizar los espines de una muestra. Esta secuencia requiere el control tanto del centro NV como del electrón de superficie para crear una dinámica de flip-flop efectiva. Al incorporar ecos de espín para mejorar la robustez de la secuencia, se transforma en una PulsePol de doble canal, conocida por su resistencia a los errores de control. El proceso habilitado por esta secuencia—específicamente, la transferencia de polarización mediada por un electrón de superficie—podría mejorar significativamente la transferencia de polarización a la superficie en la Polarización Nuclear Dinámica (DNP) basada en el centro NV.

En el capítulo 6, utilizamos la interacción ZZ en el ámbito de la RMN a microescala. Para ello, introducimos una nueva secuencia que recibe el nombre de Amplitude Encoded Radio Induced Signal (AERIS). Dicha secuencia sirve para codificar los desplazamientos de la frecuencia objetivo, causados por los desplazamientos químicos y los acoplamientos J en una muestra, en la amplitud de una oscilación controlada de la magnetización nuclear. Esta técnica facilita la RMN a microescala basada en NV a campos magnéticos arbitrariamente altos, accediendo así a un régimen experimental crítico. De esta manera, detallamos el desarrollo de esta secuencia y evaluamos su eficacia mediante simulaciones en un experimento de RMN de hidrógeno con etanol. Cabe resaltar que este enfoque no solo demuestra la versatilidad de la interacción ZZ, sino que también amplía las aplicaciones potenciales de la tecnología de RMN basada en NV.

Por otra parte, en el capítulo 7, ampliamos la aplicación de la RMN basada en la interacción ZZ incorporando la secuencia Lee-Goldburg 4 (LG4) sobre la muestra estudiada. La secuencia LG4 ofrece buenos resultados en relación a la eliminación de las interacciones homonucleares dipolo-dipolo, lo que es crucial para extraer información de muestras en estado sólido o de movimiento lento. Utilizaremos la señal longitudinal emitida por la muestra para desarrollar un protocolo que funciona a campos magnéticos arbitrariamente altos, contando con el apoyo de un marco geométrico que ayuda a identificar los tiempos de pulso adecuados para un contraste óptimo.

El capítulo final se dedica a resumir las conclusiones de la tesis. Le siguen los apéndices complementarios correspondientes y la bibliografía que ha sido utilizada.

\cleardoublepage


\section*{Acknowledgements}
\phantomsection
\addcontentsline{toc}{section}{Acknowledgements}


I feel fortunate and grateful to have found myself in the Basque Country, where I have traversed a long, challenging, and fascinating path.

I want to begin by expressing my deepest gratitude to my supervisor, Dr. Jorge Casanova. He has taught me the art of navigating through interaction pictures and approximations. His creative mind and confidence provided a perfect balance to my own pessimism, and I am proud of what we have accomplished together. I would also like to thank my tutor, Prof. Iñigo Egusquiza, for his help and for being a source of immense wisdom.

During the development of this thesis, I had the opportunity to visit the research group of Prof. Martin Plenio and develop a collaboration that is included in this thesis. I want to express my gratitude for allowing me to be part of his top-tier research group for a brief period.

I would also like to thank Prof. Dominik Bucher for giving me the opportunity to visit his top-level group and participate in their activities. This visit provided me with invaluable experience in the experimental aspects of NV centers.

Over the past four years, I have had the privilege of being part of the NQUIRE group. The good atmosphere within the group is truly unique, and I leave with not only many good memories, but also with many good friends.

Finally, I would like to thank my family for their love and support, and for enduring my endless (and often imaginative) rants about science as a child. Special thanks to Hanse's flatmate, for being who you are.

\cleardoublepage


\section*{List of Publications}
\phantomsection
\addcontentsline{toc}{section}{List of Publications}
{\bf This thesis is based on the following publications and preprints:}
\begin{enumerate}
\item {\bf \underline{C. Munuera-Javaloy}}, Y. Ban, X. Chen, and J. Casanova, \\
{\it Robust Detecttion of High-Frequency Signals at the Nanoscale}, \\
\href{http://doi.org/10.1103/PhysRevApplied.14.054054}{Physical Review Applied {\bf 14}, 054054 (2020)}.

\item {\bf\underline{C. Munuera-Javaloy}}, R. Puebla, B. D'Anjou, M. B. Plenio, and J. Casanova,\\
{\it Detection of molecular transitions with nitrogen-vacancy centers and electron-spin labels}, \\
\href{https://doi.org/10.1038/s41534-022-00653-w}{npj Quantum Information {\bf 8}, 140 (2022)}.

\item H. Espin\'os, {\bf \underline{C. Munuera-Javaloy}}, I. Panadero, P. Acedo, R. Puebla, J. Casanova, and E. Torrontegui \\
{\it Enhancing polarization transfer from nitrogen-vacancy centers to external nuclear spins via dangling bond mediators}, \\
\href{https://doi.org/10.1038/s42005-024-01536-6}{Communication Physics {\bf 7}, 42 (2024)}.

\item {\bf\underline{C. Munuera-Javaloy}}, A. Tobalina, and J. Casanova, \\
{\it High-Resolution NMR Spectroscopy at Large Fields with Nitrogen Vacancy Centeres}, \\
\href{https://doi.org/10.1103/PhysRevLett.130.133603}{Physical Review Letters {\bf 130}, 133603 (2023)}.

\item {\bf \underline{C. Munuera-Javaloy}}, A. Tobalina, and J. Casanova, \\
{\it High-Field Microscale NMR Spectroscopy with NV Centers in Dipolarly-Coupled Samples}, \\
\href{https://arxiv.org/abs/2405.12857}{arXiv:2405.12857  (2024)}.
\end{enumerate}

\newpage

{\bf Other publications and preprints not included in this thesis:}
\begin{enumerate}[resume]
\item {\bf \underline{C. Munuera-Javaloy}}, I. Arrazola, E. Solano, and J. Casanova, \\
{\it Double quantum magnetometry at large static magnetic fields}, \\
\href{https://doi.org/10.1103/PhysRevB.101.104411}{Physical Review B {\bf 101}, 104411 (2020)}.

\item {\bf \underline{C. Munuera-Javaloy}}, R. Puebla, and J. Casanova, \\
{\it Dynamical decoupling methods in nanoscale NMR}, \\
\href{https://iopscience.iop.org/article/10.1209/0295-5075/ac0ed1}{Europhysics Letters {\bf 134}, 30001 (2021)}.

\item M. G. Algaba, M. Ponce-Martinez, {\bf \underline{C. Munuera-Javaloy}}, V. Pina-Canelles, M. J. Thapa, B. G. Taketani, M. Leib, I. de Vega, J. Casanova, and H. Heimonen, \\
{\it Co-Design quantum simulation of nanoscale NMR}, \\
\href{https://doi.org/10.1103/PhysRevResearch.4.043089}{Physical Review Research {\bf 4}, 043089 (2022)}.

\item A. Biteri-Uribarren, P. Alsina-Bol\'ivar, {\bf \underline{C. Munuera-Javaloy}}, R. Puebla, and J. Casanova, \\
{\it Amplified nanoscale detection of labeled molecules via surface electrons on diamond}, \\
\href{https://doi.org/10.1038/s42005-023-01484-7}{Communication Physics {\bf 6}, 359 (2023)}.

\item P. Alsina-Bol\'ivar, A. Biteri-Uribarren, {\bf \underline{C. Munuera-Javaloy}}, and J. Casanova, \\
{\it J-coupling NMR Spectroscopy with Nitrogen Vacancy Centers at High Fields}, \\
\href{https://doi.org/10.48550/arXiv.2311.11880}{arXiv:2311.11880 (2023)}.

\item B. Varona-Uriarte, {\bf \underline{C. Munuera-Javaloy}}, E. Terradillos, Y. Ban, A. Alvarez-Gila, E. Garrote, and J. Casanova, \\
{\it Automatic Detection of Nuclear Spins at Arbitrary Magnetic Fields via Signal-to-Image AI Model}, \\
\href{https://link.aps.org/pdf/10.1103/PhysRevLett.132.150801}{Physical Review Letters {\bf 132}, 150801 (2024)}.

\end{enumerate}

\cleardoublepage

%
%


\section*{Abbreviations and \\ conventions}
\phantomsection
\addcontentsline{toc}{section}{Abbreviations and conventions}

We use the following abbreviations throughout the thesis
\begin{enumerate}[leftmargin=5cm]
	\item [\bf AERIS]{Amplitude Encoded Radio Induced Signal}
	\item [\bf DD]{Dynamical Decoupling}
	\item [\bf DEER]{Double Electron-Electron Resonance}
	\item[\bf DFT]{Discrete Fourier transform}
	\item[\bf DNP]{Dynamical Nuclear Polarization}
	\item [\bf ESR]{Electron Spin Resonance}
	\item [\bf LG]{Lee-Goldburg}
	\item [\bf NMR]{Nuclear Magnetic Resonance}
	\item [\bf NV]{Nitrogen-Vacancy}
	\item [\bf ODMR]{Optically Detected Magnetic Resonance}
	\item [\bf STA]{Shortcuts To Adiabaticity}
	\item [\bf RWA]{Rotating Wave Approximation}
	
 \end{enumerate}

Every Hamiltonian in this thesis is divided by $\hbar$.

\cleardoublepage


\renewcommand{\headrulewidth}{0.5pt}
\fancyfoot[LE,RO]{\thepage}
\fancyhead[LE]{\rightmark}
\fancyhead[RO]{\leftmark}


%
%
%

\titleformat{\section}[display]
{\vspace*{190pt}
\bfseries\sffamily \huge}
{\begin{picture}(0,0)\put(-50,-25){\textcolor{grey}{\thesection}}\end{picture}}
{0pt}
{\textcolor{white}{#1}}
[]
\titlespacing*{\section}{80pt}{10pt}{50pt}[20pt]


\pagenumbering{arabic}

\section[Introduction]{Introduction}


\thiswatermark{\put(1,-280){\color{l-grey}\rule{70pt}{42pt}}
\put(70,-280){\color{grey}\rule{297pt}{42pt}}}

%
%

\vfill
\lettrine[lines=2, findent=3pt,nindent=0pt]{R}{ecent} advancements in classical technology, such as precise control of radio and microwave fields, lasers, and cryogenic devices, allow us to manipulate the quantum properties of matter with unprecedented precision \cite{Haroche13,Wineland13}. This level of control extends across diverse systems such as superconducting circuits \cite{Blais21}, semiconductor quantum dots \cite{Burkard23}, trapped ions \cite{Monroe21}, and Rydberg atoms \cite{Bluvstein24} among several others. These advances have triggered a global race to develop the new wave of quantum technologies where quantum computing and quantum communications stand out, capturing significant media attention and investment due to their huge potential impact in the future.

However, another domain within quantum technology that is gaining increased recognition is quantum sensing. Although less well-known than the other two areas, quantum sensing is demonstrating practical applications by leveraging the intrinsic sensitivity of quantum systems to external disturbances, such as forces, rotations, or environmental magnetic fields \cite{Degen17}. This approach is paving the way for the development of devices with unmatched sensitivity. These devices not only achieve, and sometimes surpass, the sensitivity of the best classical sensors, such as superconducting quantum interference devices (SQUIDs) \cite{Fagaly06} for measuring magnetic fields, but are also expected, in some cases, to reach Heisenberg-limited measurements. These advancements are creating invaluable tools that are driving progress across a broad spectrum of fields, including material science, biomedical imaging, and geology \cite{Degen17}.

Quantum sensing employs quantum phenomena such as superposition and entanglement to detect physical quantities. Additionally, this method offers the advantage of basing measurements on fundamental physical constants, significantly reducing the need for calibration of classical devices. Quantum sensors have promising applications across a wide range of fields, such as highly sensitive magnetometers \cite{Fagaly06}, advanced radars \cite{Assouly23}, gravitational wave detectors \cite{Ligo}, and neuroimaging \cite{Aslam23}.

Various platforms can be utilized for quantum sensing, including spin systems \cite{Kitching11}, superconducting circuits \cite{Fagaly06}, trapped ions \cite{Maiwald09}, and atomic vapors \cite{Kominis03}. Among these, the nitrogen-vacancy (NV) center in diamond \cite{Doherty13, Schirhagl14} is a particularly popular choice as it is an atomic-size detector that can be operated at room temperature. The NV center and its role as a magnetic field detector is the focus of this thesis.

The NV center is a color defect in the diamond lattice, characterized by the substitution of two adjacent carbons by a nitrogen atom and a vacancy, see Fig.~\ref{fig:1ch1}(a). The NV center can be precisely manipulated and probed with microwave and laser fields, enabling accurate coherent control and readout of its quantum state. Its exceptional coherence times--$T_1$ on the order of milliseconds and $T_2$ of hundreds of microseconds--maintained even at room temperature, make the NV a particularly advantageous device for quantum sensing applications. Furthermore, the host diamond matrix, renowned for its strength, stability, and biocompatibility, offers an excellent medium for the NV quantum sensor, enhancing its utility across various applications, such as magnetic field sensors in diamond anvils for high-pressure experiments \cite{Doherty14}, contrast agents for magnetic resonance imaging (MRI) applications \cite{Waddington19}, and intracellular quantum sensors \cite{Kucsko13}.

Predominantly utilized for magnetic field detection, NV centers have also demonstrated other uses as sensors of temperature \cite{Kucsko13}, strain \cite{Doherty14}, and electric fields \cite{Dolde11}. Thus, their potential applications are diverse, ranging from serving as magnetic tip sensors \cite{Tetienne13} and probes for nuclear magnetic resonance (NMR) \cite{Glenn18} to acting as mediums for hyperpolarization \cite{London13}. In addition, some NV-based devices reached commercialization. Examples of these are NV-based diamond probes for imaging magnetic fields at nanoscale resolution at, e.g., Qnami's Quantilever \cite{Qnami} and QZabre's QS Tips \cite{Qzabre}.

The exploration of NV centers as a unique quantum system began in 1997 when Gruber et al.~\cite{Gruber97} first demonstrated optically detected magnetic resonance (ODMR) in a single NV center. In 2008, Maze et al. \cite{Maze08} and Balasubramanian et al. \cite{Balasubramanian08}, independently demonstrated detection of magnetic fields using single NV centers, shifting the dialogue around NVs from qubits for quantum information processing to quantum sensors.

The range of NV center applications quickly broadened. In 2011, Dolde et al.~\cite{Dolde11} introduced electric field sensing with NV centers, further expanding the set of target physical magnitudes. Kucsko et al. in 2013~\cite{Kucsko13} achieved temperature sensing using NV centers, performing thermometry within a single living cell using nanodiamonds. In 2014, Doherty et al.~\cite{Doherty14} demonstrated high-pressure ODMR in diamond anvils.

More recently, in 2015, Shi et al. \cite{Shi15} performed detection of a single molecule tagged with an electron-spin label using a single NV center, while Staudacher et al. \cite{Staudacher15} performed the first measurements of statistically-polarized protons on the diamond surface utilizing a correlation spectroscopy protocol. In 2018, Glenn et al. \cite{Glenn18} achieved a milestone with the first demonstration of high-resolution NMR using NV centers through a continuous sampling protocol. In subsequent years, they enhanced the experiment by employing hyperpolarization techniques to increase the method's sensitivity \cite{Bucher20, Arunkumar21}. Also in 2018, I. Schwartz et al. \cite{Schwartz18} developed and demonstrated the PulsePol sequence, improving the potential of NV centers as hyperpolarization sources.

These experimental achievements have been complemented by theoretical advancements, which have introduced new sequences and deepened our understanding of the physics of NV centers. However, significant challenges  remain to fully exploit the potential of NV centers as quantum sensors. A significant one is the limitation on the maximum AC frequencies NV centers can measure, typically capped at approximately 40 MHz due to the limitation of driving rates over NVs. This frequency limit significantly hinders applications such as NV-based NMR, where large magnetic fields--desirable for increasing polarization and enhancing chemical shifts--result in fast-rotating Larmor signals. Overcoming this frequency limit, which we will refer to as the {\it high-frequency problem}, is the central focus of this thesis.

\subsection{What you will find in this thesis}

In this thesis, we investigate control schemes in NV centers  to address the high-frequency problem for nanoscale and microscale applications. This exploration spans multiple protocols and applications, potentially enhancing quantum sensing capabilities in both single NV centers and in NV ensembles.

The thesis is structured into seven chapters, with the first chapter being the introduction.

In Chapter 2, we introduce the fundamental concepts and mathematical tools related to the NV center that are essential for understanding the rest of the thesis. We begin by examining the energy structure of the electronic ground state Hamiltonian of the NV center. Next, we discuss the basics of controlling NV centers using external microwave driving fields, with a focus on techniques such as the spin echo and the Carr-Purcell-Meiboom-Gill (CPMG) sequence. We continue by outlining the technical challenges related to control sequences in the presence of high-frequency targets, detailing their implications and setting the stage for the solutions developed in this thesis. In the final part of Chapter 2, we introduce the fundamentals of NV-based NMR spectroscopy to which we will dedicate the final chapters.

In Chapter 3, we develop a technique using pulse shaping to mitigate the loss of sensitivity in scenarios where pulse width cannot be ignored. We present a method based on Shortcuts to Adiabaticity (STA) that effectively compensate for the effect of finite pulse width while, in addition, leads to robust pulses.

In Chapter 4, we tackle the high-frequency problem by focusing on the ZZ interactions between NV centers and target spins, as these interactions are not suppressed by high Larmor frequencies. In this chapter, we explore the scenario of measuring the coupling between two electronic labels attached to a molecule on the surface of a diamond via an NV center. This approach could help to elucidate molecular structure and dynamics, providing a method for investigating biochemical behaviors.

In Chapter 5, we continue exploring the ZZ interaction and develop a protocol to transfer polarization from the NV center to an electron spin on the surface, which can subsequently be used to polarize a nuclear sample. This sequence requires control over both the NV center and the surface electron to create effective flip-flop dynamics. By incorporating spin echoes to enhance the robustness of the sequence, it transforms into a double-channel PulsePol, renowned for its resilience against control errors. The process enabled by this sequence—specifically, the polarization transfer mediated by a surface electron—could significantly enhance polarization transfer to the surface in NV center-based Dynamic Nuclear Polarization (DNP).

In Chapter 6, we utilize the ZZ interaction within the realm of microscale NMR. We introduce a novel sequence, termed Amplitude-Encoded Radio-Induced-Signal (AERIS), which encodes target frequency shifts, caused by chemical shifts and J-couplings in a sample, into the amplitude of a controlled oscillation of the sample nuclei. This technique facilitates microscale NV-based NMR at arbitrarily high magnetic fields, thereby accessing a critical experimental regime. We detail the development of this sequence and evaluate its efficacy through simulation in an ethanol proton NMR experiment. This approach not only demonstrates the versatility of the ZZ interaction but also expands the potential applications of NV-based NMR technology.

In Chapter 7, we expand on the application of ZZ interaction-based NMR by incorporating the Lee-Goldburg 4 (LG4) over a target sample. The LG4 sequence is effective in removing homonuclear dipole-dipole interactions, a crucial factor in extracting information from solid-state or slowly moving samples. In this chapter we develop a protocol to perform NV-based NMR in dipolarly coupled samples at arbitrarily high magnetic fields, supported by a geometric framework that helps to identify the right pulse timings for optimal contrast.

The final chapter is dedicated to summarizing the conclusions of the thesis. This is followed by complementary appendices and the bibliography.


\section{Fundamentals of the NV}
\label{chapter1}


\thiswatermark{\put(1,-280){\color{l-grey}\rule{70pt}{42pt}}
\put(70,-280){\color{grey}\rule{297pt}{42pt}}}

%
%

\begin{figure*}[t]
\centering
\hspace{0.0 cm}\includegraphics[width=1\columnwidth]{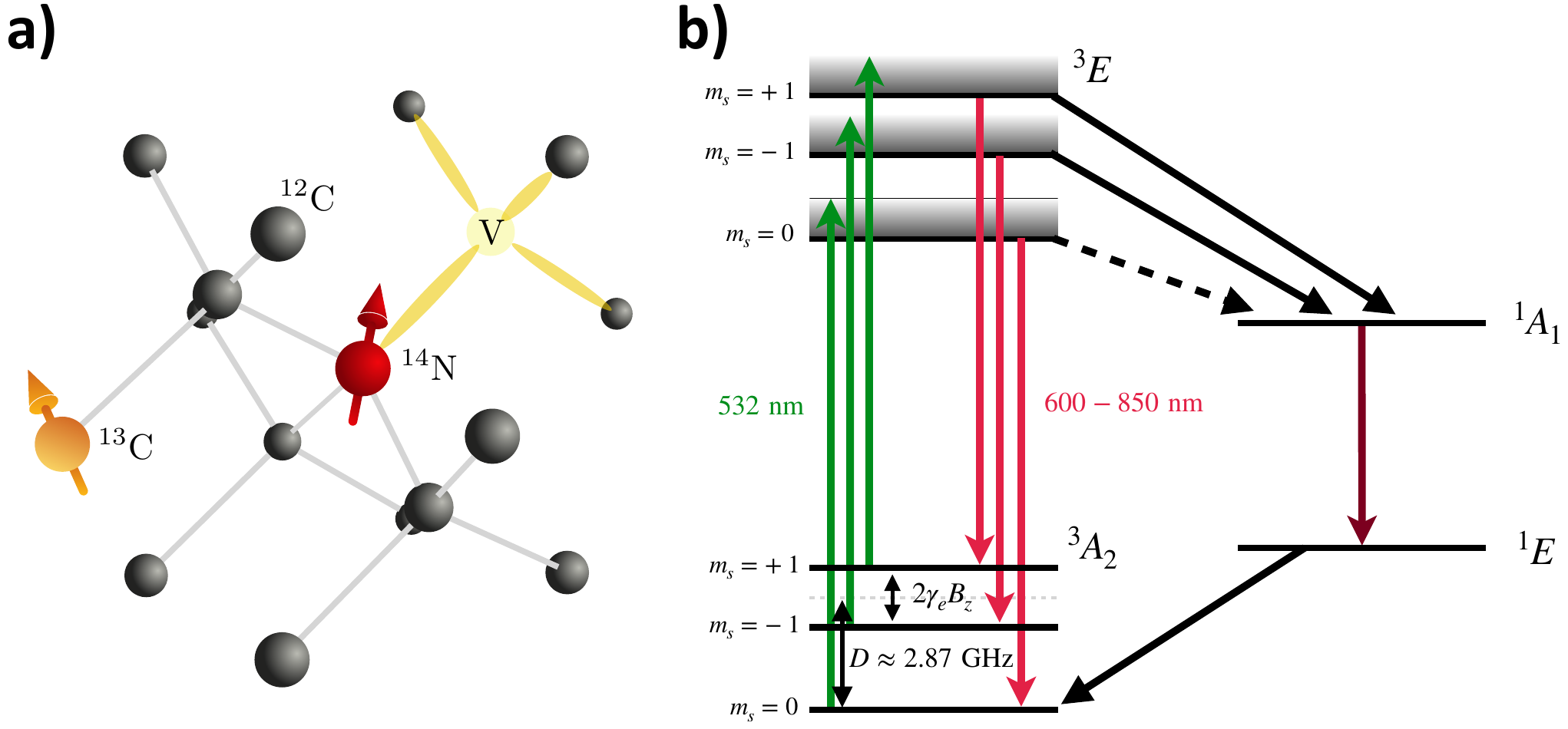}
\caption{(a) Diamond unit cell containing an NV color center. Red spin represent a $^{14}$N isotope while, in yellow, we have the vacancy.  Other spins can also appear in the structure, such as a $^{13}$C nucleus, depicted in orange. (b) Simplified energy levels of the NV center, illustrating both the radiative transitions (colored arrows) and the non-radiative transitions (in black) to the metastable state.}
\label{fig:1ch1}
\end{figure*}

\vfill
\lettrine[lines=2, findent=3pt, nindent=0pt]{T}{he} NV center is a defect in diamond characterized by a nitrogen atom adjacent to a vacancy, as shown in Figure \ref{fig:1ch1}(a). The NV can exist in three charge states: positive, neutral, and negative. The negatively charged state is the most extensively studied since its fluorescence is directly linked to its spin state \cite{Schirhagl14}. In this thesis, we will focus on this negatively charged state which we will denote simply as the NV. 

At room temperature, the NV center can be approximated to an 8-level energy system, as depicted in Figure \ref{fig:1ch1}(b) \cite{Schirhagl14}.
The energy structure features a triplet ground state, designated as $^3A_2$, and a triplet excited state, identified as $^3E$, together with a singlet metastable state comprising two intermediate levels, $^1A_1$ and $^1E$. The transition between the $^3A_2$ and $^3E$ states is radiative, with a zero-phonon line at 637 nm. This transition is spin-conserving and exhibits high brightness and exceptional photostability, even under off-resonant excitation. In experiments conducted at room temperature, a 532 nm green laser is utilized to excite the NV, leveraging the presence of phonon sidebands. Fluorescence is subsequently collected in the 600-850 nm range, enabling experimentalists to effectively filter out reflected green laser light.

The zero-field splitting, $D$, between the $m_s = \pm 1$ and $m_s = 0$ states originates from electronic spin-spin interaction and is sensitive to changes in temperature, pressure, and electric field. For the electronic ground state $^3A_2$,  the zero-field splitting is approximately 2.87 GHz, and for the excited state $^3E$, it is around 1.42 GHz \cite{Doherty13}. The degeneracy of the $m_s = \pm 1$ states can be lifted by Zeeman splitting when an external magnetic field is applied along the NV center's crystallographic axis, which is defined by the vector from the vacancy to the nitrogen nucleus. This sensitivity of the $m_s = \pm 1$ levels to an external magnetic field is the primary mechanism employed for magnetic field sensing.

In the excited state $^3E$, the $m_s = \pm 1$ states have a significant chance to decay non-radiatively via the singlet state, subsequently transitioning to the $^3A_2$ $m_s = 0$ state. Conversely, the excited $m_s = 0$ state primarily decays through the radiative transition. This difference in decay pathways facilitates the reinitialization of the NV center to the $m_s=0$ state. Moreover, it creates an optical contrast of about 30{\%} between states $m_s=0$ and $m_s = \pm 1$ for single NVs, providing a simple mechanism for spin state readout through fluorescence. 

We now outline the general strategy for magnetometry to be employed throughout this thesis. Optical initialization and readout are scheduled at the beginning and end of each protocol, respectively. Each designed sensing sequence will be executed exclusively in the NV electronic ground state through coherent manipulation of the $m_s=0,-1,1$ hyperfine levels. These are characterized by a $T_1$ relaxation time of the order of milliseconds at room temperature and a coherence time $T_2$ of several hundreds of microseconds in pure crystals \cite{Schirhagl14}. 

In the following section, we explore the fundamental principles of coherent control and the underlying Hamiltonians.

\subsection{Control sequences}
Coherent control of the NV center (more specifically of the $m_s=0,-1,1$ levels) is fundamental in any sensing sequence. This is achieved by applying customized microwave (MW) radiation patterns that induce transitions among the  $m_s=0,-1,1$ levels  in the NV. These transformations modify the NV center dynamics to encode a targeted physical quantity (e.g., a magnetic field) onto the NV spin state at the end of the sequence. A prevalent strategy for addressing the quantum control challenge in NV centers is dynamical decoupling \cite{Munuera21}. This method enables to couple the NV to the target while simultaneously protects the NV against control errors and external disturbances. 

To lay the groundwork for this discussion, we introduce the Hamiltonian hat describes the NV center under the influence of an external magnetic field, $B_z$, aligned with the crystallographic axis $\hat z$:
\begin{equation}
H_0 = D S_z^2 - \gamma_e B_z S_z,
\end{equation}
where $D\approx (2\pi)\times2.87$ GHz is the zero-field splitting, $S_{\mu \in {x, y, z}}$ denotes  spin-1 matrices, and $\gamma_e\approx(2\pi)\times$28.025  GHz/T is the electron's gyromagnetic ratio. As stated in the previous section, this external magnetic field is introduced to lift the degeneracy between the $m_s = \pm 1$ states, allowing us to individually target the transitions  $0\leftrightarrow 1$ and \mbox{$0\leftrightarrow -1$}.

To coherently control  the NV, we apply a microwave field that interacts with the NV center via the Zeeman effect such that the resulting Hamiltonian is: 
\begin{equation}\label{eq:driving}
H = H_0 + |\gamma_e| B_d S_x \cos(\omega t - \phi),
\end{equation}
where $B_d$ is the amplitude of the driving field, $\omega$ its frequency, and $\phi$ its  phase. In order to analyze this Hamiltonian, it is standard to employ an interaction picture transformation with respect to $H_0$, yielding:
\begin{equation}
H_I = \Omega(\ket{1}\bra{0}e^{i p_+ t}+\ket{-1}\bra{0}e^{i p_- t}+\text{H.c.})\cos(\omega t + \phi),
\end{equation}
where $\Omega = |\gamma_e| B_d/\sqrt{2}$, $p_{\pm} = D \pm |\gamma_e| B_z$, and H.c. denotes the Hermitian conjugate terms. To explore the NV dynamics under resonant driving, we tune the frequency $\omega$ to match one of the transition frequencies, for instance $\omega = p_+$, leading to:
\begin{equation}
\begin{split}
H_I = &\frac{\Omega}{2}\big[\ket{1}\bra{0}e^{i 2 p_+ t+i\phi}+\ket{-1}\bra{0}e^{i (p_-+p_+) t+i\phi}\\&+\ket{1}\bra{0}e^{i\phi}+\ket{-1}\bra{0}e^{i (p_--p_+) t+i\phi}+\text{H.c.}\big].
\end{split}
\end{equation}
The rotating wave approximation (RWA) is  applied to average out terms oscillating much faster than their magnitude. For example, terms like $\frac{\Omega}{2}\ket{1}\bra{0}e^{i 2 p_+ t+i\phi}$ can be neglected if $2p_+>>\frac{\Omega}{2}$. This approximation is valid for all non-resonant terms in the previous equation, as every oscillating frequency significantly exceeds $\Omega$. Consequently, we reach the simplified Hamiltonian:
\begin{equation}
H = \frac{\Omega}{2}\left(|0\rangle\langle1| e^{-i\phi} + |1\rangle\langle0| e^{i\phi} \right).
\end{equation}
Note that this effective Hamiltonian does not involve the $\ket{-1}$ state. By limiting our analysis to driving frequencies resonant with only one of the transitions, we can effectively operate within a two-dimensional subspace spanned by $\{\ket{0},\ket{1}\}$. This simplification is appropriate for the scope of this thesis, and from here on, we will often use the Pauli spin-$\frac{1}{2}$ matrices to describe NV Hamiltonians. Note that the same procedure can be applied if the driving is resonant with the $0\leftrightarrow -1$ transition.

This Hamiltonian illustrates the effective two-level system dynamics under a resonant driving, displaying coherent transitions between the $|0\rangle$ and $|1\rangle$ states. Such a mechanism, where state transfer is induced by resonant driving, can be effectively utilized as a basic sensing sequence in ODMR~\cite{Oort90}, see next section.

\subsubsection{ODMR}
ODMR  applied to NV centers represents a special case of electron spin resonance (ESR) that utilizes fluorescence for detection of the NV spin state. This feature facilitates the acquisition of electron resonance spectra from individual NV spins, a capability unique to a limited number of systems~\cite{Schirhagl14}, such as other solid-state defects~\cite{Hepp14, Christle17} and certain organic molecules~\cite{Kohler99}.
\begin{figure}[h]
\centering
\hspace{0.0 cm}\includegraphics[width=0.9\columnwidth]{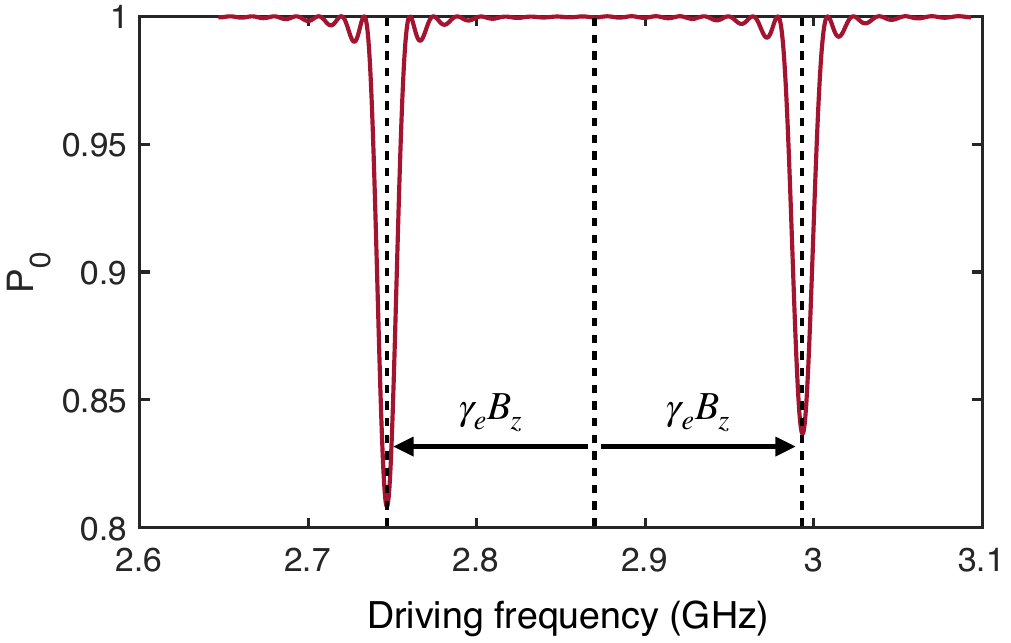}
\caption{Simulated ODMR experiment depicting the two characteristic dips corresponding for a magnetic field $B_z=44$ G. Each point is obtained by computing the survival probability $P_0$ of an initial state $\ket{0}$ after being driven for 200 oscillations of the corresponding frequency.}
\label{fig:2ch1}
\end{figure}

In ODMR experiments, laser and microwave (MW) radiation is directed to the NV center while its fluorescence is monitored. Sweeping the MW radiation frequency produces a spectrum with a noticeable decrease in fluorescence indicating the NV center departure from the $m_s=0$ state, see Fig.~\ref{fig:2ch1}. In the presence of an external magnetic field, the spectrum displays two fluorescence dips corresponding to the $0\rightarrow 1$ and $0\rightarrow -1$ transitions. The separation between these transitions, equal to $2\gamma_e B_z$, offers a direct measure of the magnetic field at the NV center's location.

Due the relative simplicity of the experiment, the ODMR has been widely utilized to measure static magnetic fields across diverse studies such as vector magnetic field sensing \cite{Maertz10, Steinert10}, wide-field imaging of bacteria \cite{Sage13}, and tip magnetometry \cite{Tetienne13, Rondin13}. Additionally, the average resonance frequency (dashed line in Fig.~\ref{fig:2ch1}) reveals shifts in the zero-field splitting, providing information about changes in electric fields \cite{Dolde11}, temperature \cite{Kucsko13}, and strain \cite{Doherty14} at the NV location.

Despite its utility, the scope and sensitivity of ODMR has limitations. More sophisticated methods have been inherited from classic NMR, and involve the application of stroboscopic MW driving composed of short, intense MW radiation bursts, known as pulses. Pulses are less error prone than soft continuous drivings, and allow us to think in terms of discrete transformations that modify the free Hamiltonian. Furthermore, pulsing techniques allow the implementation of dynamical decoupling, enhancing system robustness against noise and errors.

\subsubsection{Spin echo}
The most frequently utilized pulses in control sequences are the $\pi$ and $\pi/2$ pulses. These pulses, with durations defined as $t_{\pi} = \frac{\pi}{\Omega}$ and $t_{\frac{\pi}{2}} = \frac{\pi}{2\Omega}$ respectively, serve as the building blocks for NV quantum state manipulation. The propagators corresponding to these operations are defined as follows:
\begin{align}
&U_{\phi, \pi} = e^{-i \frac{\Omega}{2} \sigma_{\phi} t_{\pi}} = -i\sigma_{\phi}, \
&U_{\phi, \frac{\pi}{2}} = e^{-i \frac{\Omega}{2} \sigma_{\phi} t_{\frac{\pi}{2}}} = \frac{\mathbb{I}-i\sigma_{\phi}}{\sqrt{2}},
\end{align}
with $\sigma_\phi=\left(|0\rangle\langle1| e^{-i\phi} + |1\rangle\langle0| e^{i\phi} \right)$.
In the Bloch sphere representation, these pulses effectuate rotations around axes determined by the phase 
 $\phi$. The $\pi/2$ pulse is primarily employed to create the initial NV superposition state (e.g., from $\ket{0}$ to $\ket{+}$). Additionally, it is used to convert the relative phase accumulated between states $\ket{0}$ and $\ket{1}$ during the experiment back into a measurable NV population difference. On the other hand, the $\pi$ pulse inverts the spin state about an specific axis and is the basic component for refocusing noise in dynamical decoupling techniques, the spin echo method being the simplest example.
\begin{figure}[t]
\centering
\hspace{0.0 cm}\includegraphics[width=1\columnwidth]{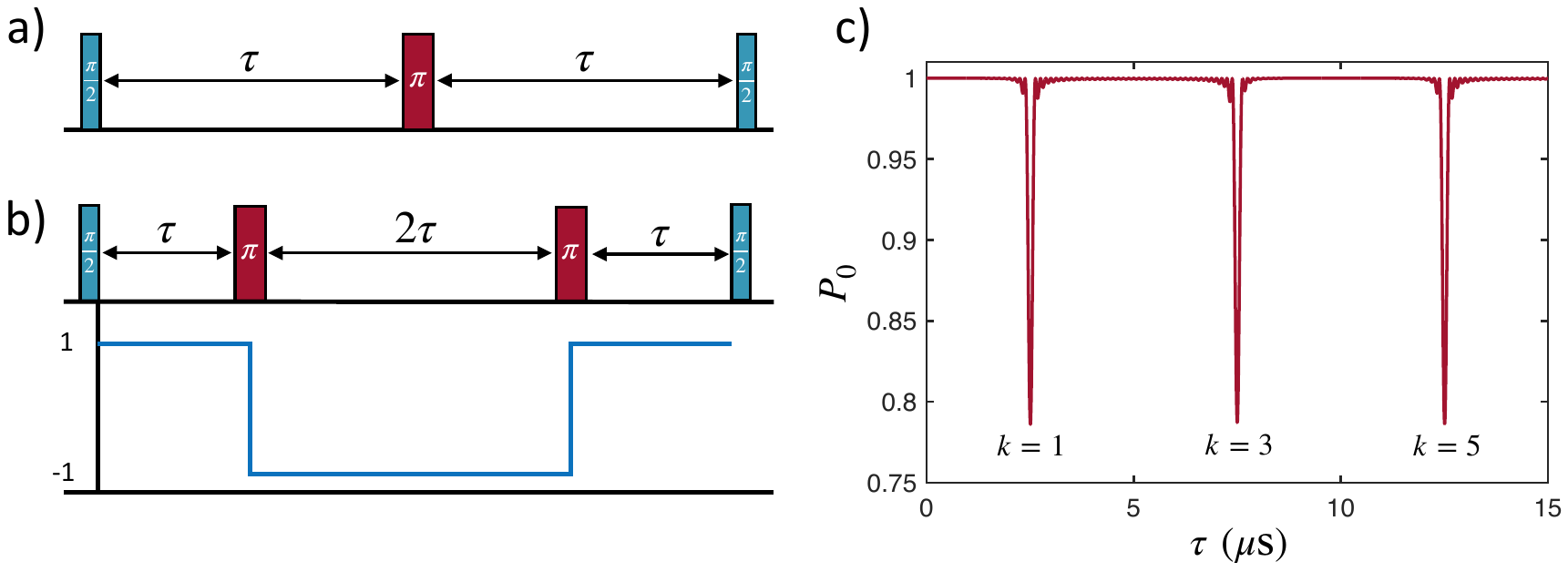}
\caption{(a) Spin-echo sequence illustrating the initialization and readout $\pi/2$ pulses (depicted in blue) along with the refocusing $\pi$ pulse (shown in red). (b) A single block of a CPMG sequence. The modulation function $F(t)$ considering instantaneous pulses is depicted below. (c) Simulated spectrum of a CPMG sequence sensing an external signal with an amplitude of $\Gamma/2=(2\pi)\times 1.2$ KHz and a signal frequency $\omega_s = 100$ KHz. The survival probability of an initial state $\ket{0}$ is computed with respect to the free time $\tau$ for ten consecutive blocks. The dips corresponding to the first three harmonics ($k=1,3,5$) can be seen.}
\label{fig:3ch1}
\end{figure}

The spin echo sequence \cite{Hahn50} begins with a $\pi/2$ pulse, creating a coherent superposition state. The spin is then allowed to evolve freely for a time period $\tau$, during which external noise introduces dephasing. Then, a $\pi$ pulse is applied, which effectively reverses the impact of static and slow-varying noises. This inversion leads the spins to rephase, rewinding the decoherence process after an additional period $\tau$. At this point, another $\pi/2$ pulse may be introduced, setting the NV for the final optical measurement, see Fig.~\ref{fig:3ch1} (a). This refocusing technique can significantly extend the coherence time by up to two orders of magnitude, increasing $T^*_2$ from around 1 $\mu$s to $T_2$ of several hundred $\mu$s in some cases for both single NV centers and NV ensembles \cite{Childress06, Bauch18}. We remark that the spin echo proves particularly beneficial in NV ensembles by counteracting the effects of inhomogeneous fields. Since the spin echo makes the system insensitive to DC magnetic fields, it is predominantly utilized for AC signal detection \cite{Maze08}. 

The spin echo mechanism makes the NV sensitive to frequencies that align with the refocusing interval. We  study this process in the next section where, in addition, we explore a more advanced sequence that consists of multiple nested spin echoes.

\subsubsection{The CPMG sequence}\label{seq:cpmg}
To understand the effect of a spin echo train, consider an example where the NV center is subjected to an external AC signal along the $\hat{z}$ axis, which we aim to detect. The free Hamiltonian of the NV center in an interaction picture with respect to $H_0= D S_z^2 - \gamma_e B_z S_z$ is described by:

\begin{equation}
H = \frac{\Gamma}{2} \sigma_z \cos(\omega_s t+\varphi),
\end{equation}
where $\Gamma = B_s |\gamma_e|$ represents the interaction strength, $B_s$ is the  amplitude of the signal, $\omega_s$ its frequency, and $\varphi$ its initial phase. In scenarios where the interaction strength $\Gamma$ is significantly smaller than the signal frequency, thus holding the RWA, the signal's influence on NV dynamics becomes negligible, making it undetectable.

A spin echo train, specifically tuned to resonance with $\omega_s$, circumvents this issue and enables signal detection by synchronizing the NV center with the targeted signal. A classic example is the CPMG \cite{Purcell54, Meiboom58} pulse sequence. This sequence starts with an initial $\pi/2$ pulse applied along the $y$ axis, placing the NV in an initial superposition state $\ket{\psi_0}=\ket{+}$. This is followed by a succession of $\pi$ pulses applied along the $x$ axis, each spaced by a free time $\tau$. This series of pulses systematically accumulates a relative phase $\theta$ on the NV state as $\ket{\psi_t}=\left(\ket{0}e^{i\theta}+\ket{1}\right)/\sqrt{2}$. The sequence concludes with a final $\pi/2$ pulse, translating the accumulated phase into observable state population changes, see Fig.~\ref{fig:3ch1} (b).
The Hamiltonian including the controls and setting (with $\varphi=0$) is
\begin{equation}
H =  \frac{\Gamma}{2}   \sigma_z \cos{\left(\omega_s t\right)} + \Omega(t) \frac{\sigma_\phi}{2}.
\end{equation}
In a rotating frame with respect to the control term $\Omega(t) \frac{\sigma_\phi}{2}$  we get 
\begin{equation}
\label{eq:cpmg_u0}
H_I(t) =  \frac{\Gamma}{2}   U_0^\dagger\sigma_z U_0 \cos{\left(\omega_s t\right)} 
\end{equation}
where $U_0 = e^{-i\int_0^t \Omega(t')\frac{\sigma_x}{2}dt'}$ is the propagator of the control term. Assuming instantaneous pulses, we can rewrite the previous Hamiltonian as
\begin{equation}
\label{eq:cpmg_f}
H_I(t) = F(t)  \frac{\Gamma}{2} \sigma_z \cos{\left(\omega_s t\right)},
\end{equation}
where $F(t)$ is the so called {\it modulation function}, which is valued $+1(-1)$ after an even (odd) number of pulses, see Fig.~\ref{fig:3ch1} (b). The modulation function can be decomposed via Fourier series leading to
\begin{equation}
F(t)=\sum_{k=1}^\infty\frac{4(-1)^ {\frac{k-1}{2}}}{\pi k}\cos{\left(k\omega_m t\right)} \text{ for odd index } k,
\end{equation}
where $\omega_m = 2\pi/T$ with $T=2\tau$ being the period of the CPMG sequence. We refer to the index $k$ as {\it harmonics}. Resonance occurs whenever $\omega_m k$ matches the frequency of  the signal $\omega_s$. This condition reveals the presence of a resonance for each harmonic $k$, see Fig.~\ref{fig:3ch1} (c). Choosing a resonant harmonic $k$ and applying the RWA, we obtain the effective Hamiltonian
\begin{equation}
\label{eq:eff_cpmg}
H_I = \frac{f_k}{4}\Gamma \sigma_z,
\end{equation}
where the coefficient $f_k = \frac{4(-1)^ {\frac{k-1}{2}}}{\pi k}$ modulates the interaction strength between the NV center and the signal based on the selected harmonic. We can extend this analysis, generalizing to an arbitrary initial signal phase $\varphi$. In that case, Hamiltonian \eqref{eq:eff_cpmg} is modified to
\begin{equation}
\label{eq:eff_cpmg_phi}
H_I = \frac{f_k}{4}\Gamma \sigma_z\cos{(\varphi)},
\end{equation}
demonstrating sensitivity of the method to the initial phase of the signal, which will be relevant for the last section of this chapter. 

The CPMG sequence is one of the simplest pulse trains in a broader family of sequences. More sophisticated pulse trains, such as the widely utilized XY4 and \mbox{XY8} sequences~\cite{Viola99, Souza12}, select the phase of each delivered $\pi$ pulse in a manner that allows the sequence to self-correct pulse errors, significantly enhancing robustness. The sensing mechanism of these advanced sequences follows the same derivation as presented in this section.

\subsection{The high-frequency problem}
\begin{figure*}[t]
\centering
\hspace{0.0 cm}\includegraphics[width=0.9\columnwidth]{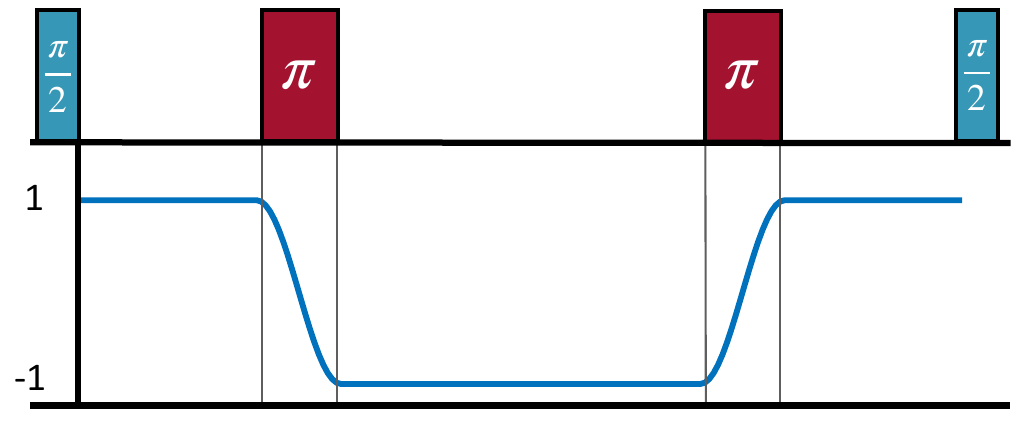}
\caption{A depiction of a single block of a CPMG sequence, with its corresponding modulation function $F(t)$ accounting for finite-width pulses.}
\label{fig:4ch1}
\end{figure*}
The AC sensing mechanism of the CPMG sequence highlights the challenge of coupling to high-frequency targets. One harmonic of the sequence must match the frequency of the external AC signal for the NV to accumulate phase. Considering the pulse duration $t_\pi$, the free evolution adjusts to $\tau'=\tau-t_\pi/2$. At sufficiently high frequencies, in particular when $t_\pi/2>\tau$, the pulses overlap, making the sequence ineffective. This overlap sets an upper frequency limit for the first harmonic equal to the pulse strength, around 50 MHz. To circumvent this limitation, one potential strategy is to select a higher harmonic. This choice effectively extends $\tau$, though it simultaneously diminishes the interaction strength, as indicated by the coefficient $f_k$.

Another subtler issue arises when the pulse width $t_\pi$ is comparable to the period of the signal, even before reaching the limit of pulses overlapping \cite{Casanova18}. When this happens, the modulation function is no longer a square wave (see Fig.~\ref{fig:4ch1}), and its Fourier series must be recalculated. The new coupling strength modulation factor $f_k$ scales inversely with the square of the ratio between the pulse duration and the Larmor period, which can significantly reduce the coupling to the signal. In the following chapters, we will explore several approaches to address these problems, ensuring robust and effective control over NV centers in various experimental settings.

\subsection{Fundamentals of microscale NMR}
\begin{figure*}[t]
\centering
\hspace{0.0 cm}\includegraphics[width=0.9\columnwidth]{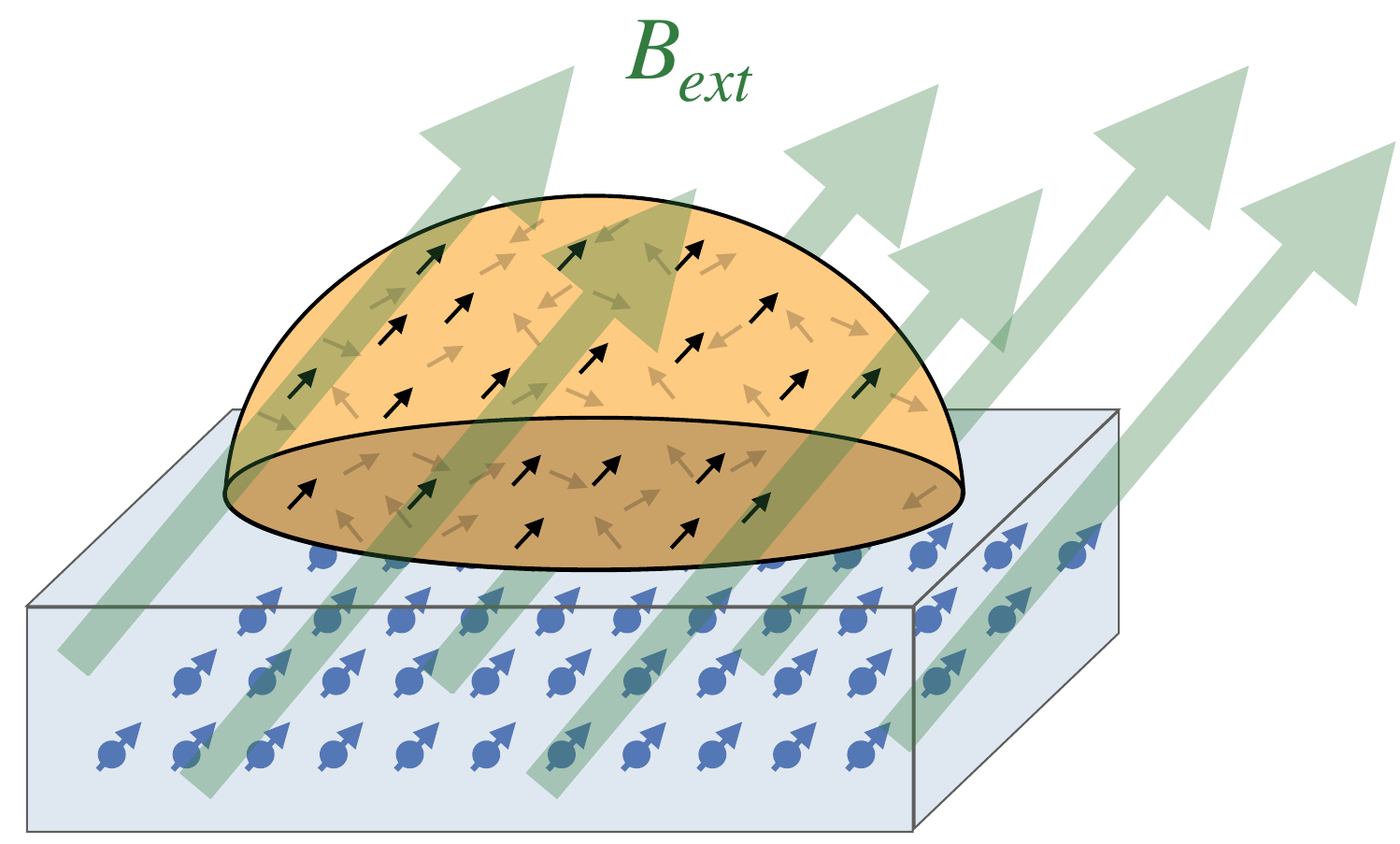}
\caption{Illustration of the experimental setup, showing a liquid sample positioned on top the diamond crystal. The external magnetic field (depicted by green arrows) is aligned with the NV ensemble crystallographic axis.}
\label{fig:5ch1}
\end{figure*}
NV-based NMR applications are significantly affected by the high-frequency problem. Operating in a high-field regime, thus under large Larmor frequencies, offers several advantages such as enhanced spectral clarity, increased resolution due to the linear dependence of chemical shifts on the external magnetic field's strength, and stronger signals resulting from enhanced thermal polarization. The final two chapters of this thesis are dedicated to addressing the high-frequency problem in NV-based NMR spectroscopy. This section will explore the fundamental concepts of microscale NMR.

Consider a sample containing $^{1}$H spins described by the  Zeeman Hamiltonian:
\begin{equation}
\label{eq:basic_zeeman}
H=-\gamma_H B_z I_z,
\end{equation}
where $\gamma_H\approx 2\pi\times42.58$ MHz/T is the hydrogen gyromagnetic ratio, and $I_z=\sigma_z/2$ is the nuclear spin operator. The initial state of the nuclei in thermal equilibrium at room temperature is given by the mixed state \cite{Levitt08}:
\begin{equation}\label{eq:fun_thermal}
\rho \approx \frac{1}{2}\mathbb{I} + \frac{1}{2} \mathcal{B} I_z,
\end{equation}
where $\mathbb{I}$ is the identity operator, and $\mathcal{B}=\frac{\hbar \gamma_H B_z}{k_B T}$ with $k_B$ the Boltzmann constant and $T$ the temperature. This formula highlights how increasing the magnetic field $B_z$ enhances the polarization $P=\bra{0}\rho\ket{0}-\bra{1}\rho\ket{1}=\frac{1}{2}\mathcal{B}$ of the sample. In a typical NMR experiment, a $\pi/2$ pulse resonant with the nuclear spins is applied to tilt the magnetization into the orthogonal plane. This manipulation shifts the polarization of the nuclei, allowing their subsequent evolution under Hamiltonian \eqref{eq:basic_zeeman} to be observed. As the spins precess about the magnetic field direction at a frequency of $\gamma_H B_z$, their collective motion generates a detectable RF signal, analogous to classical gyroscopic precession. This RF signal, produced by the transverse magnetization, is detected and then Fourier-transformed to identify the constituent frequencies. In this simple model, the only spectral peak would correspond to the Larmor frequency. However, real-world scenarios reveal a more complex spectrum influenced by several phenomena, including chemical shifts and J-couplings. Analyzing these (chemical shifts and J-couplings) provides information about the molecular structure of the sample.

In NV-based NMR spectroscopy, traditional detection probes (or coils) are replaced with a diamond containing an ensemble of NV centers. The liquid sample is placed directly above the diamond surface (see Fig.~\ref{fig:5ch1}), allowing the  RF signal to be detected by the NV centers using, for instance, a sequence such as the CPMG or the XY4 \cite{Staudacher15}. However, the spectral linewidth in these measurements is influenced by the duration of the measurement, this is $\Delta_f\propto\frac{1}{t}$, as dictated by the Fourier transform. This measurement duration is inherently limited by the NV centers' $T_2$ relaxation time. This imposes a resolution limit on the order of tens of kHz, insufficient for many NMR applications that require better spectral resolution.

To enhance resolution, heterodyne techniques are employed~\cite{Glenn18, Schmitt17, Boss17}. The simplest example of the latter is the continuous application of a CPMG sequence. These echo pulse trains are sensitive to both the amplitude and the phase of the signal as depicted in Eq.~\eqref{eq:eff_cpmg_phi}. By tuning the CPMG sequence close to resonance, each successive block measures the signal's phase at the beginning of that block. By chaining several such blocks together, the signal's phase is sampled at multiple points. After applying a Fourier transform to these phase data, the spectrum is reconstructed. This approach allows resolution to be governed not by the NV coherence time but by that of the signal. In liquid  $^1$H NMR with NVs, the application of heterodyne techniques translates into linewidths on the order of 1 Hz, improving performance with respect to using a single CPMG sequence by several orders of magnitude.


\section[High-frequency signal detection]{High-frequency signal\\detection}
\label{chapter2}

%
%

\thiswatermark{\put(1,-302){\color{l-grey}\rule{70pt}{60pt}}
\put(70,-302){\color{grey}\rule{297pt}{60pt}}}

%
%

\vfill
\lettrine[lines=2, findent=3pt,nindent=0pt]{I}{n} this chapter, we develop a technique originally introduced by Casanova et al. in Ref.~\cite{Casanova18}, which combines high harmonics with shaped pulses to create sequences able to operate at high frequencies. Furthermore, we demonstrate that the incorporation of quantum control techniques based on Shortcuts to Adiabaticity (STA) leads to a significant enhancement of the pulse robustness to errors. This improvement is achieved through a reparametrization of the NV quantum state into Bloch sphere angles and a geometrical phase. By deriving the equations of motion for these parameters and introducing an appropriate ansatz, we optimize the pulse resilience against detuning and control errors. This chapter contains the work developed in Ref. \cite{Munuera-Javaloy20}.

Section \ref{seq:sta_mod_modulation} examines the impact of pulse width on the filter function and its effect on the coupling between the NV center and the target signal. In Section \ref{seq:sta_ham}, we introduce a generalized driving Hamiltonian to facilitate pulse shaping. Following this, Section \ref{sta:reparam} details the parametrization of the NV state and the derivation of the corresponding equations of motion. Section \ref{sta:coupling} adapts the solution proposed by Casanova et al. to our new parametrization, and Section \ref{sta:ansatz} presents our proposed ansatz for pulse shaping. Finally, Section \ref{seq:sta_res} provides numerical results supporting the effectiveness of our STA-based pulses.

\subsection{Modified modulation function}\label{seq:sta_mod_modulation}
In this section, we explore the impact of finite-width pulses on the modulation function $F(t)$. Analyzing $F(t)$ through its Fourier series reveals various frequency components, which also indicates those with which the sequence can interact. The coupling strength between the NV center and a target signal is proportional to the Fourier coefficient of the corresponding resonant harmonic $f_k$.
The explicit expression of the coefficient $f_k$ is
\begin{equation}
\begin{split}
 f_k = & \frac{2}{T}\left[\int_{0}^{t_1} \cos(k \omega_m t) dt+\int_{t_1}^{t_2} F(t) \cos(k \omega_m t) dt- \right. \\
 & - \int_{t_2}^{t_3} \cos(k \omega_m t) dt+\int_{t_3}^{t_4} F(t)\cos(k \omega_m t) dt+ \\ 
 & + \left. \int_{t_4}^{T} \cos(k \omega_m t) dt\right],
 \end{split}
 \end{equation}
where $\omega_m = 2\pi/T$ with $T$ the period of the employed sequence, and $t_2-t_1=t_4-t_3\equiv t_\pi$. With instantaneous pulses where $t_\pi=0$, one finds $|f_k|=|\frac{4}{\pi k}|$ for odd k. However, when the pulse width $t_\pi$ becomes significant, the coefficient modifies to
\begin{equation}
f_k(\alpha)=\frac{4(-1)^{(k+1)/2}\cos{(\alpha \pi)}}{(4\alpha^2-1)k\pi},
\end{equation}
with $\alpha = k t_\pi / T$ denoting the ratio of pulse width to the signal period. The $\alpha^{-2}$ dependency, substantially diminishes the coefficient's magnitude as $\alpha$ increases, making the sequence ineffective.
\begin{figure}[h]
\center
\hspace{-0. cm}\includegraphics[width=1\columnwidth]{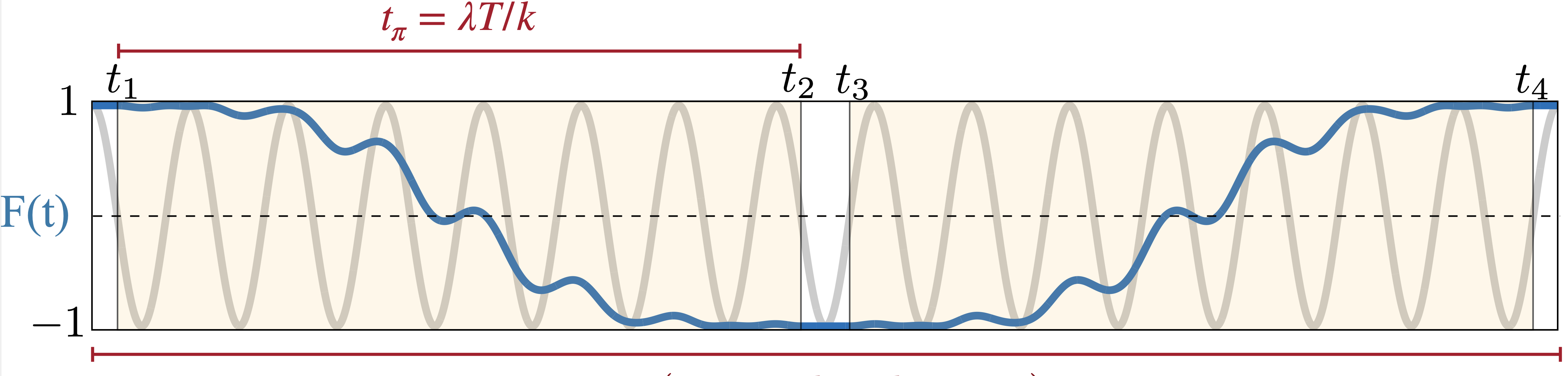}
\caption{Modulation funciton $F(t)$ corresponding to a $\pi$ pulse generated using our method (blue line) and the distribution of times for $\pi$ pulses. The areas in which the pulses are being applied are displayed in yellow.}
\label{sta_results}
\end{figure}
To address this challenge, as proposed in Casanova et al.~\cite{Casanova18}, one can employ a shaped pulse to tailor $F(t)$ such that 
\begin{equation}\label{eq:coupling}
\int_{t_1}^{t_2} F(t) \cos(k \omega_m t) dt = \int_{t_3}^{t_4} F(t)\cos(k \omega_m t) dt =0,
\end{equation}
 which eliminates the contribution of $F(t)$ during the pulses.
 
Under these conditions, the coefficient becomes
\begin{equation}
\begin{split}
f_k = &\frac{2}{T}\left[\int_{0}^{t_1} \cos(k \omega_m t) dt-\int_{t_2}^{t_3} \cos(k \omega_m t) dt + \right. \\
&\left. +\int_{t_4}^{T} \cos(k \omega_m t) dt\right]= \frac{4}{k\pi}\sin\left(\frac{k\pi}{T}\right)\cos\left(\frac{k\pi}{T}t_\pi\right).
\end{split}
\end{equation}
With this expression, the coefficient for instantaneous pulses $|f_k| = \frac{4}{k\pi}$ is recovered whenever $t_\pi = \frac{\lambda T}{k}$, where $\lambda$ is an integer. 

The rest of the chapter deals with a new design for the modulation function and corresponding pulse shapes to improve the robustness of the previous method.

\subsection{General driving Hamiltonian}\label{seq:sta_ham}
In order to have enough flexibility for pulse shaping, we will increase the degrees of freedom in our driving Hamiltonian \eqref{eq:driving}
\begin{equation}
H = D S_z^2 + |\gamma_e| B_z S_z + |\gamma_e| B_d S_x \cos\left[\omega t +\Delta(t)- \phi\right],
\end{equation}
where an extra time-dependent phase $\Delta(t)$ has been added. Going to an interaction picture with respect to $H_0=D S_z^2 + |\gamma_e| B_z S_z$, and setting the driving frequency to $\omega=D+|\gamma_e| B_z$ we reach the following Hamiltonian after applying the RWA
\begin{equation}
H = \frac{\Omega}{2}\left\{|0\rangle\langle1| e^{i[\Delta(t)-\phi]} + |1\rangle\langle0| e^{-i[\Delta(t)-\phi]} \right\}.
\end{equation}
Next we add and subtract the term $\frac{\delta(t)}{2}\sigma_z$, where $\sigma_z\equiv\left(\ket{0}\bra{0}-\ket{1}\bra{1}\right)$ and $\delta(t)$ is defined such that $\Delta(t)=\int^t_0\delta(t')dt'$. Going to an interaction picture with respect to $H_0=-\frac{\delta(t)}{2}\sigma_z$, we reach our final control Hamiltonian:
\begin{equation}
\label{eq:sta_control}
H_c= \frac{\Omega(t)}{2} \sigma_{\phi} + \frac{\delta(t) }{2} \sigma_z.
\end{equation}
This driving Hamiltonian  includes controls over the amplitude $\Omega(t)$, the phase $\phi$ and the detuning $\delta(t)$ of the MW field. The full Hamiltonian combines the control in Eq.~(\ref{eq:sta_control}) with the target Hamiltonian $H_T$—which represent an external classical signal—as shown below
\begin{equation}\label{eq:sta_ham}
H= H_T+\frac{\Omega(t)}{2} \sigma_{\phi} + \frac{\delta(t) }{2} \sigma_z.
\end{equation}

\subsection{State reparametrization}\label{sta:reparam}
To design the pulses, we start by parametrizing the NV spin state evolution during a pulse as \cite{Daems13}
\begin{equation}
\label{wavefuction}
|\phi (t) \rangle = \left[\cos\left(\frac{\theta}{2}\right) e^{i\frac{\beta}{2}} |0\rangle + \sin \left(\frac{\theta}{2}\right) e^{-i\frac{\beta}{2}} |1\rangle \right] e^{i \gamma},
\end{equation}
with $\theta\equiv\theta(t)$ and $\beta\equiv\beta(t)$ being  the polar and azimuthal angles on the Bloch sphere, and $\gamma\equiv\gamma(t)$ a geometric phase. When inserting Eq. (\ref{wavefuction}) into
the time-dependent Schr\"{o}dinger equation governed by the control Hamiltonian \eqref{eq:sta_control}, we get the following auxiliary equations 
\beqa
\label{dottheta}
\dot{\theta} &=& \Omega(t)\sin(\beta),
\\
\label{dotbeta}
\dot{\beta} &=&  \dot{\theta} \cot(\theta)\cot(\beta)-\delta(t),
\\
\dot{\gamma} &=& - \dot{\theta}\cot(\beta)/(2 \sin (\theta))\label{dotgamma}.
\eeqa
For the sake of simplicity, in the previous equations we have particularized to the case $\sigma_\phi = \sigma_x$, but the formalism is equally applicable to  $\sigma_\phi$. Equations~(\ref{dottheta}, \ref{dotbeta}, \ref{dotgamma})  connect the Rabi frequency $\Omega(t)$ and the detuning $\delta(t)$ with the $\theta$ and $\beta$ angles. 
Note that, similar expressions to Eqs.~(\ref{dottheta}, \ref{dotbeta}, \ref{dotgamma}) can be derived from a dynamical invariant \cite{Lu13,Ruschhaupt12}.  

To achieve a $\pi$ pulse, e.g. from $|0\rangle$ at $t=0$ to $|1\rangle$ at $t=t_\pi$, the state described by Eq~(\ref{wavefuction}) must satisfy the following boundary conditions 
\begin{equation}
\theta(0) = 0, ~ \theta(t_\pi) = \pi.
\label{Boundary}
\end{equation}
A possible parametrisation for $\theta$ and $\beta$ is: $\theta= \pi t/t_\pi$ and $\beta= \pi/2$ leading to $\Omega(t) = \pi/t_\pi$ and $\delta(t) = 0$. This simple trajectory corresponds to a top-hat pulse. On the other hand, we note that there exists much freedom to tailor the functions $\theta$ and $\beta$. This flexibility allows for the design of pulse sequences that not only enable coupling with rapidly precessing nuclei but also exhibit robustness against control errors.
\subsection{Signal coupling}\label{sta:coupling}
To adapt the coupling condition in Eq.~\eqref{eq:coupling} to our  state parametrization, we revisit the original appearance of the modulation function in Eq.~\eqref{eq:cpmg_u0} and Eq.~\eqref{eq:cpmg_f} to derive
\begin{equation}
F(t)=\bra{0}U_0^\dag\sigma_zU_0\ket{0},
\end{equation}
with $U_0=\hat T e^{-i\int^{t_{\pi}}_{t_0}H_c(t')dt'}$. Assuming that $\ket{0}=\ket{\phi (0)}$ and substituting from Eq.~\eqref{wavefuction}, we obtain:
\begin{equation}
\label{eq:sta_f}
F(t)=\bra{\phi (0)}U_0^\dag\sigma_zU_0\ket{\phi (0)}=\bra{\phi (t)}\sigma_z\ket{\phi (t)}=\cos{(\theta)}.
\end{equation}
To achieve the same coupling as in the case of instantaneous pulses, we substitute Eq.~\eqref{eq:sta_f} into the previous coupling condition $\int_{t_0}^{t_\pi} F(t) \cos(k \omega_m t) dt=0$ which leads to
\begin{equation}
\label{CouplingCondition}
\int_{0}^{t_\pi}\cos (\theta) \cos(k\omega_m t)dt = 0.
\end{equation}
Hence, Eq.~(\ref{CouplingCondition}) establishes the first requirement for the $\theta$ function. From now on we will refer to this equation as the {\it coupling condition}.

Further constraints have to be imposed in the dynamics of $|\phi (t) \rangle$ to cancel  control errors during the pulse. Typically, these errors are: $(i)$ Deviations in the Rabi frequency, i.e. $\Omega(t)  \rightarrow\Omega(t)(1+\xi_\Omega)$, as a consequence of MW power variations denoted by $\xi_\Omega$. And, $(ii)$, errors in the $\delta(t)$ function (i.e. $\delta(t) \rightarrow \delta(t)+\xi_\delta$) with $\xi_\delta$ being a frequency offset that appears owing to, e.g., undetermined stress conditions in the diamond and/or because of nearby electronic impurities leading to NV energy shifts. 

We use perturbation theory over  $|\phi (t) \rangle$ during the $\pi$ pulse, and calculate the transition probability $P(t_\pi)$ of having an NV spin-flip driven by an imperfect $\pi$ pulse (up to second order in $\xi_\Omega$ and $\xi_\delta$). 

This reads $P(t_\pi) \approx 1-\frac{1}{4}\left|\int_{0}^{t_\pi}dt e^{i 2 \gamma}\left(\xi_\Delta\sin(\theta) - i2\xi_\Omega\dot{\theta}\sin^2(\theta)\right)\right|^2$. See Appendix~\ref{app:sta_err} for more details regarding the derivation of $P(t_\pi)$. In this manner, the second requisite for $\theta$ and $\gamma$ is the {\it error cancelation condition} that eliminates control errors during the NV spin-flip. This reads
\begin{equation}\label{errorcancelation}
\left|\int_{0}^{t_\pi}dt e^{i 2 \gamma}\left(\xi_\Delta\sin(\theta) - i2\xi_\Omega\dot{\theta}\sin^2(\theta)\right)\right| = 0.
\end{equation}

Once we get expressions for $\theta$ and $\gamma$ (and consequently to $\beta$ as Eq.~(\ref{dotbeta}) relates $\beta$ with $\theta$ and $\gamma$) one can find the control parameters $\Omega(t)$ and $\delta(t)$ by solving \mbox{Eqs.~(\ref{dottheta}, \ref{dotbeta})}.
\subsection{Ansatz}\label{sta:ansatz}
In order to interpolate a function for $\theta$, we use an ansatz inspired by the Blackman function~\cite{Blackman58}. This is 
\begin{equation}
\label{theta}
\theta(t) = \alpha_0 + \alpha_1 \cos\left(\frac{\pi}{t_\pi}t\right)+\alpha\sin\left(\frac{2\pi\lambda}{t_\pi} t\right),
\end{equation}
where $\lambda$ is a free tunable parameter that regulates the $\pi$ pulse length as $t_\pi = \lambda T/k$, see Appendix A.  In addition, $\alpha_0$, $\alpha_1 $, and $\alpha$ are parameters that we adjust to hold the previously commented conditions. In particular, when the boundaries in Eq.~(\ref{Boundary}) are applied to $\theta(t)$, we get $\alpha_0=-\alpha_1=\pi/2$, while the additional parameter $\alpha$ will be selected to fulfill the coupling condition in Eq.~(\ref{CouplingCondition}).  

Now, we pose the following ansatz for $\gamma(t)$
\begin{equation}
\label{gamma}
\gamma(t) = \theta + \eta_1\sin(2\theta)+\eta_2\sin(4\theta),
\end{equation}
that introduces two additional free parameters $\eta_{1}$ and $\eta_{2}$. The expression for $\gamma(t)$ can be combined with Eq.~(\ref{dotgamma}) leading to
\begin{equation}
\label{beta}
\beta =\cos^{-1}\left(\frac{-2M\sin(\theta)}{\sqrt{1+4M^2\sin^2(\theta)}}\right), 
\end{equation}
where $M = 1+2\eta_1\cos(2\theta)+4\eta_2\cos(4\theta)$. We will use $\eta_1$ and $\eta_2$ to achieve  Eq.~(\ref{errorcancelation}) over some reasonable error interval. In this manner, undesired NV transitions caused  by  errors in the Rabi frequency and detuning get cancelled up to second order. 

\subsection{Numerical results}\label{seq:sta_res}
We demonstrate the performance of our method with numerical simulations in Nanoscale NMR escenarios. In particular, we have computed the evolution of an NV under an XY8 sequence in the presence of a nearby $^{13}$C nuclear spin, as well as under the influence of a classical electromagnetic wave modelling a $^1$H nuclear spin cluster. In both cases we consider a strong  magnetic field $B_z = 3$~T~\cite{Aslam17}, which leads to high-frequency Larmor precession of approximately 32 MHz and 128 MHz, respectively. We compare the obtained Nanoscale NMR spectra  in situations involving: Standard top-hat $\pi$ pulses, extended  $\pi$ pulses that follow the scheme in~\cite{Casanova18}, and $\pi$ pulses designed with our method. 

\subsubsection*{Carbon spins}
\begin{figure}[h]
\center
\hspace{-0. cm}\includegraphics[width=1\columnwidth]{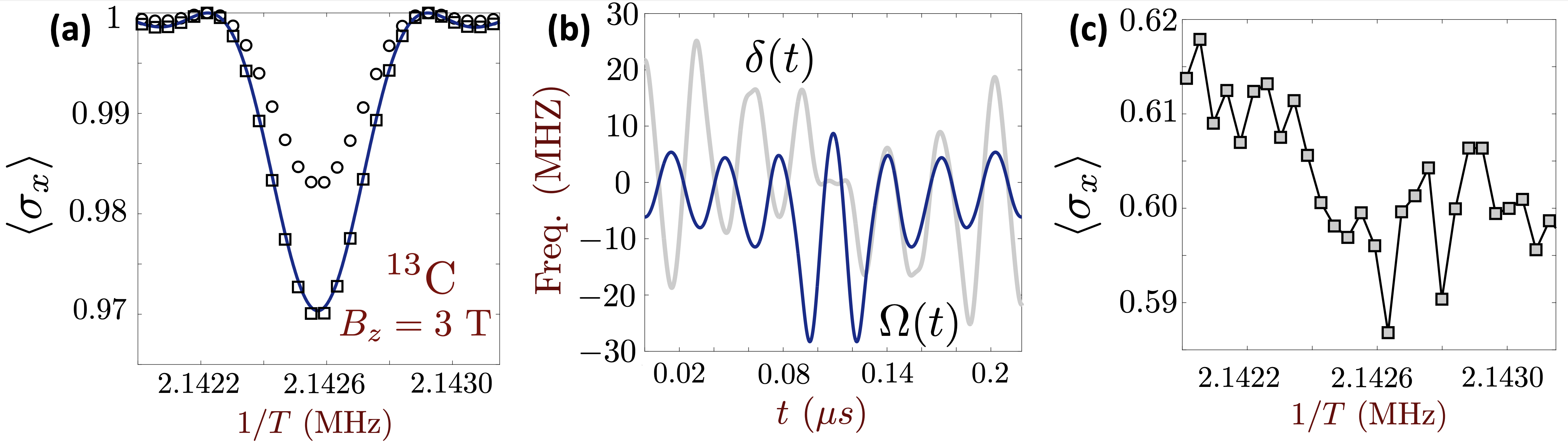}
\caption{Distinct Nanoscale NMR scenarios at a strong magnetic field $B_z = 3$ T involving a $^{13}$C nucleus. In (a) we show the Nanoscale NMR spectrum, i.e. the $\langle\sigma_x\rangle$ of the NV, in three different situations:  Solid-blue curve corresponds to an ideal case involving instantaneous $\pi$ pulses. The squares represent the spectrum obtained with our method, while circles uses standard top-hat pulses. In all cases we have repeated the XY8 sequence 102 times. This implies that 816 $\pi$ pulses have been employed leading to a final sequence time $\approx 0.19$ ms. In (b) we show the controls $\Omega(t)$ (dark curve) and $\delta(t)$ (clear curve) used in (a) for computing the spectrum including squares. Note $\Omega(t)$ and  $\delta(t)$ are in units of frequency, while the $\pi$ pulse induced by these controls has a duration $\approx 0.22\ \mu$s. (c)  Obtained spectrum with the extended $\pi$ pulses in Ref.~\cite{Casanova18}.}
\label{sta_results}
\end{figure}
The results are presented in Fig.~\ref{sta_results}. In (a) we show the computed spectra (encoded in the expectation value  $\langle\sigma_x\rangle$ of the NV center) of a problem involving an NV coupled to a nearby $^{13}$C nucleus (adding the term $H_T = -\gamma_N B_z I_z  +S_z  \vec{A}\cdot \vec{I}$ to the system Hamiltonian \eqref{eq:sta_ham}). The nucleus is at a distance of $1.1$ nm from the NV, such that its hyperfine vector $\vec{A} = (2\pi)\times[-4.81, -8.331, -26.744]$ KHz. The solid-blue line corresponds to the spectrum that would appear if instantaneous pulses (this is $\pi$ pulses with infinite MW energy) were delivered to the system. In addition, this solid-blue line has been obtained without introducing control errors, representing an idealized experimental scenario. The spectrum represented by the squares in Fig.~\ref{sta_results} (a) has been calculated by using our method. The particular values for the control parameters $\Omega(t)$ and $\delta(t)$ are shown in Fig.~\ref{sta_results} (b), and have led to a $\pi$ pulse of length $t_{\pi} = 0.21 \ \mu$s.  A detuning error of $\xi_\delta = (2\pi)\times1$ MHz, as well as a Rabi frequency deviation of $\xi_\Omega = 0.5\%$ are included in our numerical simulations. 

Even in these conditions involving significant errors, the spectrum produced by our method (squares) overlaps well with the ideal one (solid-blue). On the other hand, the spectrum represented by circles in Fig.~\ref{sta_results} (a) has been computed with standard top-hat $\pi$ pulses with a Rabi frequency ($\Omega_{\rm th}$) that equals the maximum of $\Omega(t)$ in our method, see Fig.~\ref{sta_results} (b). More specifically, this is $\Omega_{\rm th} \approx (2\pi) \times 30$ MHz. It is noteworthy to mention that the spectral contrast achieved by top-hat pulses (this is the peak depth of the spectrum with circles) is significantly lower than the one achieved by our method, which demonstrates the better performance of the latter. In Fig.~\ref{sta_results} (c) we show the spectrum computed with the extended pulses in Ref.~\cite{Casanova18} which include the same errors on the controls ($\xi_\delta = (2\pi)\times1$ MHz, and $\xi_\Omega = 0.5\%$). Notably, the extended pulses in Ref.~\cite{Casanova18} produce a completely distorted spectrum that does not allow to identify the resonance of the $^{13}$C. As a further comment, in absence of control errors our method and the one in Ref.~\cite{Casanova18} lead to similar results.\\

\subsubsection*{Hydrogen cluster}
\begin{figure}[h]
\center
\hspace{-0. cm}\includegraphics[width=1\columnwidth]{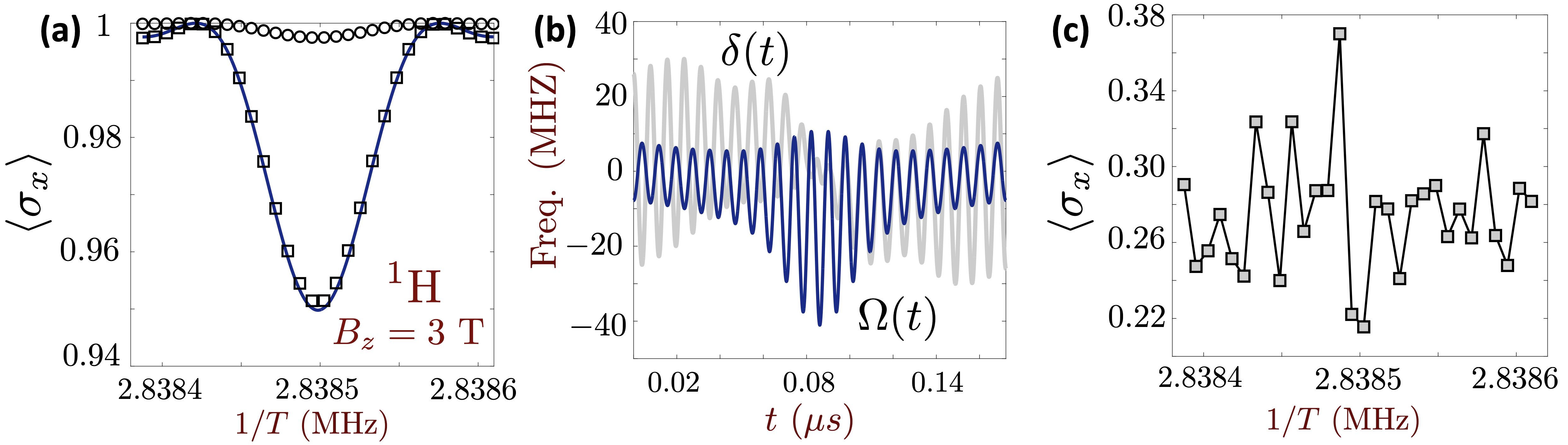}
\caption{Analogous to Fig.~\ref{sta_results} but interacting with a cluster of $^1$H nuclei. (b) Controls used for finding the spectrum (squares) in (a), in this case leading to a $\pi$ pulse duration of $\approx 0.17\ \mu$s. (c) Spectrum obtained with the extended $\pi$ pulses. We repeat the XY8 sequence 102 times, which leads to a final sequence time $\approx 0.14$ ms.}
\label{sta_results2}
\end{figure}
In Fig.~\ref{sta_results2} (a) we present the spectra that result of averaging the response of several NVs, each of them with a different detuning error, whilst they are all coupled to the same classical electromagnetic wave. Thus, we add to the system Hamiltonian the term $H_T = \Gamma S_z \cos(\omega_s t)$, where we employ $\Gamma = (2\pi)\times 28$ kHz in the simulations. This scenario describes, for instance, an NV ensemble  used as a detector for a $^1$H spin cluster out of the diamond sample~\cite{Aslam17}. As in the previous case, the ideal solid-blue curve in Fig.~\ref{sta_results2} (a) has been obtained by delivering instantaneous $\pi$ pulses, and in absence of control errors. In the same figure, the squares represent the signal obtained with our method, i.e. by using  the controls in Fig.~\ref{sta_results2} (b) and averaging the responses of of 10 NVs where the detuning error has been randomly taken from a Gaussian distribution centered at $\xi_\delta = 0$ and with a width of 1 MHz. More specifically, we have used the following values $\xi_\delta = (2\pi)\times[ 0.5376, 1.8338, -2.2588, 0.8622, 0.3188, -1.3076,\\ -0.4336, 0.3426, -2.7784, 2.1694]$ MHz, while the Rabi frequency deviation is $\xi_\Omega = 1\%$ for all cases.

One can observe that this average spectrum fully overlaps with the ideal NV response, which demonstrates the good performance of our method. The circles in Fig.~\ref{sta_results2} (a) denotes the signal obtained with top-hat $\pi$ pulses with a Rabi frequency  $\Omega_{\rm th} = (2\pi)\times 40$ MHz, i.e. equal to the maximum amplitude of $\Omega(t)$ in Fig.~\ref{sta_results2} (b).  Again, the signal-contrast produced by standard top-hat $\pi$ pulses is much lower than the one achieved by our method which further confirm the advantages of the latter. Finally, in Fig.~\ref{sta_results2} (c) we plot the average signal obtained with the $\pi$ pulses in Ref.~\cite{Casanova18}, and for the same errors in Fig.~\ref{sta_results2} (a). We can observe that the spectrum in Fig.~\ref{sta_results2} (c) cannot  offer any information regarding the scanned sample while, with our method, we can clearly observe a resonance peak that meets the ideal response leading to reliable identification. \\

In this chapter, we have explored how integrating Shortcuts to Adiabaticity (STA) techniques into the design of
$\pi$ pulses enhances the performance of control sequences for detecting high-frequency signals. This flexible approach enables the design of pulse shapes that not only improve coupling to the target signal but also offer robustness against detuning and Rabi frequency errors. We demonstrate the effectiveness of our technique for two scenarios at a high magnetic field of $B=3$ T.


\section{Electron-spin labels}
\label{chapter3}


\thiswatermark{\put(1,-280){\color{l-grey}\rule{70pt}{42pt}}
\put(70,-280){\color{grey}\rule{297pt}{42pt}}}

%
%

\vfill
\lettrine[lines=2, findent=3pt,nindent=0pt]{A}{} straightforward strategy to address challenges associated with high frequencies involves the use of ZZ-type interactions. These are interaction terms in the Hamiltonian that contain only spin operators in the $\hat z$ direction (i.e. in the direction of the externally applied magnetic field). These interactions commute with the Larmor terms, thereby eliminating the need for strong control drivings to compensate for frequency mismatches. This method, for instance, is employed in Double-Electron-Electron Resonance (DEER) \cite{Milov81}, where a dual control scheme effectively manages these interactions to measure distances between spin labels in macroscopic samples.

Elucidating molecular structure and dynamics through Electron Spin Resonance (ESR) by tagging specific sites with electron radicals is crucial for studying complex biomolecules. Traditional methods, however, require large sample sizes and high purity, rendering them ineffective for investigating individual molecules. In contrast, single NV centers can interact directly with electron tags within a single molecule, overcoming these limitations. This capability has already been demonstrated with the detection of a single nitroxide electron-spin label attached to a protein using an NV center \cite{Shi15}. Attaching multiple nitroxides to a single target macromolecule could enable the exploration of distances and potentially reveal insights into internal dynamics at room temperature using an NV center. However, this approach faces challenges as spin labels' resonance frequencies and interaction strengths are highly dependent on their orientation relative to the external magnetic field. Identifying these parameters is further complicated by inevitable molecular motion and environmental noise impacting both the NV and the electron-spin labels.

In this chapter, we conduct a theoretical analysis suggesting that a DEER-like pulse sequence, when applied to a single pair of electron-spin labels and a shallow NV, can detect single-molecule conformational transitions at room temperature by selecting the appropriate control and magnetic field values. Additionally, we develop a model that enables us to estimate not only the coupling but also the distance between labels using Bayesian inference. Interestingly, we find that the minor residual effect of tumbling, rather than being detrimental, actually facilitates the extraction of the distance between labels. The content of this chapter corresponds to the work developed in Ref. \cite{Munuera-Javaloy22}.

The organization of this chapter is as follows: In Section \ref{nitro:ham}, we introduce the system Hamiltonian. In Section \ref{nitro:approx_model}, we derive an approximate model to better understand the system's dynamics and prepare it for Bayesian analysis. Section \ref{nitro:sequence} details the employed pulse sequence. Section \ref{nitro:results} discusses the results of our numerical simulations. Finally, Section \ref{nitro:bayes} describes the Bayesian analysis performed on these results

\subsection{System Hamiltonian}\label{nitro:ham}
\begin{figure}[t]
\center
\hspace{0.0 cm}\includegraphics[width=1.0\columnwidth]{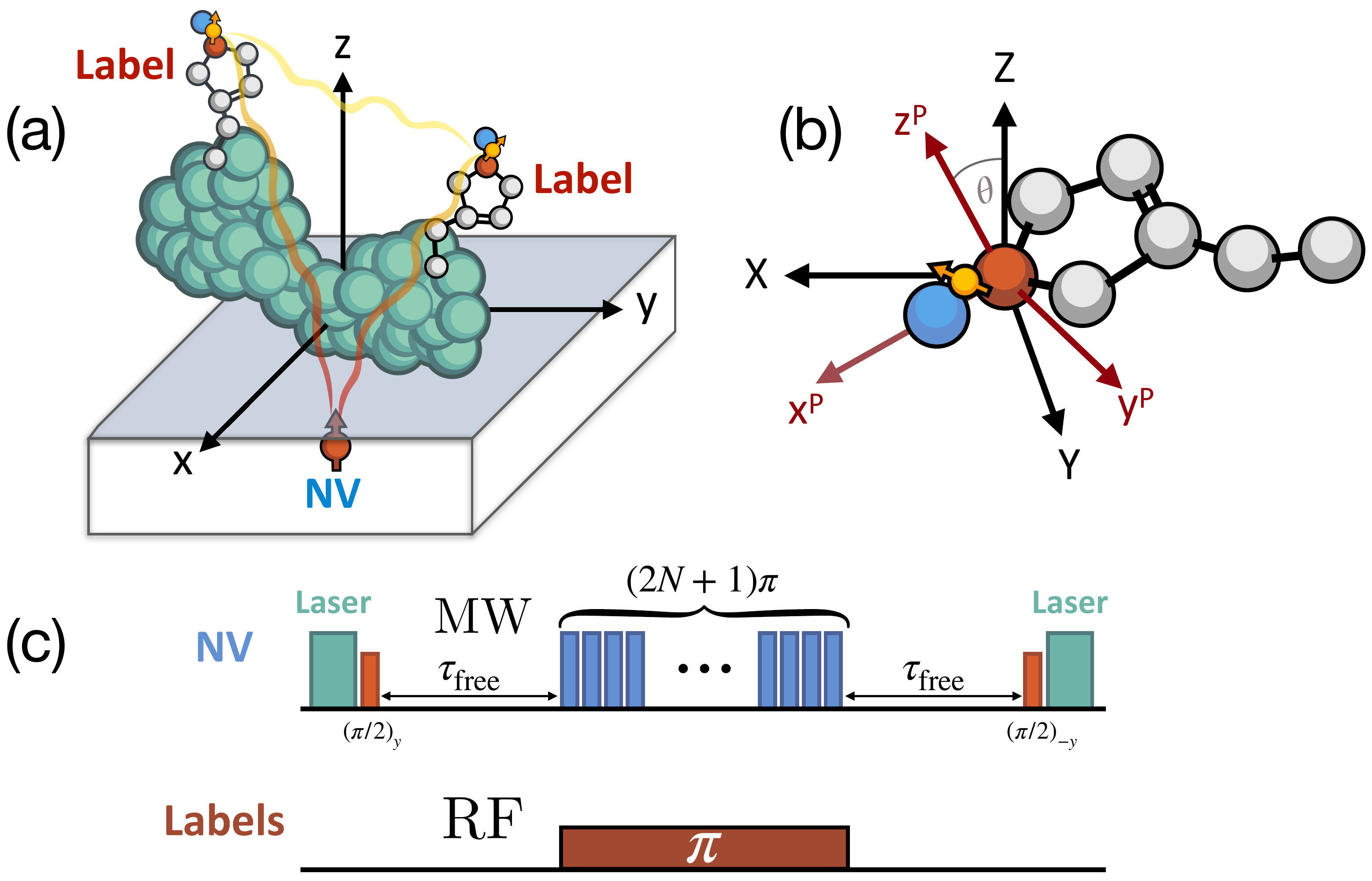}
\caption{(a) A protein with two attached labels is near a diamond surface that contains a shallow NV. (b) Nitroxide molecule including the electron spin (yellow). The principal axes are shown in red, while the laboratory frame is shown in black. The azimuth $\theta$ (the angle between the z principal axis and the z laboratory axis) determines the energy-transition branches of the label. (c) MW and RF driving scheme. This includes initialization and readout with laser and microwave pulses, two free evolution stages of duration $\tau_{\rm free}$, $2N+1$ MW $\pi$-pulses on the NV, and a simultaneous RF $\pi$-pulse on the labels. \label{fig:system}}
\end{figure}

The Hamiltonian of the NV, the two nitroxide electron-spin labels, and the microwave (MW) and radiofrequency (RF) driving fields reads
\begin{align}
\begin{aligned}
H = &\frac{1}{2}\left(\mathbb{I} + \sigma^z \right)\left(\vec{A}_1 \cdot \vec{J}_1+\vec{A}_2 \cdot \vec{J}_2\right) \\
&+ H_{n_1}+H_{n_2} + g_{12}\left[J_1^z J_2^z-\frac{1}{4}\left(J_1^+ J_2^- + J_1^- J_2^+ \right)\right] \\
&+\frac{\Omega_{\rm MW}}{2} \sigma^x+2 \Omega_{\rm RF} \left(J_1^x+J_2^x\right) \cos(\omega_{\rm RF} t). \label{eq:main}
\end{aligned}
\end{align}
Here, $\sigma^z$ is the Pauli $z$ operator for the NV two-level system and $\vec{J}_i$ are spin-$1/2$ operators for each label ($i=1,2$). The Hamiltonian of the $i$th nitroxide label is $H_{n_i}$ and the coupling of each label to the NV is mediated by the vectors $\vec{A}_i$. Moreover, $J_i^\pm = J_i^x \pm i J_i^y$ are electron-spin ladder operators and $g_{12}$ is the coupling constant between labels. The last line in Eq.~\eqref{eq:main} describes a MW field resonant with the NV, leading to the term $\frac{\Omega_{\rm MW}}{2} \sigma^x$ in a suitable interaction picture. In addition, an RF field of frequency $\omega_{\rm RF}$ and Rabi frequency $\Omega_{\rm RF}$ excites the electronic resonances of the nitroxide labels. Equation~\eqref{eq:main} is derived in Appendix~\ref{app:nitro_ham} and a sketch of the system is presented in Fig.~\ref{fig:system}(a).

The term $H_{n_i}$ models each nitroxide label as~\cite{Marsh15, Marsh19}
\begin{align}
\begin{aligned}
H_{n_i} = \mu_B B^z \hat{z}\cdot \mathbb{C}_i \cdot \vec{J}_i+ \gamma_N B^z I_i^z+\vec{I}_i \cdot \mathbb{Q}_i \cdot \vec{I}_i+\vec{J}_i \cdot \mathbb{G}_i \cdot \vec{I}_i, \label{eq:nitro}
\end{aligned}
\end{align}
where $\mu_B$ is the Bohr magneton, $B^z$ is a magnetic field applied along the $z$ axis, $\gamma_{\rm N}$ is the nuclear gyromagnetic ratio equal to $2\pi \times 3.077$\ kHz/mT for $^{14}$N and to $2\pi \times -4.316$\ kHz/mT for $^{15}$N, $\vec{I}_i$ is the nuclear spin operator for the $i$th nitroxide label, and $\mathbb{C}_i$, $\mathbb{Q}_i$, and $\mathbb{G}_i$ are respectively the Land\'e, quadrupolar, and electron-nucleus interaction tensors~\cite{Marsh19}. We note that $\vec{I}_i$ is a spin-1 (spin-$1/2$) operator when describing a $^{14}$N ($^{15}$N) nuclear spin. The components of the $\mathbb{C}_i$, $\mathbb{Q}_i$, and $\mathbb{G}_i$ tensors in a general reference frame depend on the frame's relative orientation with respect to the principal axes of the nitroxide, see Fig.~\ref{fig:system}(b).

In the principal frame, $\mathbb{C}_i$, $\mathbb{Q}_i$, and $\mathbb{G}_i$ are diagonal~\cite{Marsh19}. In particular, the Land\'e tensor in the principal frame is $\mathbb{C}_i^{(P)} = {\rm diag}\left(C^x,C^y,C^z\right)$, with $C^x\approx C^y\approx 2.007 \equiv C^\perp$ and $C^z \approx 2.002 \equiv C^\parallel$. The quadrupolar tensor for $^{14}$N is $\mathbb{Q}_i^{(P)} = {\rm diag}\left(Q^x,Q^y,Q^z\right)$, with $Q^x \approx 2\pi \times 1.26$ MHz, $Q^y \approx 2\pi \times 0.53$ MHz, and $Q^z \approx 2\pi \times -1.79$ MHz. The quadrupolar tensor vanishes for $^{15}$N. Finally, the tensor that mediates electron-nucleus interactions in each nitroxide is $\mathbb{G}_i^{(P)} = {\rm diag}\left(G^x,G^y,G^z\right)$, where $G^x\approx G^y\approx 2\pi \times 14.7 \ {\rm MHz} \equiv G^\perp$ and $G^z \approx 2\pi \times 101.4 \ {\rm MHz} \equiv G^\parallel$ for $^{14}$N, while $G^x\approx G^y\approx 2\pi \times 27 \ {\rm MHz} \equiv G^\perp$ and $G^z \approx 2\pi \times 141 \ {\rm MHz} \equiv G^\parallel$ for $^{15}$N~\cite{Marsh19}.

\subsection{Approximate model}\label{nitro:approx_model}
\begin{figure}[t!]
\hspace{-0.5cm}\includegraphics[width=0.9\columnwidth]{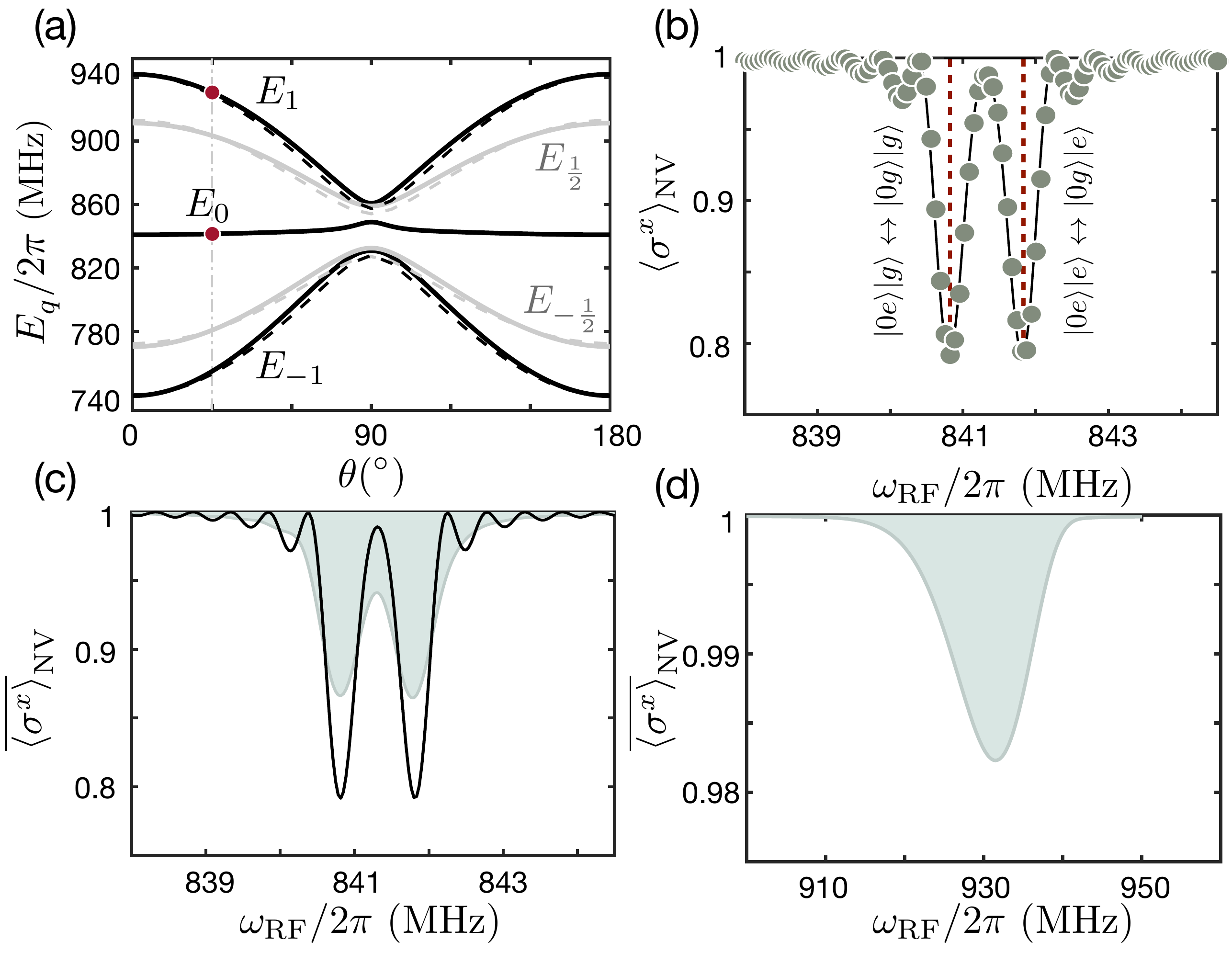}
\centering
\caption{(a) Energy-transition branches of an electron spin label at $B^z = 30$\ mT for $^{14}$N ($^{15}$N). Solid black (grey) lines show the transition energies as a function of the azimuth $\theta$ obtained via numerical diagonalization of Eq.~(\ref{eq:nitro}). Dashed black (grey) lines correspond to $E_{1,0,-1}$ in Eq.~(\ref{eq:trace1}) [$E_{1/2,-1/2}$ in Eq.~(\ref{eq:trace2})]. The vertical line highlights the case $\theta_i = 30^\circ$, with red dots indicating the transition energies relevant in the following plots. (b) NV spectrum $\langle\sigma^x\rangle_{\rm NV}$ as a function of the RF frequency applied near $E_0^1$. The spectrum was obtained by unitary propagation of Eq.~\eqref{eq:main} with $H_{n_i}$ given by Eq.~\eqref{eq:nitro} (black line) and by $E_0^i J_i^z$ (grey circles). The first nitroxide has azimuth $\theta_1 = 30^\circ$ while the second nitroxide has azimuth $\theta_2 = 91.7^\circ$. The splitting of the peaks is induced by the inter-label coupling $g_{12}\approx 2\pi\times 1$ MHz. The positions of the resonances are indicated by the dashed red lines. In (c, d), shaded areas show the average NV spectra $\overline{\langle\sigma^x\rangle}_{\rm NV}$ in the presence of tumbling of the first (near-resonant) nitroxide. Tumbling was mimicked by averaging over a random azimuth following a Gaussian distribution centered at the equilibrium value $\theta_{1,{\rm eq}}=30^\circ$. The orientation of the other nitroxide was kept fixed for simplicity. The standard deviation was chosen to be $\sigma_\theta = 6.25^\circ$. In (c), we observe that tumbling leads to a loss of contrast and to signal broadening near $E_0^1$. Nevertheless, the spectrum still shows two clearly separated peaks. For comparison, the case without tumbling in (b) is superimposed (black line). (d) Average NV spectrum $\overline{\langle\sigma^x\rangle}_{\rm NV}$ near $E_1^1$. Here, the energy splitting cannot be resolved due to nitroxide tumbling. Also note that tumbling severely reduces contrast. For that reason, the spectrum in the absence of tumbling has much higher contrast and is not shown. \label{fig:energies}}
\end{figure}

We obtain approximate nitroxide transition frequencies via a perturbative treatment of the nuclear degrees of freedom in Eq.~(\ref{eq:nitro}), see Appendix~\ref{app:nitro_simple}. For a nitroxide hosting
$^{14}$N, we find

\begin{align}
\begin{aligned}
H_{n_i} \approx \left[ E_1^i |\widetilde{1}\rangle\langle \widetilde{1}|_i + E_0^i |\widetilde{0}\rangle\langle \widetilde{0}|_i + E_{-1}^i |-\widetilde 1\rangle\langle -\widetilde{1}|_i \right] J_i^z. \label{eq:trace1}
\end{aligned}
\end{align}
The three {\it energy-transition branches} exhibit transition energies given by
\begin{eqnarray*}
E_{1,-1}^i&=& \mu_B B^z C(\theta_i) \pm \frac{1}{\sqrt{2}} \sqrt{\left[(G^{\parallel})^2- (G^{\perp})^2\right] \cos (2 \theta_i )+(G^{\perp})^2+(G^{\parallel})^2},\\
E_0^i&=& \mu_B B^z C(\theta_i) + \frac{1}{2 \mu_B B^z C(\theta_i)} \frac{2 (G^{\perp} G^{\parallel})^2 +\left[(G^{\perp})^4-(G^{\perp} G^{\parallel})^2\right] \sin^2(\theta_i)}{(G^{\perp})^2 \sin^2(\theta_i)+(G^{\parallel})^2 \cos^2(\theta_i)}.
 \end{eqnarray*}
Here, $\theta_i$ is the azimuth angle between the applied magnetic field and the principal $z$ axis of the nitroxide [see Fig. \ref{fig:system}(b)], $C(\theta_i) =\frac{1}{2} [(C^\parallel - C^\perp) \cos (2 \theta_i)+ C^\perp+ C^\parallel]$, and $| \widetilde{1} \rangle_i$, $| \widetilde{0} \rangle_i$, and $| -\widetilde{1} \rangle_i$ are the states of the $i$th nitroxide nucleus dressed by the hyperfine interaction with the nitroxide electron, see Appendix~\ref{app:nitro_14n} for a detailed derivation. Similarly, for a nitroxide hosting $^{15}$N, we find
\begin{align}
\begin{aligned}
H_{n_i} \approx \left[ E_{1/2}^i |\widetilde{1/2}\rangle\langle \widetilde{1/2}|_i + E_{-1/2}^i |-\widetilde{1/2}\rangle\langle -\widetilde{1/2}|_i \right] J_i^z. \label{eq:trace2}
\end{aligned}
\end{align}
There are now two energy-transition branches with transition energies given by (see Appendix~\ref{app:nitro_15n}). \begin{equation}
E_{1/2,-1/2}^i=\mu_B B^z C(\theta_i) \pm \frac{1}{2\sqrt{2}}\sqrt{(G^\perp)^2+(G^\parallel)^2+\left[(G^\parallel)^2-(G^\perp)^2\right]\cos(2\theta_i)}.
\end{equation}
Fig.~\ref{fig:energies}(a) compares the energy-transition branches in Eqs.~(\ref{eq:trace1}) and (\ref{eq:trace2}) (dashed lines) with numerical diagonalization of Eq.~(\ref{eq:nitro}) (solid lines). Note that since we will target the $E_0^i$ branch for label detection, we have further developed its functional form to include second-order corrections in the electron-nitrogen coupling. Fig.~\ref{fig:energies}(a) focuses on a single nitroxide (the index $i=1$ is removed for clarity) and shows that our expression for $E_0$ is in excellent agreement with numerical diagonalization.
The inter-label coupling leads to an additional splitting $\propto g_{12}$ of the energy-transition branches in Eqs.~(\ref{eq:trace1}, \ref{eq:trace2}). This term typically simplifies as 
\begin{equation}
g_{12}\left[J_1^z J_2^z-\frac{1}{4}\left(J_1^+ J_2^- + J_1^- J_2^+ \right)\right]\longrightarrow g_{12} J_1^z J_2^z, 
\end{equation}
since labels with different orientations have different resonance frequencies, thus eliminating the flip-flop contribution. Note that while this simplification is frequently valid when using two $^{14}$N, it is guaranteed to be valid when using nitroxides hosting different nitrogen isotopes since $E_0$ (for $^{14}$N) differs from $E_{1/2,-1/2}$ (for $^{15}$N) for any value of the azimuth (see Fig.~\ref{fig:energies}(a)]) For $^{14}$N, the available resonances of, e.g., the first label are then determined by the modified nitroxide Hamiltonian $H'_{n_1} = H_{n_1}+g_{12} J_1^z J_2^z$
\begin{align}
\begin{aligned}
H'_{n_1} &= \left[ E_1^1 |\widetilde{1}\rangle\langle \widetilde{1}|_1 + E_0^1 |\widetilde{0}\rangle\langle \widetilde{0}|_1 + E_{-1}^1 |-\widetilde 1\rangle\langle -\widetilde{1}|_1 \right] J_1^z + g_{12} J_1^z J_2^z \\
&=\sum_{q=1,0,-1}\left[\left(E_q^1 + \frac{g_{12}}{2}\right) |\widetilde{q} e\rangle\langle \widetilde{q} e| + \left(E_q^1 - \frac{g_{12}}{2}\right) |\widetilde{q} g\rangle\langle \widetilde{q} g|\right] J_1^z.
\end{aligned}
\end{align}
That is, any energy-transition branch $E_q^1$ (corresponding to the $|\widetilde{q}\rangle$ nuclear state of the first nitroxide) splits as
\begin{align}
\begin{aligned}
E_q^1 \longrightarrow \left(E_q^1 + \frac{g_{12}}{2}, E_q^1 - \frac{g_{12}}{2}\right) \label{eq:interLabelSplitting}
\end{aligned}
\end{align}
depending on the electronic state $|e\rangle$, $|g\rangle$ of the second nitroxide. Appendix~\ref{app:nitro_14n} includes a complete energy diagram of an electron spin in a nitroxide.

\subsection{Sequence}\label{nitro:sequence}
Our DEER protocol simultaneously delivers MW and RF pulses to detect energy shifts in the NV spectrum due to inter-label coupling. The sequence contains two free-evolution stages of duration $\tau_{\rm free}$ separated by a driving stage (see Fig.~\ref{fig:system}(c)). The NV is prepared in $\ket{+}$ and the nitroxides are assumed to be in a fully thermalized state. During free evolution, the NV accumulates a phase due to the term $\sigma^z \left(\vec{A}_1 \cdot \vec{J}_1+\vec{A}_2 \cdot \vec{J}_2\right)$ in Eq.~\eqref{eq:main}. This term can be approximated as $\sigma^z \left(A_1^z J_1^z+A_2^z J_2^z\right)$ in the RWA since the $J_i^{x,y}$ contributions are suppressed by the large electronic precession frequencies of the nitroxides. Indeed, the energies $E_{1,0,-1}^i$ reach several hundreds of MHz even for moderate values of $B^z$ (we use $B^z = 30$\ mT in our simulations). By contrast, the NV-label coupling takes values of hundreds of kHz for typical NV-label distances of several nanometers.

During MW/RF irradiation, the NV undergoes an overall flip $\sigma^z \rightarrow -\sigma^z$ via $2N+1$ contiguous $\pi$-pulses arising from the same continuous MW drive. At the same time, a single weaker RF $\pi$-pulse on the labels induces $J_i^z\rightarrow-J_i^z$ if it is near-resonant with an electronic transition (i.e., with $E_{1,0,-1}^i$ for $^{14}$N or with $E_{1/2,-1/2}^i$ for $^{15}$N). 

The simultaneous flipping of the NV and spin label ensures a constructive and coherent accumulation of the NV phase during free evolution. Meanwhile, the use of $2N+1$ $\pi$-pulses on the NV ensures that the Rabi frequencies of the two drives are well separated, which effectively averages out spurious NV-label interactions during irradiation. Consequently, scanning the RF frequency near nitroxide transitions yields clean resonance peaks in the NV response. As long as $\Omega_{\rm RF}< g_{12}$, the resonance peaks are split by $g_{12}$. Fig.~\ref{fig:energies}(b) shows the NV spectrum (the expectation value of $\sigma^x$) associated with the transitions $|0g\rangle |g\rangle \leftrightarrow |0e\rangle |g\rangle$
and $|0g\rangle |e\rangle \leftrightarrow |0e\rangle |e\rangle$. Each resonance is split by $\sim g_{12}$ as expected. The solid black line in Fig.~\ref{fig:energies}(b) was obtained by numerically propagating Eq.~\eqref{eq:main} with $H_{n_i}$ given by Eq.~\eqref{eq:nitro}, while grey circles were obtained assuming the simplified $H_{n_i}$ in Eq.~\eqref{eq:trace1}. The two spectra are in excellent agreement. Simulations in Fig.~\ref{fig:energies}(b) assume $B^z = 30$\ mT, which is in the range of values that ensure the stability of the $E_0^i$ energy-transition branch as discussed in the next paragraph. Moreover, the parameters of the pulse sequence are $\tau_{\rm free} = 1.3\ \mu{\rm s}$, $\Omega _{\rm RF} = 2\pi \times 250$\ kHz, and $\Omega_{\rm MW} = 31\times \Omega_{\rm RF}$, for a total sequence time of $4.6\ \mu$s. In addition, the labels are separated from the NV by the distances $d_1 \approx 6.9$ nm and $d_2 \approx 7.3$ nm, leading to $\vec{A}_1 \approx 2\pi \times \left(128, -132, -223\right)$ kHz and $\vec{A}_2 \approx 2\pi \times \left(-22, -16, -264\right)$ kHz, while the distance between labels is $d_{12} \approx 3.26$ nm.

The expressions for the energy-transition branches in Eqs.~(\ref{eq:trace1}, \ref{eq:trace2}) reveal a dependence on the azimuth angle, i.e., $E_q^i\equiv E_q^i(\theta_i)$. Consequently, unavoidable protein motion during the irradiation stage leads to a distortion in the spectrum and to a difficult interpretation of the signal. However, it is important to note the distinct nature of the $E_0^i$ branch, which shows a much weaker dependence on $\theta_i$ than $E_{1,-1}^i$. This makes the $E_0^i$ branch particularly well suited for robust detection of the energy splitting [see Figs.~\ref{fig:energies}(c,d)]. To maximize the robustness, we chose the magnetic field to be large enough to energetically suppress the effect of the anisotropic hyperfine interaction, but small enough that the anisotropy of the Land{\'e} tensor does not become significant, see Appendix~\ref{app:nitro_14n}.

\subsection{Numerical simulations}\label{nitro:results}
We now illustrate our method in realistic ambient conditions including decoherence leading to NV dephasing with decoherence time $T_2 = 20\ {\rm\mu s}$ (for a 4 nm NV depth~\cite{Shi15}), electron-spin label relaxation with $T_1 = 4\ {\rm \mu s}$~\cite{Shi15}, and molecular tumbling. The dissipative model used in this section is described in Appendix~\ref{app:nitro_dissi}. and captures the main decoherence mechanisms identified in Ref.~\cite{Shi15}. Note that for the specific protocol considered here, decoherence mainly manifests as a reduced line contrast and not as an increased linewidth. This is because we keep the sequence duration fixed and sweep $\omega_{\rm RF}$ to find the resonances, as opposed to varying the sequence length at fixed $\omega_{\rm RF}$ to observe an echo signal. The molecular tumbling is modelled as a random rigid rotation of both nitroxides around an axis parallel to the laboratory $x$ axis, see Appendix~\ref{app:nitro_tumbl}. Our simulations use a rotation angle $\delta$ that is Gaussian-distributed with standard deviation $\sigma_\delta = 6.25^\circ$. This is somewhat smaller but comparable to the fluctuations observed in Ref.~\cite{Shi15}. Note that immobilizing proteins in more rigid matrices~\cite{Meyer15,Lira16} or attaching them to the diamond surface~\cite{Lovchinsky16} through multiple rigid covalent bonds could further reduce tumbling. 

\begin{figure}[t!]
\hspace{-0.5 cm}\includegraphics[width=.9\columnwidth]{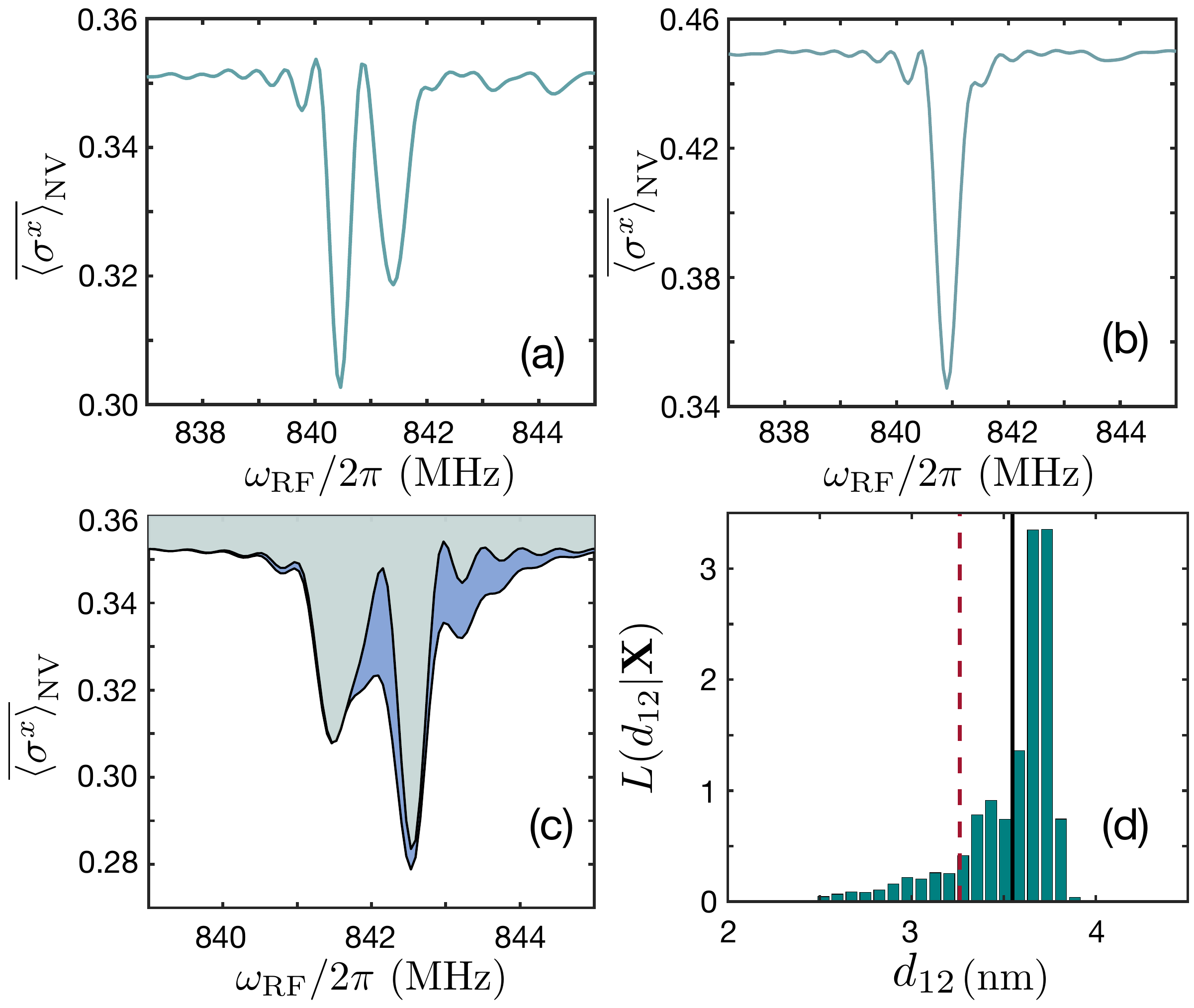}
\centering
\caption{(a) Simulated average NV spectrum $\overline{\langle\sigma^x\rangle}_{\rm NV}$. Two resonance peaks due to $g_{12}$ are observed. (b) Similar to (a) but with lower $g_{12}$. (c) Simulated average spectrum $\overline{\langle\sigma^x\rangle}_{\rm NV}$ for overlapping label resonances with two $^{14}$N isotopes (blue shaded area) and with distinct $^{14}$N and $^{15}$N isotopes (grey shaded area). (d) Marginal distribution $L(d_{12}|{\bf X})$ of the inter-label distance for simulated data acquisition using the spectrum shown in (a). The ``true'' value of the distance and its posterior expectation, $3.26$ nm and $3.54(25)$ nm, are indicated with vertical lines. The error is the standard deviation of the marginal posterior. \label{fig:tilting}}
\end{figure}

The resulting tumbling-averaged spectra $\overline{\langle\sigma^x\rangle}_{\rm NV}$ are shown in Figs.~\ref{fig:tilting}~(a,b,c). The parameters $\tau_{\rm free}$, $\Omega_{\rm MW}$, $\Omega_{\rm RF}$, and $\vec{A}_i$ are the same as in Fig.~\ref{fig:energies}. The details of the equilibrium nitroxide configurations are given in Appendix~\ref{app:nitro_tumbl}. In Fig.~\ref{fig:tilting}(a), the labels are separated by $d_{12} \approx 3.26$ nm and the inter-label coupling is $g_{12} \approx 2 \pi \times 1$ MHz. When the frequency of the RF field is set close to $E_0^1$, we clearly identify two resonance peaks in spite of tumbling. We have verified that the two resonances are still visible even if we choose $\sigma_\delta$ to be four times larger. Since $g_{12}\propto d_{12}^{-3}$, a change in the relative position of the labels significantly modifies the observed spectrum. This is shown in Fig.~\ref{fig:tilting}(b) where the second label was displaced such that $d_{12} \approx 4.03$ nm, leading to the disappearance of the splitting. This change in the spectrum certifies a change in the relative position between labels and, by extension, a conformational change in the host molecule.

So far, all simulations were performed for nitroxides hosting $^{14}$N and having distinct transition energies. If both nitroxides have similar transition energies (e.g., due to similar azimuths), the spectra of the two nitroxides overlap and the inter-label flip-flop terms cannot be neglected. This complicates the interpretation of the spectrum. As shown in Fig.~\ref{fig:tilting}(c), this problem is avoided by using distinct $^{14}$N and $^{15}$N isotopes in each nitroxide. In this case, the $E_0$ branch of the $^{14}$N nitroxide never overlaps with the branches of the $^{15}$N nitroxide and flip-flop terms can be safely neglected. As a result, the interpretation of the spectrum is much simpler. We emphasize that the purpose of using such ``orthogonal labels'' is not merely to resolve their distinct spectral signatures~\cite{Galazzo22}: rather, it is to simplify the form of the inter-label interaction itself for all label orientations.

\subsection{Bayesian inference}\label{nitro:bayes}
Finally, we show how to infer the inter-label distance $d_{12}$ under realistic ambient conditions. We first simulate the experimental acquisition of $\overline{\langle\sigma^x \rangle}_{\rm NV}$. The simulated dataset has the form ${\bf X}=\{ X_1,\ldots,X_M\}$, where $X_j$ is an experimental estimate of the probability of measuring $\sigma^x = + 1$ after $N_m$ measurements at RF frequency $\omega_{{\rm RF},j}$. Second, we use a simplified model to efficiently extract information from ${\bf X}$. The model captures the relevant features of the tumbling-averaged spectrum $\overline{\mathcal{S}}(\omega_{\rm RF})$. More precisely, we derive the approximate expression 
\begin{equation}
\mathcal{S}(\omega_{\rm RF},\theta,\beta)=\mathcal{S}_0 - \sum_{s=+,-} \mathcal{C}_s \left[\frac{\Omega_{\rm RF}}{\Omega_s(\omega_{\rm RF},\theta,\beta)}\right]^2 \sin^2\left[\frac{\pi}{2}\frac{\Omega_s(\omega_{\rm RF},\theta,\beta)}{\Omega_{\rm RF}}\right],
\end{equation}
for a specific nitroxide azimuth $\theta$ and a specific angle $\beta$ between the magnetic field and the line joining the nitroxides. Here, $\Omega_\pm^2(\omega_{\rm RF},\theta,\beta) = \Omega_{\rm RF}^2 + \left[\omega_{\rm RF}-E_0(\theta) \pm g_{12}(\beta)/2 \right]^2$ and $\mathcal{S}_0$ and $\mathcal{C}_\pm$ are parameters that adjust the baseline and contrast, respectively (see Appendix~\ref{app:nitro_infer}). Moreover, $E_0(\theta)$ is the ${}^{14}$N energy-transition branch, while $g_{12}(\beta)\propto d_{12}^{-3}(1-3\cos^2\beta)$ is the inter-label coupling. Both $\theta$ and $\beta$ depend on the tumbling angle $\delta$. Averaging $\mathcal{S}[\omega_{\rm RF},\theta(\delta),\beta(\delta)]$ over a Gaussian distribution for $\delta$ gives the tumbling-averaged spectrum $\overline{\mathcal{S}}(\omega_{\rm RF})$. Assuming that the baseline $\mathcal{S}_0$ is known, our model contains eight free parameters denoted by ${\bf V}$, see Appendix~\ref{app:nitro_infer}. These include $d_{12}$ and the unknown standard deviation $\sigma_\delta$ of the tumbling. It must be emphasized that the light tumbling dependence of the $E_0$ branch enables the determination of the inter-label distance. 

Indeed, a simple measurement of the line splitting $g_{12} \propto d_{12}^{-3}(1-3\cos^2\beta)$ cannot give independent access to the distance $d_{12}$ and the angle $\beta$. This is because $\beta$ is unknown {\it a priori}. However, light tumbling fluctuations allow the NV to probe different geometric configurations of the nitroxides. This results in a small distortion of the line shape that yields information beyond that contained in the splitting $g_{12}$. This in turn enables the independent extraction of the distance $d_{12}$ and of the angle $\beta$. With the help of our model, we can therefore obtain the posterior probability of the parameters ${\bf V}$ using Markov Chain Monte Carlo sampling~\cite{Gilks96}. Assuming a uniform prior for ${\bf V}$, the posterior probability for ${\bf V}$ is 
\begin{equation}
L({\bf V}|{\bf X})=\prod_{j=1,M}e^{-[X_j-(\overline{\mathcal{S}}(\omega_{{\rm RF},j})+1)/2]^2/2\sigma_m^2}/\sqrt{2\pi \sigma_m^2}. 
\end{equation}
Here, the noise variance $\sigma_m^2 \approx \left(1-\overline{\langle\sigma^x\rangle}_{\rm NV}^2\right)/4 N_m$ is assumed to be approximately constant and known. In Fig.~\ref{fig:tilting}(d) we show the resulting marginal $L(d_{12}|{\bf X})$ of $d_{12}$ for a dataset ${\bf X}$ simulated from the spectrum shown in Fig.~\ref{fig:tilting}(a). Here, ${\bf X}$ was obtained by taking $N_m = 2\times 10^4$ ideal measurements for each of $M=25$ frequencies ranging from $\omega_{\rm RF}/2\pi=839$ to $843$ MHz (we estimate that the same accuracy is achieved with $N_m \approx 5\times10^5$ for imperfect NV detection efficiency~\cite{Wan18}, see Appendix~\ref{app:nitro_dis}). The expectation of the marginal posterior is $d_{12}=3.54(25)$ nm, close to the ``real'' value $3.26$ nm (see Fig.~\ref{fig:tilting}(d)).


Our results open many interesting avenues for future investigation. In particular, it would be of great interest to extend our scheme to molecules with more than two attached spin labels. This would yield a stronger NV response, leading to a more detailed observation of conformational changes as well as to an enhancement in the range of inter-label distances to be estimated. The scheme could also be improved through the use of multiple NVs to better triangulate the label positions. In addition, our analytical understanding of the NV response could enable fast Bayesian inference of molecular dynamical properties. Our findings open up new possibilities for using magnetic resonance tools to observe the conformational dynamics of individual proteins.


\section{Dangling bonds}
\label{sec4}
\label{chapter4}


\thiswatermark{\put(1,-280){\color{l-grey}\rule{70pt}{42pt}}
\put(70,-280){\color{grey}\rule{297pt}{42pt}}}

%
%

\vfill
\lettrine[lines=2, findent=3pt,nindent=0pt]{A}{} fundamental aspect in NMR protocols is the level of spin polarization within the sample, as the strength of the NMR signal (thus, the sensitivity of each NMR protocol) is proportional to this factor. However, at room temperature --and even under large magnetic fields of 3 T-- the nuclear spin polarization is weak (typically about $10^{-5}$). Dynamical nuclear polarization (DNP) protocols address this issue by transferring the high polarization of electron spins—due to their gyromagnetic ratio being, typically,  three orders of magnitude larger than that of nuclei, see Eq.~\eqref{eq:fun_thermal}—to nuclear spins in the sample, thus significantly enhancing the NMR signal strength.

The NV center in diamond is a promising polarization source for DNP applications due to its easy polarization above 90\% at room temperature via green laser irradiation \cite{Waldherr11}. Distributing NV polarization among $^{13}$C nuclei within the diamond lattice has already been achieved \cite{London13}, while transferring this polarization out of the diamond present numerous challenges. Among these, the physical distance between the NV centers and the external spins substantially slows the polarization transfer, and electron spins on the diamond surface can interfere with the polarization process. 

In this chapter, we propose a protocol that addresses both challenges by utilizing surface electrons as polarization repeaters. This method enhances polarization transfer to surface spins, referred to as dangling bonds, by leveraging the significantly larger gyromagnetic ratio of electrons. Traditional polarization protocols, such as PulsePol \cite{Schwartz18}, struggle to transfer polarization to target spins with high-frequency Larmor precession because of the high-frequency problem, limiting their ability to polarize electrons even at moderate magnetic fields. To overcome these limitations, we have developed a sequence that effectively utilizes ZZ-interactions to facilitate polarization transfer. The content of this chapter is based on the work published in Ref. \cite{Espinos24}.

The organization of this chapter is as follows: In Section \ref{sec:double_protocol}, we analyze the dynamics of the protocol. Section \ref{section:II} explores the simultaneous transfer of polarization from NVs to dangling bonds to nuclear spins. Finally, Section \ref{sec:results} presents a numerical analysis of the method performance.
\subsection{Dual-channel protocol}\label{sec:double_protocol}
\begin{figure*}[t!]
    \centering
    \includegraphics[width=1\linewidth]{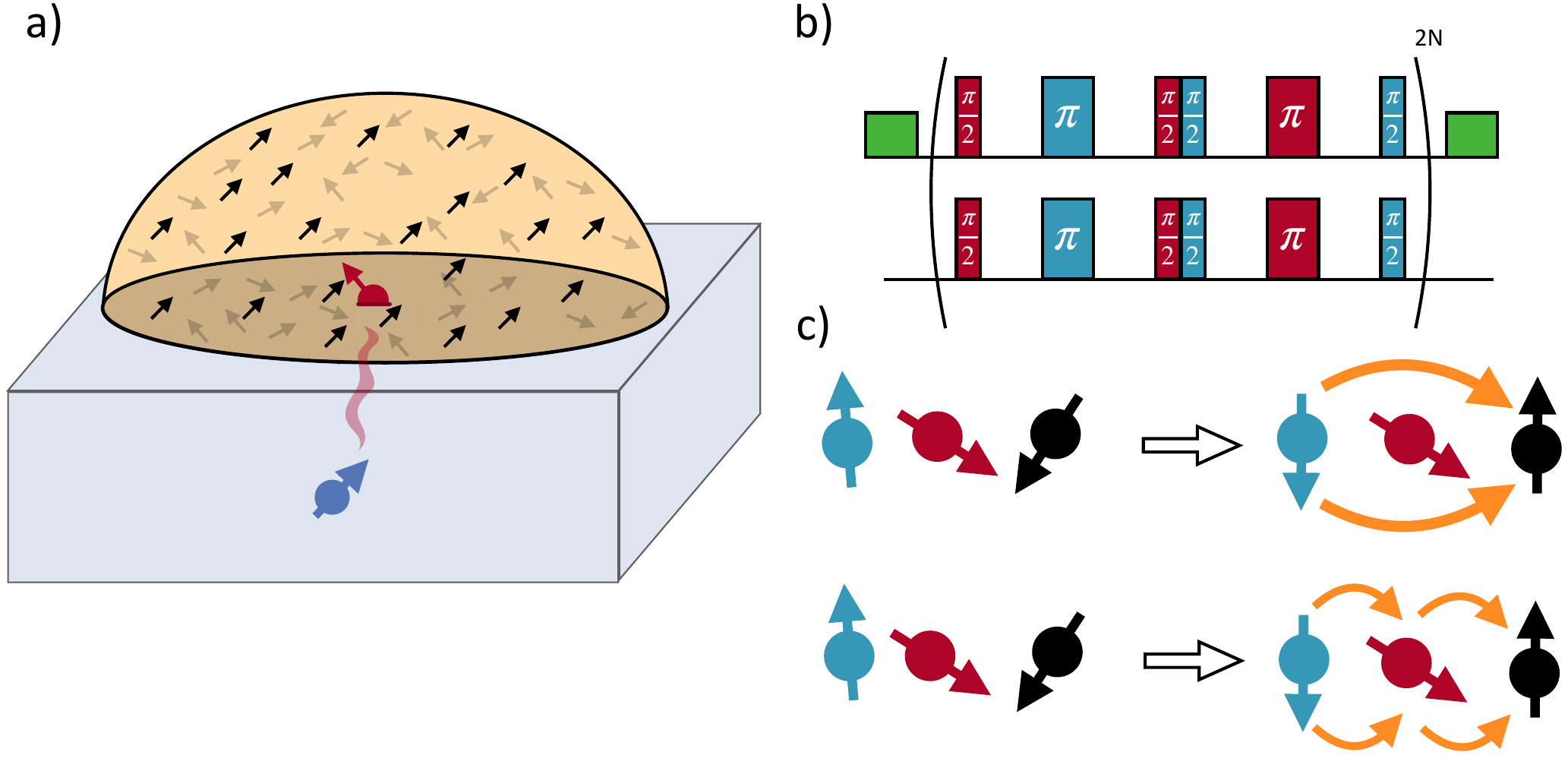}
\caption{Outline of the protocol. (a) The NV center transfers its polarization via an intermediary electron located at the diamond surface. (b) Diagram of the MW pulse sequence applied to both the NV and the electron spin for the polarization transfer. (c) Comparison between the traditional PulsePol method and our approach. In the latter, the electron coherently interacts with both the NV and adjacent nuclear spins to facilitate the polarization process. Utilizing the electron as a mediator enables faster transfer due to larger coupling, leading to accelerated protocol execution.}
  \label{fig:sequence}
\end{figure*}
We examine the Hamiltonian for an NV center interacting with a surface electron spin, both subject to external  MW drivings. Under the assumption of a ZZ-type dipolar interaction, as simplified by the RWA, the Hamiltonian is
\begin{equation}\label{eq:H_RWA}
H_T = D S_{z}^2 - \gamma_{e} B S_{z} - \gamma_{e} B J_{z} + A_z S_z J_z + H_{D},
\end{equation}
where $A_z$ represents the coupling strength, $J_i$ is the $i$th electron spin operator and $H_D$ are the drivings. In a rotating frame with respect to $H_0 = D S_{z}^2 - \gamma_{e} B S_{z} - \gamma_{e} B J_{z}$ and assuming resonant drivings, the Hamiltonian becomes
\begin{equation}\label{eq:H_RWA2} 
H_T = \frac{A_z}{2} \left(\sigma_z J_z+J_z\right) + \Omega_1(t) \frac{\sigma_\phi}{2}+ \Omega_2(t) J_\phi,
\end{equation}
where the $\ket{-1}\bra{-1}$ term is omitted from the NV subspace, applying the transformation $S_z\rightarrow\frac{\left(\sigma_z+\mathbb{I}\right)}{2}$, with $\sigma_z$ representing a Pauli operator acting on the reduced subspace. This exclusion is justified as the $\ket{-1}$ state does not get populated, given that the driving detuning is assumed to be several orders of magnitude greater than the driving strength, effectively suppressing it through the RWA. In this Hamiltonian, $\Omega_1(t)$ and $\Omega_2(t)$ are the time-dependent Rabi frequencies of the MW driving fields acting on the NV center and the surface electron spin, respectively. Our polarization protocol converts the ZZ interaction described in Eq.~\eqref{eq:H_RWA} into a flip-flop interaction, comprising the sum of $\sigma_{x}J_{x}$ and $\sigma_{y}J_{y}$ interactions. This is implemented by a sequence block executed simultaneously on both the NV center and the electron, as illustrated in Fig. \ref{fig:sequence}(b). The sequence is specified by
\begin{equation}\label{eq:sequence}
    \left[\left(\frac{\pi}{2}\right)_{Y}\rule{10pt}{0.1pt}\left(\pi\right)_{-{X}}\rule{10pt}{0.1pt}\left(\frac{\pi}{2}\right)_{Y}\left(\frac{\pi}{2}\right)_{X}\rule{10pt}{0.1pt}\left(\pi\right)_{Y}\rule{10pt}{0.1pt}\left(\frac{\pi}{2}\right)_{X}\right]^{2},
\end{equation}
where $\left(\phi\right){\pm {X,Y}}$ represents $\phi$-pulses around the x/y axes, with the pulse duration set by $t'_{1,2}=\phi/\Omega_{1,2}$. Importantly, and unlike the PulsePol sequence applied solely to one component, the interval between pulses here is entirely flexible ($\tau/4$ in Fig. \ref{fig:sequence}(b)), allowing for an arbitrary free evolution time of both the NV and the electron. Our sequence is organized into two sections. In each section, $\pi/2$-pulses are simultaneously applied to both the NV and the electron, converting the ZZ-interaction into either an XX or a YY interaction. A $\pi$-pulse, placed in the middle of each block, eliminates the $\frac{A_z}{2}J_{z}$ component from Hamiltonian~\eqref{eq:H_RWA2} and compensates for other detuning effects that arise during free evolution, such as strain or static noise. Consequently, the evolution under Hamiltonian~\eqref{eq:H_RWA2} can be reformulated as detailed in Section S2 of the Supplementary Material.
\begin{equation}\label{eq:evolution_operator}
    U_\text{seq} = \exp\left(-i\frac{\tau}{2}\frac{A_z}{2}\sigma_{x}J_{x}\right)\exp\left(-i\frac{\tau}{2}\frac{A_z}{2}\sigma_{y}J_{y}\right),
\end{equation}
where $\tau$ denotes the duration of one sequence cycle. When only one NV center and one electron are present, or if $\tau$ is significantly small compared to $1/A_z$ (i.e., $\tau \ll 1/A_z$), the evolution of the system simplifies to:
\begin{equation}\label{eq:evolution_operator2}
U_\text{seq}= \exp\left[-i\frac{\tau}{2}\frac{A_z}{2}\left(\sigma_{x}J_{x}+\sigma_{y}J_{y}\right)\right].
\end{equation}

In a configuration where the NV starts polarized in the $\ket{0}$ state and the electron is in a thermal state, the polarization $P$ after a sequence is determined by:
\begin{equation}\label{eq:pol1}
P(\tau) = \text{Tr}\left[U_\text{seq} \rho(0) U_\text{seq}^\dagger 2J_{z}\right] = -\sin^2\left(\frac{A_z \tau}{4}\right).
\end{equation}
Here, polarization $P$ is calculated as $P = p_\uparrow - p_\downarrow$, where $p_\uparrow$ and $p_\downarrow$ represent the probabilities of the electron spin being in the $\ket{\uparrow}_{e}$ and $\ket{\downarrow}_{e}$ states, respectively. These states are the eigenstates of the $J_{z}$ operator. 

\subsection{Polarization transfer to an external nucleus}\label{section:II}

The polarization transfer method outlined in the previous section is effective for any target particle, regardless of its Larmor frequency. This method is robust, as we will see in the next section, while obviating the need for specific resonance conditions in the interpulse spacing. This flexibility allows for the timing to be adjusted such that the surface electron spin couples with the nuclear spin, facilitating a two-stage polarization process. This protocol is depicted in Figure \ref{fig:sequence}. Due to its location on the diamond surface, the electron exhibits stronger coupling with external nuclei compared to an NV center embedded within the diamond lattice. The incorporation of flexible pulse spacing from the previous protocol, combined with the PulsePol resonance condition, enables simultaneous spin state transfers from the NV center to the electron and from the electron to the external nucleus in a single application of our sequence. This results in an improved polarization transfer rate, outperforming direct transfers from the NV center to the external nucleus.

The system under consideration is depicted in Fig.\ref{fig:sequence}(a). The governing Hamiltonian including the Zeeman term for the nucleus and the dipole coupling between the electron and the nucleus is: 

\begin{eqnarray}
\label{eq:H2}
H_T &=& D S_{z}^2 - \gamma_{e} B S_{z} - \gamma_{e} B J_{z} + \gamma_{n} B I_{z} + \\
&&+ A_z S_z J_z + J_z \vec{G}\cdot \vec{I} + H_{D}. \nonumber
\end{eqnarray}

Here, $\vec{I}$ represents the spin--$\frac{1}{2}$ operator of the external nucleus, $\gamma_{n}$ is the gyromagnetic factor of the nucleus, and $G$ is the coupling vector describing the interaction between the electron and the nucleus. The interaction between the NV center and the nucleus is omitted due to the substantially larger coupling between the electron and the nucleus, given their closer proximity. In particular, if we assume the nucleus is located 1 nm from the electron and 4 nm from the NV center, the interaction between the nucleus and the electron would be 64 times stronger than that between the nucleus and the NV center, given the same relative orientation. The drivings $H_{D}$ are applied over the electron and the NV, as in the previous case. Choosing the condition
\begin{equation}\label{eq:tau}
\tau = \frac{n\pi}{\gamma_{n} B},
\end{equation}
where $n$ is an odd number, we get an effective flip-flop dynamics between the electron and the nucleus $(H_{\rm flip-flop} \propto J_xI_x+J_yI_y)$.
Given that the gyromagnetic ratio of nuclei is relatively smaller than that of electron spins, the condition set in Eq.~\eqref{eq:tau} can be fulfilled even under moderate magnetic field strengths. This is because the Larmor frequency of nuclei is much lower than that of electrons. For instance, in the case of hydrogen, the Larmor frequency is approximately 600 times less than that of an electron at the same magnetic field, resulting in a much longer precession period. In contrast, for electrons, the precession period would be extremely short, making it difficult to satisfy the condition under similar magnetic field strengths due to the rapid oscillations. To accommodate the finite duration of the sequence pulses, it is necessary to adjust the free evolution time $t_{\text{free}}$ within the sequence. For simplicity, we assume that the driving pulses over NV and dangling bond share the same Rabi frequency, $\Omega_1=\Omega_2\equiv\Omega$,
\begin{equation}\label{eq:t_free}
t_{\text{free}} = \tau - 2\frac{\pi}{\Omega} - 4\frac{\pi}{2\Omega} = \tau - 4\frac{\pi}{\Omega}.
\end{equation}
In a frame rotating with respect to $H_0= D S_{z}^2 - \gamma_{e} B S_{z} - \gamma_{e} B J_{z} + \gamma_{n} B I_{z}$, and omitting fast-rotating terms, the effective Hamiltonian under the condition $\tau \ll 1/A_z$, and Eqs.~(\ref{eq:tau},~\ref{eq:t_free}) \cite{Schwartz18} is given by:
\begin{equation}\label{eq:Heff}
H_{\text{eff}}= \frac{A_z}{4}\left(\sigma_{x}J_{x}+\sigma_{y}J_{y}\right)+\frac{\alpha G_{\perp}}{2}\left(J_{x}I_{x}+J_{y}I_{y}\right),
\end{equation}
where $\alpha<1$ is a coefficient defined by the filter function produced by the PulsePol sequence \cite{Schwartz18} and the coupling term $G_{\perp}$ is expressed as:
\begin{equation}\label{eq:Bzx}
G_{\perp} = \frac{3\hbar\mu_0\gamma_{e}\gamma_{n}}{4\pi\abs{\vec{r}'}^3}(\sin\Theta\cos\Theta),
\end{equation}
with $\vec{r'}$ representing the position vector of the nucleus relative to the electron, and $\Theta$ the angle between this vector and the magnetic field. The optimal resonance condition occurs at $n=3$, yielding $\alpha \approx 0.72$. The shortest resonance ($n=1$) results in $\alpha \approx 0.37$, for more details see the Supplementary Material of Ref.~\cite{Schwartz18}. Under these dynamics, assuming the NV is initially polarized and both the electron and the nucleus start in thermal states, the polarization of the nucleus $P'$ after a sequence evolves as:
\begin{align}\label{eq:pol_transfer_NV_e_n}
    P'(\tau) &= \Tr\left[e^{-iH_{\text{eff}}\tau}\rho(0)e^{iH_{\text{eff}}\tau}2I_{z}\right] \nonumber\\    &=\frac{4 A_z^2 \left(\alpha G_{\perp}\right)^2 }{\left[A_z^2 + \left(\alpha G_{\perp}\right)^2\right]^2}\sin^4\left[\frac{\sqrt{A_z^2 + \left(\alpha G_{\perp}\right)^2} }{8}\tau\right].
\end{align}
The final polarization state of a nucleus, whether it is in the $\ket{\uparrow}_n$ or $\ket{\downarrow}n$ state (the eigenstates of the $I_{z}$ operator), is determined by the harmonic $n$ selected. At the same time, the electron's polarization ($P_{{e}^-}$) oscillates as described by:
\begin{align}\label{eq:pol_transfer_NV_e}
    P_{{e}^-}(\tau) &= \Tr\left[e^{-iH_{\text{eff}}\tau}\rho(0)e^{iH_{\text{eff}}\tau}2J_{z}\right] \nonumber\\
    &=-\frac{A_z^2}{A_z^2 + \left(\alpha G_{\perp}\right)^2}\sin^2\left[\frac{\sqrt{A_z^2 + \left(\alpha G_{\perp}\right)^2} }{4}\tau\right].
\end{align}

As the nucleus reaches maximum polarization, the electron returns to an unpolarized state. The efficiency of polarization transfer, as detailed in Equation~\eqref{eq:pol_transfer_NV_e_n}, is maximized when the interaction strength between the NV and the electron matches that between the electron and the nucleus, scaled by the constant $\alpha$. If an electron is positioned directly above the NV at a distance $\abs{\vec{r}}$ such that the angle is ($\theta=0$), optimal polarization transfer occurs when a nucleus is situated at a distance $\abs{\vec{r}'}$ from the electron, calculated as
\begin{equation}
\abs{\vec{r}'} \approx \left(\frac{3\alpha\gamma_{n}}{4\gamma_{e}}\right)^{1/3}\abs{\vec{r}},
\end{equation}
where the $\Theta$-dependent term has been averaged over all possible orientations. This distance formula becomes critical if, for example, the interpulse spacing is set so that $n=1$ in Eq.~\eqref{eq:tau}, corresponding to $\alpha \approx 0.37$. Under these parameters, and assuming the polarization is transferred from an NV positioned 3.5 nm below the surface to a hydrogen nucleus, the protocol reaches its peak effectiveness if the nucleus is about 2.6 \r{A} away from the electron. In an organic sample with a proton density of approximately 50 nm$^{-3}$ (i.e., similar to water), typically two protons are found within a semisphere of radius 2.6 \r{A} around the bond.

Using the electron as an intermediary offers a significant advantage due to its substantially higher gyromagnetic ratio, which is 628 times greater than that of hydrogen. This higher gyromagnetic ratio, along with the electron's proximity to the surface, results in coupling constants $A_z$ and $G_\perp$ that are significantly larger—by two to three orders of magnitude—than those observed between the NV center and the nucleus in comparable setups (illustrated in Fig.~\ref{fig:sequence}(c)). This proximity and increased coupling strength allow for a much quicker polarization buildup using the intermediate electron compared to direct polarization transfers from the NV to the nucleus. Moreover, while surface electrons can cause decoherence in the standard PulsePol approach, in our sequence, these electrons are coherently managed. This could to enhanced efficiency and sensitivity in experimental applications.

\subsection{Results}\label{sec:results}
To illustrate the performance of our method, we consider a  scenario in which an NV center interacts with a single electron. In this setting, if the system maintains perfect coherence, polarization transfer between the electron and the NV occurs according to Eq.\eqref{eq:pol1}, with the rate constrained only by the strength of their magnetic dipole-dipole interaction. 

For the polarization method to be effective, robustness is essential \cite{Korzeczek}, as the protocols must withstand errors in the driving fields. These errors include resonance offsets due to varying strain conditions across NV centers, which alter the zero-field splitting parameter $D$, and amplitude fluctuations in each pulse. The former is modeled via a detuning term ($\frac{\Delta_\text{NV}}{2}\sigma_z$) in Hamiltonian~\eqref{eq:H_RWA}, while the latter leads to deviations in the Rabi frequency ($\Tilde{\Omega} = \Omega + \Omega_\text{error}$), where $\Omega$ denotes the Rabi frequency that produces ideal. Such errors induce rotations that deviate from the intended $\pi$ or $\pi/2$ pulses.

\begin{figure*}[t]
    \centering
    \includegraphics[width=0.9\linewidth]{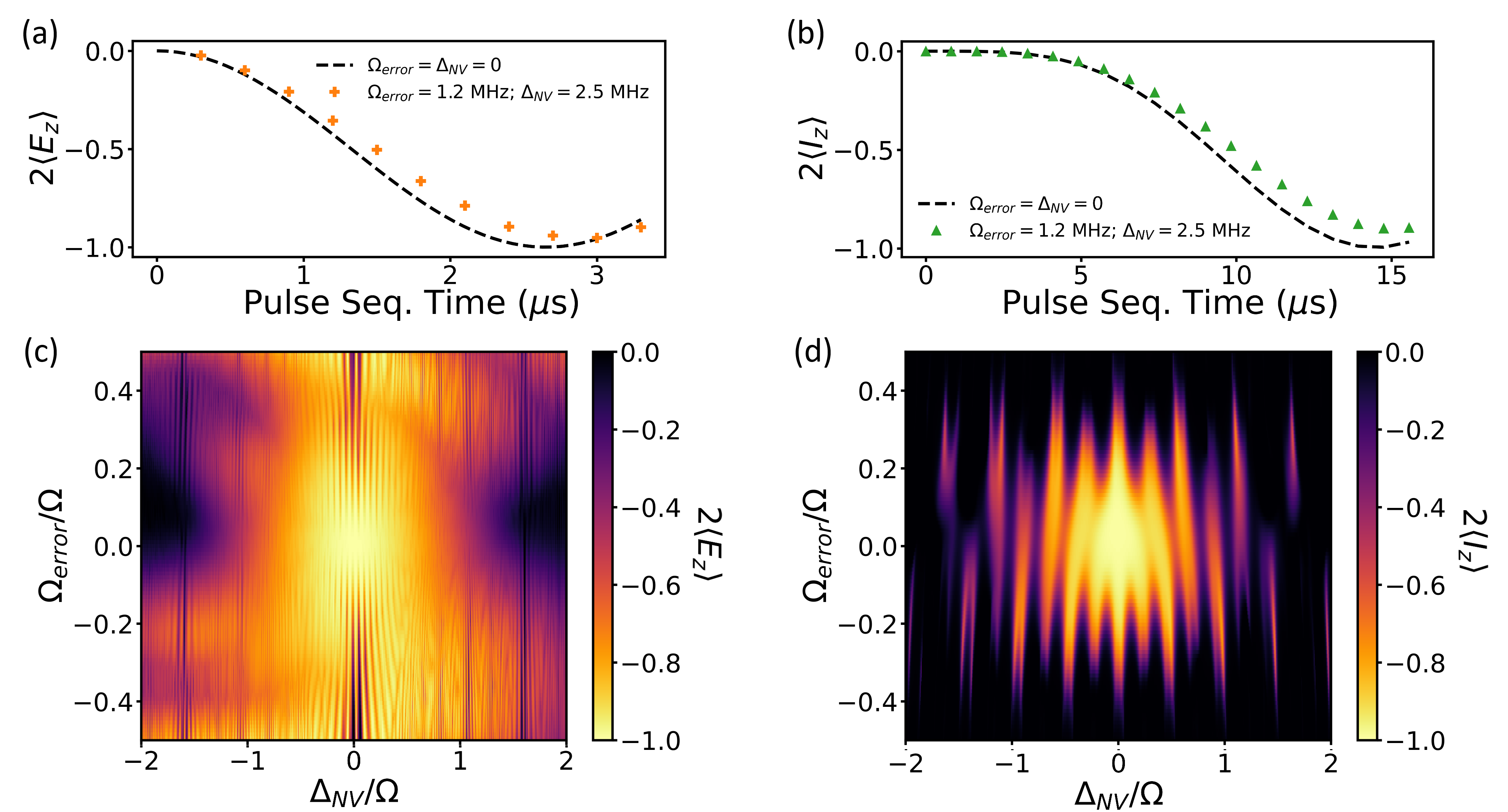}
  \caption{Robustness simulations of the double-channel PulsePol. (a) Polarization transfer to a single electron spin under perfect conditions (black dashed line) and under pulse errors (orange crosses) that include a detuning error $\Delta_\text{NV}=(2\pi)$2.5 MHz in the driving field of the NV and Rabi frequency errors $\Omega_\text{error}=(2\pi)$1.2 MHz in both driving fields. The NV-electron coupling is $A_z=(2\pi)$0.4 MHz and the Rabi frequency $\Omega=(2\pi)$20 MHz. (b) Polarization transfer to a $^1$H nucleus using an electron as mediator under perfect conditions (black dashed line) and under the same pulse errors and parameters as in (a) (green triangles). The NV-electron and the electron-nuclear couplings are $A_z=\alpha G_\perp=(2\pi)0.1$MHz. (c-d) For the same parameters as (a-b), maximum polarization transfer versus $\Delta_\text{NV}$ and $\Omega_\text{error}$.  }
  \label{fig:robustness}
\end{figure*}
Figure \ref{fig:robustness}(a) displays numerical simulations of polarization, measured by $\langle 2J_{z} \rangle$, achieved by a surface electron using our proposed method near an initially fully polarized NV spin. The simulations include the presence of errors and are contrasted with the ideal transfer (represented by a dashed line) as described in Eq.~\eqref{eq:pol1}. Both Rabi frequencies, $\Omega_1$ and $\Omega_2$, are set to the same value $\Omega$, affecting both channels equally when errors are present. The chosen parameters are detailed in the caption of Fig.~\ref{fig:robustness}. As expected from the intrinsic PulsePol structure of our sequence, these errors minimally impact the polarization transfer. In particular, the electron achieves 95\% of the total polarization, although the duration is about 14\% longer than that of the error-free sequence.

In Figure \ref{fig:robustness}(c), we validate the resilience of our protocol by demonstrating polarization transfer across different values of $\Delta_{\text{NV}}$ and $\Omega_\text{error}$. Importantly, effective polarization transfer is still possible over a broad spectrum of errors. However, as shown in Figs. \ref{fig:robustness}(b) and \ref{fig:robustness}(d), involving the electron as a mediator in the polarization process introduces additional complexities. This element makes the process more sensitive to errors, where minor deviations in driving strength or frequency may impair the sequence's efficacy. Figure \ref{fig:robustness}(b) shows that, even under the same error conditions as previously described, the sequence retains about 90\% of its effectiveness, with only a 6\% reduction in speed.

The results in Figure \ref{fig:robustness}(d) illustrate that despite increased sensitivity, the sequence remains effective for a wide range of detunings and even with imperfect pulses, affirming the robustness of our sequence for direct polarization transfer to a target spin and emphasizing the need for meticulous control of experimental parameters when aiming to polarize additional nuclear spins using the first target electron spin as the mediator.
\begin{figure}[t!]
    \centering
    \includegraphics[width=0.55\linewidth]{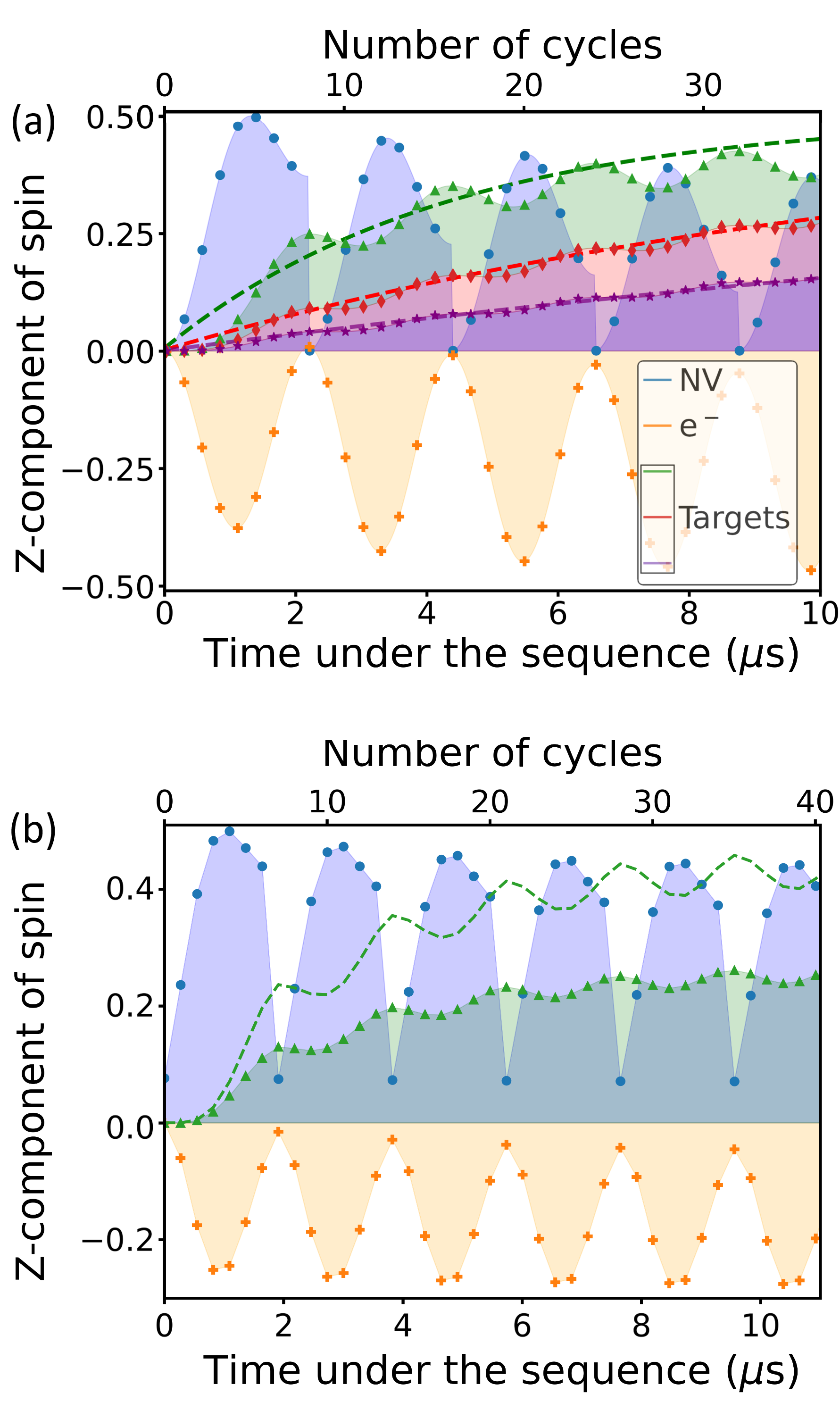}
  \caption{Theoretical overview of NV-electron-nuclei polarization transfer mechanism for the case in which all elements are treated as discrete entities. Symbols represent polarization of each element at the end of a double-channel PulsePol sequence. (a) Dynamical simulation of one NV, an electron and 3 target $^1$H nuclei. Dashed lines denote the analytical expression given by Eq.~\eqref{eq:P_exp}. The NV and electron interact with a coupling $A_z=(2\pi)0.9$ MHz, while the nuclei interact with the electron with couplings $G_\perp=$ $(2\pi)1$ MHz, $(2\pi)0.6$ MHz and $(2\pi)0.4$ MHz. The magnetic field is set to 430 G and the pulses are considered to be ideal and instantaneous. The interpulse spacing corresponds to $n=3$, and the NV is re-initialized every 8 sequences. (b) Dynamical simulation of one NV, an electron and a target $^1$H nucleus (first nucleus of the previous simulation) with decoherence. The relaxation and coherence times are $T_1=1$ ms, $T_2=10$ $\mu$s, $T_1^{e^-}=30$ $\mu$s, $T_2^{e^-}=1$ $\mu$s, $T_1^{n}=1$ s, $T_2^{n}=1$ ms. In this simulation, we assume that the NV is 80\% polarized in the $\ket{0}$ state. The green dashed line represents the polarization acquired in a fully decoherence-free environment. The inter-pulse spacing corresponds to $n=3$, and the NV is re-initialized every 7 sequences.} 
  \label{fig:model}
\end{figure}

To further explore the role of the electron as a mediator for transferring polarization beyond the diamond lattice, we consider a scenario with just one electron and multiple target spins, given that the nuclear density in the target area greatly exceeds the density of surface dangling bonds. In a system composed of an NV, an electron, and multiple target spins, polarization transfer still proceeds according to Eq.~\eqref{eq:pol_transfer_NV_e_n}, with the following modification
\begin{equation}\label{eq:A+B}
A_z^2 + \left(\alpha G_{\perp}\right)^2\mapsto A_z^2 + \left(\alpha G_0\right)^2,
\end{equation}
where
\begin{equation}
G_0^2=\sum_i^\text{nuclei}\left(G_\perp^{(i)}\right)^2.
\end{equation}

Given that only one excitation from the NV is distributed among a group of nuclear spins, frequent reinitialization of the NV is necessary to build up polarization, requiring the protocol to be executed repeatedly. To optimize the polarization transfer rate, the sequence should be performed $N$ times before the NV is reinitialized, allowing each nucleus to achieve its maximum potential polarization. Notably, continuously optically driving the NV spin into its $\ket{0}$ state could induce a quantum Zeno-like effect, effectively slowing down the polarization rate. After completing $N$ cycles, each with a duration of $T=2N\tau$—where $\tau$ represents the duration of a single sequence—each nucleus can reach its highest possible level of polarization.
\begin{equation}\label{eq:P(t+T)}
    P^{(i)}(t+T) = P^{(i)}(t) + \left[1-P^{(i)}(t)\right]T u_i(T),
\end{equation}
where $u_i(T)$ is the cooling rate of nucleus $i$, defined according to Eq.~\eqref{eq:pol_transfer_NV_e_n} as
\begin{equation}
    u_i(T)=\frac{4 A_z^2 \left(\alpha G_{\perp}^{(i)}\right)^2 }{T\left[A_z^2 + \left(\alpha G_0\right)^2\right]^2} \sin^4\left[\frac{\sqrt{A_z^2 + \left(\alpha G_0\right)^2} }{8}T\right].    
\end{equation} 

Treating Eq.~\eqref{eq:P(t+T)} as a differential equation, for $T$ sufficiently small compared to the cooling rate, the polarization buildup can be approximated by an exponential,
\begin{equation}\label{eq:P_exp}
    P^{(i)}(t)\approx P^{(i)}(0)+\left[1-P^{(i)}(0)\right]\left[1-e^{-u_i(T)t}\right].
\end{equation}

The outcomes of a simulated implementation of the sequence in a system consisting of an NV, an electron, and three nuclei are illustrated in Fig.\ref{fig:model}(a). The y-axis of this figure shows the mean z-component of the spin for each particle ($\langle S_z\rangle$, $\langle J_z\rangle$, and $\langle I_z^{(i)}\rangle$ for the NV, the electron, and the nuclei, respectively). During the polarization transfer process, the three target nuclei display varying polarization rates, which are influenced by their individual dipole-dipole interactions with the electron. The average dynamics of each nucleus are effectively captured by Eq.\eqref{eq:P_exp}, and are indicated by the dashed lines in the figure. As the nuclei reach their peak polarization levels, the NV center is reinitialized to the $\ket{0}$ state through optical pumping. The marked decreases in the NV's dynamics at specific intervals (such as around 2 $\mu$s, 4 $\mu$s, and 6 $\mu$s) result from these optical pumping events.

The previous description does not consider the relaxation rates of the target nuclei, the interactions between nuclei that could lead to polarization diffusion throughout the sample, or the decoherence rates of the NV or electron. In fact, a major challenge of this protocol is the short coherence time of the electron, on the order of 1 $\mu$s. To account for the effects of decoherence, Fig. \ref{fig:model}(b) presents a numerical simulation of the dynamics involving an NV center, an electron, and a nuclear spin, incorporating typical relaxation and dephasing rates through a master equation 
\begin{equation}
    \frac{d\rho}{dt}=-i\left[H,\rho\right] + \sum_j\Gamma_j\left(2J_j\rho J_j^\dagger - J_j^\dagger J_j\rho - \rho J_j^\dagger J_j\right),
\end{equation}
where $H$ is the system Hamiltonian describing the coherent interaction, while $J_j$ are so-called jump operators (each of which with associated rate $\Gamma_j>0$) which describes irreversible
processes such as excitations which are lost to the large environment never to come back to the system. In particular, we use $J_1 =2J_\text{z}$, $J_2 = 2J_\text{x}$, $J_3 =\sigma_\text{z}^\text{NV}$, $J_4=\sigma_\text{x}^\text{NV}$, $J_5 =2I_\text{z}$, and $J_6 = 2I_\text{x}$, with associated rates $\Gamma_1= 30$ $\mu$s, $\Gamma_2=1$ $\mu$s, $\Gamma_3=1$ ms, $\Gamma_4=10$ $\mu$s, $\Gamma_5 = 1$ s, and $\Gamma_6 = 1$ ms.

The green dashed line in Fig. \ref{fig:model}(b) indicates that the polarization achieved under ideal, decoherence-free conditions is significantly greater than that obtained under typical decoherence rates. Nonetheless, despite the presence of decoherence, our sequence still facilitates markedly better polarization transfer compared to the direct application of the single-channel PulsePol sequence to a nucleus with specific coupling alignments. Notably, the direct approach, characterized by NV-to-nucleus couplings in the range of kHz, results in polarization transfer times on the order of milliseconds. Therefore, within the simulation's timeframe, which spans just a few microseconds, the polarization achieved by the nucleus is considered negligible.

The protocol presented could significantly enhance polarization transfer from NV centers to external nuclear spins using dangling bond mediators. The study introduces a double-channel PulsePol sequence that effectively addresses high Larmor frequency challenges in polarization transfer. This method utilizes surface electron spins as mediators, enabling stronger interactions and faster polarization processes compared to direct methods. Additionally, the robustness of this sequence against control errors and its ability to operate without matching resonance conditions make it particularly advantageous. The results demonstrate an improvement in polarization transfer efficiency, offering a promising approach for enhancing NMR sensitivity and expanding its applicability in various scientific and medical fields.


\section[Amplitude Encoded Radio Induced Signal]{Amplitude Encoded\\Radio Induced Signal}
\label{sec5}
\label{chapter5}

%
%

\thiswatermark{\put(1,-302){\color{l-grey}\rule{70pt}{60pt}}
\put(70,-302){\color{grey}\rule{297pt}{60pt}}}

%
%

\vfill

\lettrine[lines=2, findent=3pt,nindent=0pt]{T}{he} integration of NV centers into NMR spectroscopy promises to enhance spatial resolution while significantly reducing the scanned sample volume. This advancement is particularly valuable for NMR analysis of mass-limited samples, with applications in areas such as surface coatings, single cells, and tissues. In traditional NMR spectroscopy, resonant RF pulses are applied to a sample, triggering signals through Larmor precession, which can then be analyzed to obtain information about molecular structure and dynamics.

For NV centers to be competitive in NMR spectroscopy, they must achieve signal detection with resolutions that exceed their coherence times. Recent advancements have led to protocols using NV probes that achieve sub-Hertz resolution in controlled signal measurements~\cite{Boss17,Schmitt17}. These pioneering approaches have also enabled the detection of J-couplings and chemical shifts at low magnetic fields scenarios when applied to micron-sized samples~\cite{Glenn18,Bucher20,Arunkumar21}.

Achieving micro-scale NMR at high static magnetic fields is particularly beneficial as it enhances sensor responses, enabling the study of low-density samples and the detection of structural information such as chemical shifts and J-couplings. However, micron-scale NMR has been limited to low-field scenarios due to the challenges of coupling quantum sensors to high-frequency nuclear signals.

To address this, we have developed a protocol—the Amplitude Encoded Radio Induced Signal (AERIS)—which leverages the ZZ interaction to overcome this limitation. In the context of NMR, AERIS interacts with the sample and with the NV-based sensor to achieve high resolution while maintaining sensitivity. Specifically, by appropriately combining RF and microwave radiation patterns, AERIS induces a forced nuclear signal that bridges the interaction between quantum sensors and rapidly precessing nuclear spins. Additionally, since the sensor is only active during the scanning of the forced signal, the AERIS protocol can extend beyond the coherence time of the quantum sensor, achieving a spectral resolution limited only by the properties of the nuclear sample. Thus, our AERIS protocol unlocks the high-field regime in micro-scale NMR spectroscopy.

The content of this chapter is based on the work published in Ref. \cite{Munuera-Javaloy23}. This chapter is organized as follows: In Section \ref{seq:aeris_protocol}, we present an analysis of the protocol and discuss the underlying intuition. In Section~\ref{seq:signal_produced} we analyze the signal produced by the sample. Section \ref{seq:aeris_results} details a simulation of the protocol applied to an ethanol sample.

\begin{figure}[t]
\center
\includegraphics[width= .8\linewidth]{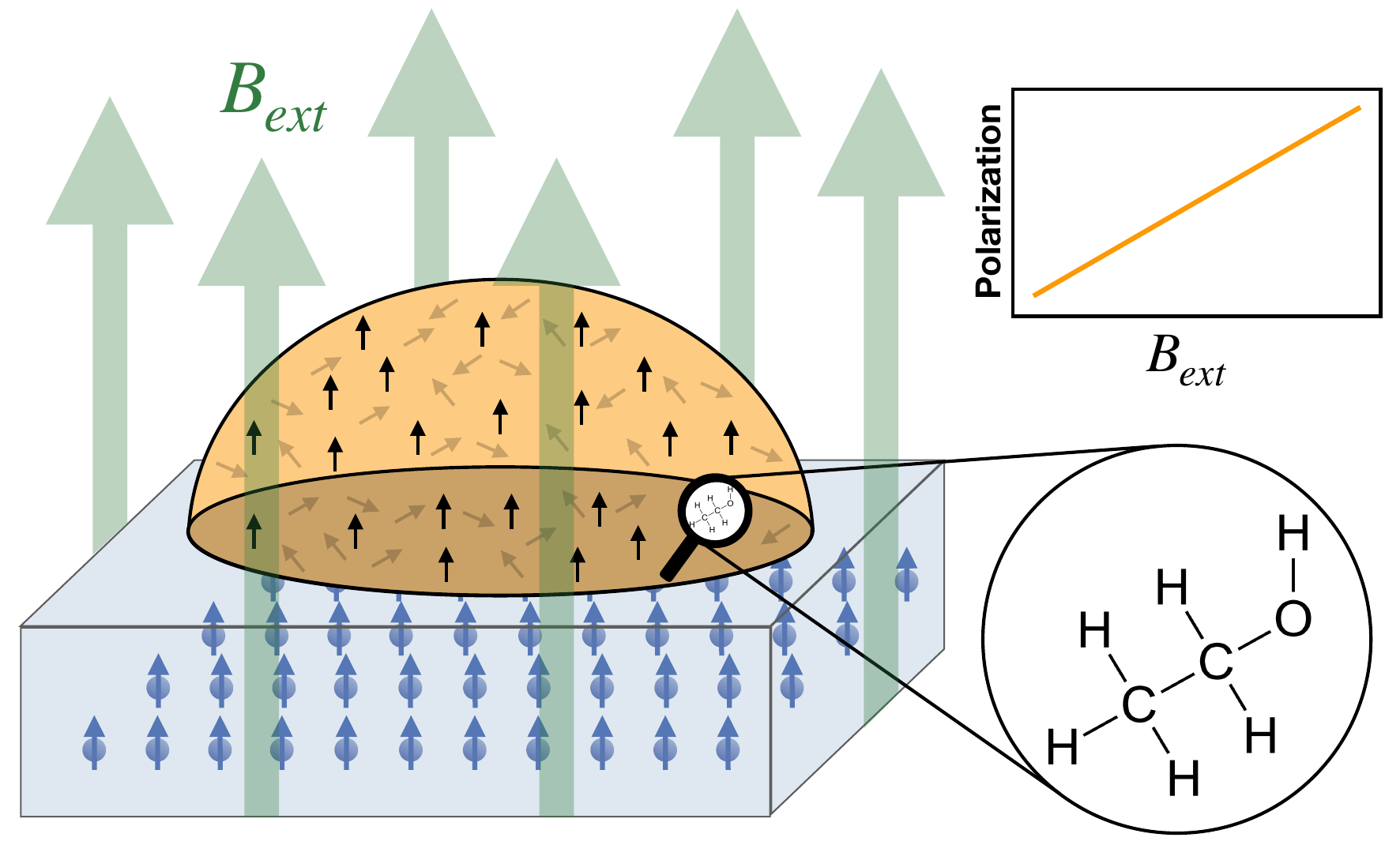}
\caption{\label{fig1} The setup consists on a picoliter  sample placed on a diamond. This contains an NV ensemble as the sensor of the magnetic signal generated by the analyte. The protons of the studied molecule emit a signal that depends on their local environment, thus carrying structural information.}
\end{figure}

\subsection{The protocol}\label{seq:aeris_protocol}
State of the art NV-based AC field magnetometry  targets the oscillating signal produced by precessing nuclear spins, whose frequency is proportional to the local field felt by the nuclei. On the one hand, this relation allows to acquire information on the molecular environment of the nuclei by unraveling the spectral composition of the signal. On the other hand, when the sample is exposed to a large magnetic field, it leads to signals that oscillate too fast to be traced by the NV.  Here we target a deliberately manufactured signal that carries the spectroscopic information of the studied sample encoded in its amplitude rather than in its frequency. 

We consider a thermally polarized sample placed on top of an NV-ensemble-based-sensor and in the presence of a large external magnetic field $B_{ext}$, see Fig.~\eqref{fig1}. The external magnetic field should exceed 1 T, as experiments with lower fields can be performed using standard heterodyne sequences~\cite{Boss17,Schmitt17,Glenn18}. The sample  contain a certain type of molecule with nuclear spins in different locations of its structure. Hereafter we use subindex $k$ (or superindex when required by the notation) to
indicate the different precession frequencies produced by distinct local magnetic fields. This scenario is similar to those reported in \cite{Glenn18,Bucher20,Arunkumar21} with the critical difference of the magnitude of $B_{ext}$, which in our case is chosen to be 2.1 T. 
\begin{figure*}[t!]
\center
\includegraphics[width=  \linewidth]{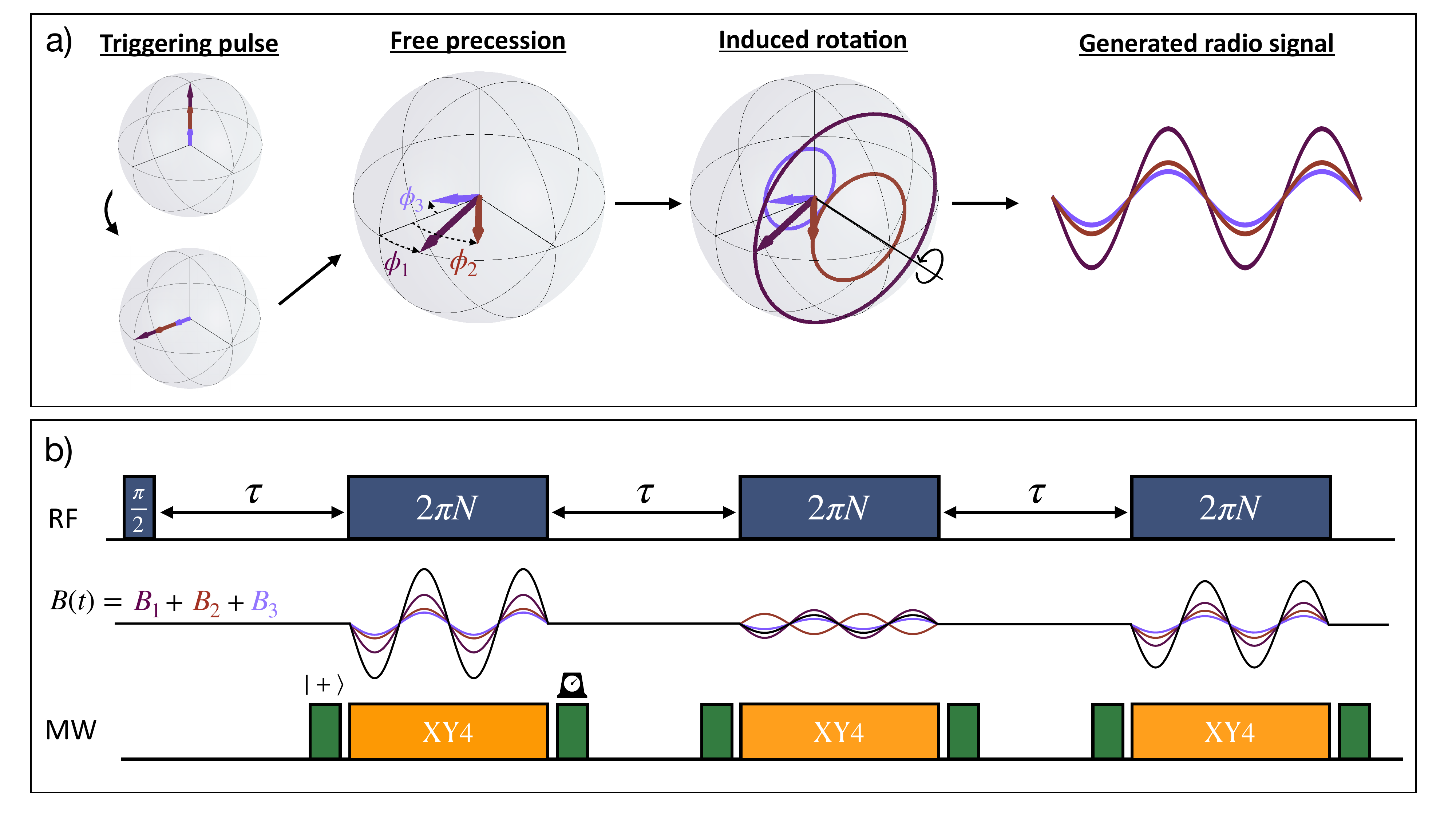} 
\caption{\label{fig2} Custom signal production and measurement. a) An initial RF $\pi/2$ pulse brings the sample thermal polarization to the orthogonal plane and triggers the AERIS protocol consisting on free precessions and induced rotations stages. For a time $\tau$ each magnetization vector $\boldsymbol{M}_k(t)$ precess according to the local field at the position of the nuclear spin. The phase covered by each $\boldsymbol{M}_k(t)$ --this is $\phi_k=\delta_k \tau$-- is encoded in the amplitude of the oscillating field generated via controlled rotations of these vectors. b) (First panel) RF control sequence with interleaved free precessions. (Second panel)  Sample emitted fields. These have different amplitudes due to the distinct projections of each rotating $\boldsymbol{M}_k(t)$ on the Z axis. The depicted case shows three $B_i$ fields as a consequence of the splitting among three magnetization vectors that spin at rates $\delta_1$, $\delta_2$, and $\delta_3$. (Third panel) MW pattern --in our case an XY4 sequence-- on each NV devised to capture the induced signal. Note that the NVs remain inactive during the long free precession stages of the sample, providing our protocol with increased spectral resolution regardless of the sensor coherence time. Prior to the MW sequence, the NV ensemble is initialized in $|+\rangle$ while once its state encodes the desired information it is optically readout in the $\sigma_y$ basis.}
\end{figure*}

Following~\cite{Levitt08} we describe the spins of our sample via the nuclear magnetization $\boldsymbol M =(M_x,M_y,M_z)$. This is a time-dependent vector proportional to the sample average nuclear magnetic moment. Its behavior during an RF pulse of intensity $\Omega$ in a frame that rotates with the frequency of the RF driving ($\omega$) is described by the Bloch equations
\beq
\label{bloch}
\frac{d}{dt}
\left(\begin{array}{c}
M_x\\
M_y\\
M_z
\end{array}\right)
=
\left(\begin{array}{ccc}
-1/T^*_2 & - \delta & \Omega \sin\phi\\
\delta & -1/T^*_2 & - \Omega \cos\phi\\
-\Omega \sin \phi & \Omega \cos \phi & - 1/T_1
\end{array}\right)
\left(\begin{array}{c}
M_x\\
M_y\\
M_z
\end{array}\right)
+
\left(\begin{array}{c}
0\\
0\\
1/T_1
\end{array}\right),
\eeq
where $\phi$ is the phase of the RF field, and $T_1$ ($T^*_2$) is the nuclear relaxation (dephasing) rate. The detuning $\delta=\omega_L-\omega$ between the RF pulse frequency and the Larmor precession rate $\omega_L$ depends on the local magnetic field at the nuclear spin site, which differs from $B_{ext}$. Hence, the sample comprises $k$ different precession frequencies $\omega_L^{\,k}$ leading to $k$  detunings $\delta_k$. 
 
Our AERIS protocol comprises two parts. The first one creates a detectable signal by exploiting the dynamics  in Eq.~\eqref{bloch}. This is achieved with an RF triggering pulse on the sample followed by an alternation among free precession periods and induced rotations as shown in Fig.~\eqref{fig2}. The second part consists on probing the produced signal with NV sensors that acquire a phase determined by its amplitude, gathering in their spin state information about the spectral composition of the signal and allowing to determine the local magnetic environment around nuclear spins.

More in detail: An RF pulse along the X axis (i.e. $\phi = 0$) of duration $\pi/(2\Omega)$, see Eq.~\eqref{bloch}, tilts the initial thermal polarization of the sample, $\boldsymbol{M}=(0,0,1)$ (note that $\boldsymbol{M}$ provides the direction of the thermal polarization when it is a unit vector, and it attenuates the polarization amount as $T_1$ and $T^*_2$ diminish its modulus) to the perpendicular plane and triggers off the protocol. Once the pulse is turned off, i.e. $\Omega=0$ in Eq.~\eqref{bloch}, the nuclear spins  precess around the external field at a rate determined by the local magnetic field at their respective locations. Similar to a clock mechanism that rotates the needles representing hours, minutes, and seconds at different speeds, the free precession stage of fixed duration $\tau$ splits the magnetization vector $\boldsymbol M(t)$ in components $\boldsymbol M_k (t)= \left( \sin( \delta_k \,t),-\cos( \delta_k t), 0\right)$. Recall, $\delta_k$ is the detuning between the driving frequency and the $k$th precession frequency in the sample. Crucially, the NV sensor remains inactive at this stage, thus $\tau$ could be significantly larger than NV coherence times leading to a high spectral resolution ultimately limited by the coherence of the nuclear sample.

A long RF pulse then continuously rotates the magnetization components at a speed $\propto \Omega$ around an axis on the xy-plane determined by $\phi$, as described by Eq.~\eqref{bloch}. The projection of the resulting field in the NV axis sets a target field  $B(t)$ with two key features. Firstly, it oscillates with frequency $\Omega \ll \omega_L$ (note that, at large magnetic fields, this relation is naturally achieved for realistic Rabi frequencies). This is a parameter that can be tuned such that $B(t)$ oscillations can be tracked by the NV ensemble regardless of the magnetic field value acting on the sample. Secondly, $B(t)$ comprises the radio signals  produced by each rotating $\boldsymbol{M}_k(t) = \left( \sin( \delta_k  \tau), -\cos( \delta_k \tau) \cos(\Omega t), -\cos( \delta_k   \tau) \sin(\Omega t) \right)$, thus it contains the footprint of each nuclear  environment (encoded in the distinct $\delta_k$ shifts). Note that, for the sake of simplicity in the presentation, we do not account for potential deviations in the rotation axes caused by each $\delta_k$ shift. However, these are included in our numerical analysis. For more details see Section~\eqref{seq:signal_produced}.

After $N$ complete rotations of the magnetization vectors, thus after N periods of $B(t)$, the RF rotation pulse is switched off and the sample advances to the next free precession stage in which each $\boldsymbol{M}_k(t)$ continues to dephase. This sequence is iterated leading to an oscillation in the amplitudes of the signals emitted during  successive induced rotation stages, whose spectrum relates directly to the various $\omega_L^{\,k}$ in the sample. 

The radio signal $B_n(t)$ produced during the $n^{th}$ induced rotation stage is captured by the NVs in the ensemble such that each NV evolves according to 
\begin{equation}\label{eq:NVHamiltonian}
H = - \gamma_e B_n(t) \frac{\sigma_z}{2} + \Omega_{\rm MW}(t) \frac{\sigma_\phi}{2}.
\end{equation}
Here $\gamma_e$ is the electronic gyromagnetic factor, $\boldsymbol{\sigma}$ are the Pauli operators of the NV two-level system, and the target signal $B_n(t)$ is expressed in Section~\eqref{seq:signal_produced}.
The control field $\Omega_{\rm MW}(t)$ is synchronized with the rotation pulse over nuclear spins, see Fig.~\eqref{fig2}, leading to an XY4 control sequence that allows the sensor to capture a phase determined by (i) The amplitude of the radio signal stemming from the sample, and (ii) The length of the RF pulse. This information is gathered by reading  the state of the sensor, with an expected result for the $n^{\text{th}}$ phase acquisition stage of 
\beq
\label{expectedsigma}
\langle \sigma_y \rangle_n = \frac{2 \gamma_e t_m }{\pi} \sum_k b_k \cos(\delta_k n \tau),
\eeq
where $b_k$ is the initial magnetic field amplitude on the NV site produced by the $k^\text{th}$ spectral component, see Section~\eqref{seq:signal_produced} for further details.

Thus, subsequent detections provide a stroboscopic record of the oscillating amplitudes, see Fig.~\eqref{fig:results} (a), whose Fourier spectrum relates to the frequency shifts of nuclei at different sites of the sample molecule. 

Let us recall that the NV ensemble sensor is only active during phase acquisition (i.e. while the dynamical decoupling sequence is active), and after that, it is optically readout and reinitialized. Therefore, the duration of our  protocol, and thus its spectral resolution, gets over the cap imposed by the coherence of the sensor, being only limited by the coherence of the nuclear fields. 

\subsection{Signal produced}\label{seq:signal_produced}
In this section, we study the signal strength and the phase accumulated by the NV. After the triggering pulse with $\phi=0$, the magnetization reads ${\boldsymbol M}=(0,-1,0)$. This is the initial configuration for the first free precession stage (of fixed duration $\tau$) that splits the magnetization in different components ${\boldsymbol M}_k$ such that the initial configuration for the $k^\text{th}$ spectral component at the $n^\text{th}$ phase acquisition stage reads ${\boldsymbol M} = (\sin(\delta_k n \tau), -\cos(\delta_k n \tau),0)$. This is obtained from the Bloch equations with $\Omega=0$. 

Now, the evolution of the magnetization during the $n^{th}$ phase acquisition stage reads 
\beq
\begin{split}
\label{solbloch}
\boldsymbol{M}_k^{\,n}(t)&=
\left(\begin{array}{ccc}
1 & 0 & 0\\
0 & \cos(\Omega t) & - \sin(\Omega t)\\
0  & \sin(\Omega t) & \cos(\Omega t)
\end{array}\right)
\left(\begin{array}{c}
\sin( \delta_k \,n \tau)\\
-\cos( \delta_k \, n \tau)\\
0
\end{array}\right)
\\&=
\left(\begin{array}{c}
\sin( \delta_k \,n \tau)\\
-\cos( \delta_k \, n \tau) \cos(\Omega t)\\
-\cos( \delta_k \, n \tau) \sin(\Omega t)
\end{array}\right).
\end{split}
\eeq 
Notice that the Bloch equations in Eq.~\eqref{bloch}, and therefore the solutions obtained from them, describe the dynamics in a frame that rotates around the external field at the RF driving frequency $\omega$. For the sake of simplicity in the presentation, the solution in Eq.~\eqref{solbloch} assumes no decoherence (i.e. $T_1,T^*_2 \rightarrow \infty$) and perfect resonance (not taking into account the natural $\delta_k$ shifts during the driving). 

In our case, the effect of the $\delta_k$ shifts on the rotation speed is, to first order, a factor of approximately $\frac{\delta_k^2}{2 \Omega^2} \approx 2 \times 10^{-5}$, which is negligible and has no significant impact on the results. If necessary (i.e., in case of facing more severe energy shifts) a modified sequence, as outlined in Section~\ref{seq:robustness}, can be used to further correct this error.

The interaction between the signal produced by the rotating $\boldsymbol{M}_k^{\,n}(t)$ and the sensor in a rotating frame w.r.t. the NV electronic-spin ground-state triplet is contained in the $B_n(t)=\sum_k B_k^n(t)$ term of Eq.~\eqref{eq:NVHamiltonian}, such that
\beq
\label{targetfieldapp}
B_k^n(t) = \frac{\hbar^2 \gamma_N^2 \mu_0 \rho_k  B_{ext}}{16 \pi k_B T}\, {\boldsymbol M}_k^{\,n}(t) \int  \big[g_x(r), g_y(r), f(r)\big] \ dV,
\eeq
\begin{figure*}[t]
\centering
\includegraphics[width=0.7 \linewidth]{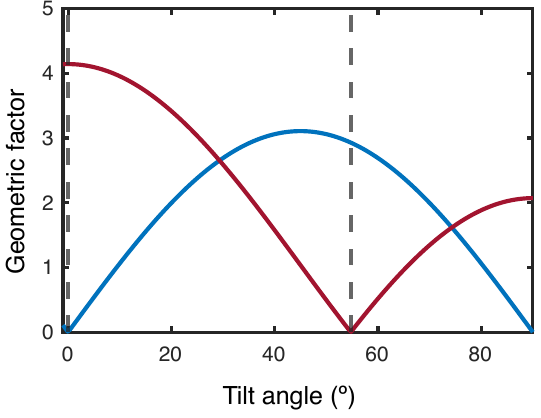}
\caption{\label{fig:tilt} Geometric factors $\mathfrak{f}=\int{f(\vec{r})dV}$ (red line) and $\mathfrak{g}=\int{g_x(\vec{r})dV}$ (blue line) as a function of the NV orientation with respect to the diamond surface. The orientations corresponding to the $(1 1 1 )$ (tilt$=0$) and $(1 0 0 )$ (tilt$=\arccos{\frac{1}{\sqrt{3}}}$) diamond cuts are marked with dashed lines.}
\end{figure*}
where $\mu_0$ is the vacuum permeability, $\rho_k$ the density of spins with the $k^{th}$ precession frequency, $\gamma_N$ is the nuclear gyromagnetic factor, $T$ is the temperature of the sample, $k_B$ is the Boltzmann constant, and $B_{ext}$ is the external magnetic field. The geometric functions $g_{x,y}(r)$ and $f(r)$ read
\beq
f(r)= \frac{1}{r^3}(3 r_z^2 -1) \hspace{1.5cm} \text{and} \hspace{1.5cm} g_{x, y}(r)= \frac{1}{r^3} (3 r_z r_{x, y}),
\eeq
with $\hat{r} = (r_x, r_y, r_z)$ being  the unitary vector joining the NV and $dV$, while $r$ represents their relative distance. The expression in~\eqref{targetfieldapp} (which can be derived from a microscopic description of a system involving NVs and nuclear spins \cite{Meriles10}) is valid provided that the external magnetic field $B_{ext}$ is greater than the coupling strength, which allows ignore the backaction of the sensor in the sample \cite{Reinhard12}. As we are in a large field regime, this condition is met.

As we are interested in the $\hat z$ component of ${\boldsymbol M}_k^{\,n}(t)$, as it doesn't oscillate with the Larmor frequency, we need the geometric factor $\mathfrak{f}=\int{f(\vec{r})dV}$ to be non-zero. In Fig.~\ref{fig:tilt}, we can observe that for the widespread $(1 0 0)$ diamond cut this number goes to zero. For this reason, the protocol is optimal at the $(1 1 1)$ diamond cut, where the term is maximized. Although less common, this cut is receiving increased attention as diamonds grown in this orientation have all the NV centers aligned in the same direction, increasing fluorescence contrast \cite{Hughes24}.

The MW control implements an XY4 dynamical decoupling sequence that modulates the interaction between target and sensor leading to 
\beq
H = \frac{\gamma_e  \sigma_z}{\pi}  \sum_k b_k \cos(\delta_k n \tau),
\eeq
where $b_k = \frac{\hbar^2 \gamma_N^2 \mu_0 \rho_k \mathfrak{f} B_{ext}}{16 \pi k_B T}  $. 

 The NV is initialized in the $|+\rangle = \frac{1}{\sqrt{2}}\left(|1\rangle+|0\rangle\right)$ state, then evolves during $t_m$, and it is finally measured such that (in the small angle regime)
\beq
\langle \sigma_y \rangle_n = \frac{2 \gamma_e t_m}{\pi}  \sum_k b_k \cos(\delta_k n \tau).
\eeq

On the other hand, an RF trigger pulse with $\phi=\pi/2$ leads to $\boldsymbol{M}_k=(1,0,0)$, which yields a splitting of the $k$ spectral components during the free precession stages described by  ${\boldsymbol M}_k = (\cos(\delta_k n \tau),\sin(\delta_k n \tau),0)$. For the same dynamical decoupling control sequence over NVs, we find   

 \beq\label{eq:sinu}
\langle \sigma_y \rangle_n = \frac{2 \gamma_e t_m }{\pi} \sum_k b_k \sin(\delta_k n \tau).
 \eeq

\subsection{Numerical results}\label{seq:aeris_results}
\begin{figure*}[t]
\centering
\includegraphics[width=1.\linewidth]{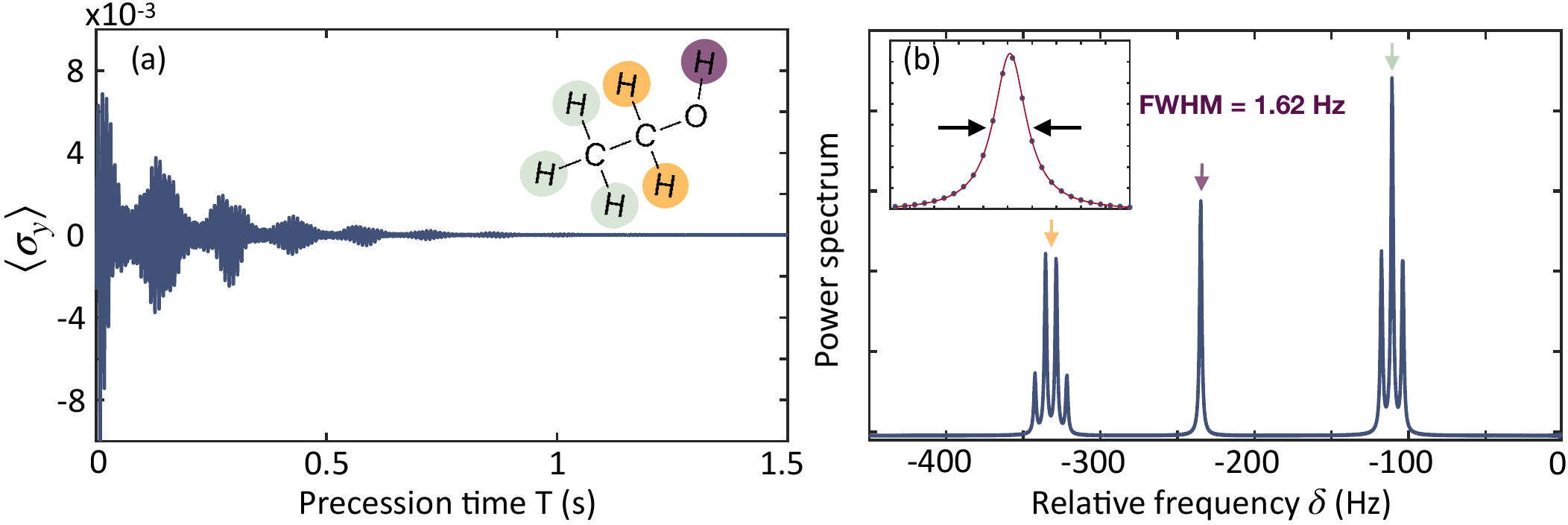}
\caption{\label{fig:results}  Measurements and spectrum obtained by considering $\delta_k = -\{342.45, 335.55, 328.65, 321.75, 234.9, 117.6, 110.7, 103.8\}$ Hz, and magnetic field amplitudes $b_k = \{106 ,320, 320, 106, 426, 320, 640, 320\}$ pT along the Z axis of a generic NV in the ensemble. (a) Simulated stroboscopic record collected by measuring $\langle \sigma_y \rangle$ on the NV as a function of the cumulated precession time, after interacting with the ethanol sample (inset). The three sites of the ethanol molecule with different chemical shifts are indicated with distinct colors. (b) Fourier transform of the measurement record (blue solid line) showing peaks in the expected values. Each peak group has its origin site/chemical shift indicated with an arrow of the corresponding color. Inset, the central peak was fitted to a Lorentzian function that exhibits a full width at half maximum (FWHM) of 1.62 Hz.}
\end{figure*}
We illustrate the AERIS protocol by simulating the evolution of 8 magnetization vectors taken from the ethanol [C$_2$H$_6$O] spectrum~\cite{Levitt08} in a scenario that comprises a magnetic field of 2.1 T, while the RF driving frequency $\omega$ is set to $(2\pi)\ \times$ 90 MHz, which is assumed to be the origin of the chemical shift scale (this is the resonance frequency of TMS~\cite{Levitt08}). Each $\delta_k$ detuning is obtained by considering the three chemical shifts of $3.66$, $2.6$, and $1.19$ ppm, as well as a J-coupling of 6.9 Hz between the CH$_3$ and the CH$_2$ groups of ethanol~\cite{Levitt08}, see caption in Fig.~\eqref{fig:results}. The average field amplitude over each NV in the ensemble is estimated to  $\approx2.56$ nT, by taking into account the proton concentration of ethanol, the geometric factor $\int  f(r) dV$ for the $(1 1 1)$ cut, the temperature and the external magnetic field of 2.1 T, see Appendix~\ref{app2} for further details. This  field amplitude is distributed in different $b_k$ according to the ethanol spectral structure, see caption in Fig.~\eqref{fig:results} and Appendix~\ref{app2}. We find the radio signal emitted by the sample by numerically solving the Bloch equations during RF irradiation (i.e. at the induced rotation stages). The free precession time is selected as $\tau = 1$ ms, and the induced rotation stage has a duration of 40 $\mu$s (corresponding approximately to 2 full rotations of the magnetization vectors) while the NV ensemble is controlled with an XY4 sequence. Furthermore, we use $\Omega_{\rm MW} = (2 \pi)\times 20$ MHz, $\Omega_{\rm RF} = (2 \pi)\times 50$ KHz, and sample coherence times  $T_1 = 2$ s and $T^*_2 = 0.2$ s. This process is repeated 1500 times, leading to the stroboscopic record of Fig.~\eqref{fig:results} (a) which follows Eq.~(\ref{expectedsigma}). 

We run again the protocol by employing an initial $\pi/2$ pulse over the Y axis leading to the sinusoidal version  Eq.~\eqref{eq:sinu}. 

\subsection{Robustness}\label{seq:robustness}
\begin{figure*}[t!]
\centering
\includegraphics[width=1 \linewidth]{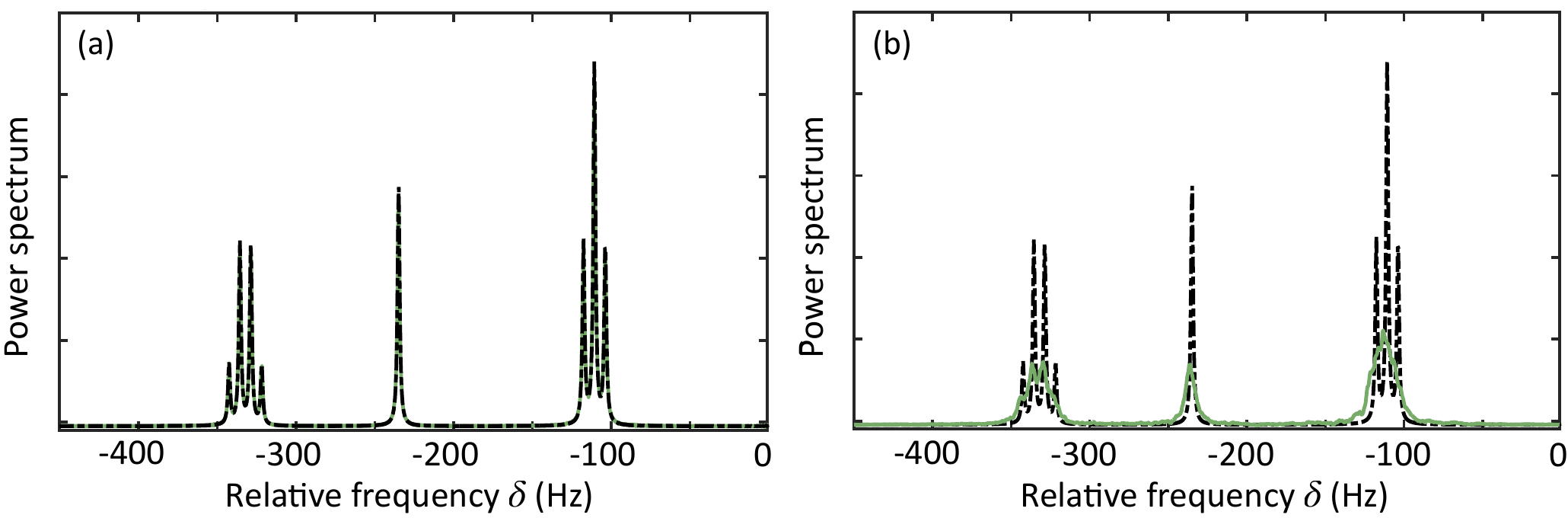}
\caption{\label{fig:appRobust1}  Spectra comparison for the AERIS sequence with perfect RF controls (black dotted line) and in the presence of OU noise (green solid line) with (a) $\sigma = 0.24\%$ and $\tau = 1$ms and (b) $\sigma = 2\%$, $\tau = 0.5$ms and an amplitude shift of 1$\%$. Both spectra were obtained averaging 200 realizations.}
\end{figure*}
For the sake of simplicity in the description of the AERIS method, the previous simulations consider perfect controls. We consider now the effect of errors in the RF control, which could be potentially detrimental for the sequence as the nuclear signal coherence has to be maintained throughout the protocol. The control error is modeled as an Ornstein-Uhlenbeck \cite{Wang45, Gillespie96} process
\beq
\epsilon_\Omega(t+\Delta t) = \epsilon_\Omega(t) e^{-\Delta t/\tau} + \sigma N(t)
\eeq
where $\tau$ is the correlation time of the noise, $N(t)$ is a normally distributed random variable, and $\sigma$ is the relative amplitude of the fluctuations. For standard expected experimental errors \cite{Boris22, Cai12}, the obtained spectrum overlaps with the case without control errors, see Fig.~\ref{fig:appRobust1} (a).

Finally, both measurement records in Eqs.~(\ref{expectedsigma}, \ref{eq:sinu}) are combined and converted, via discrete Fourier transform, into the spectrum in Fig.~\eqref{fig:results} (b). There we demonstrate that the AERIS method leads in the studied case to Lorentzian peaks with a FWHM  $\approx 1.62$ Hz (limited by the sample $T^*_2$) thus sufficient to detect the posed chemical shifts and J couplings.
\begin{figure*}[t]
\centering
\includegraphics[width=1\linewidth]{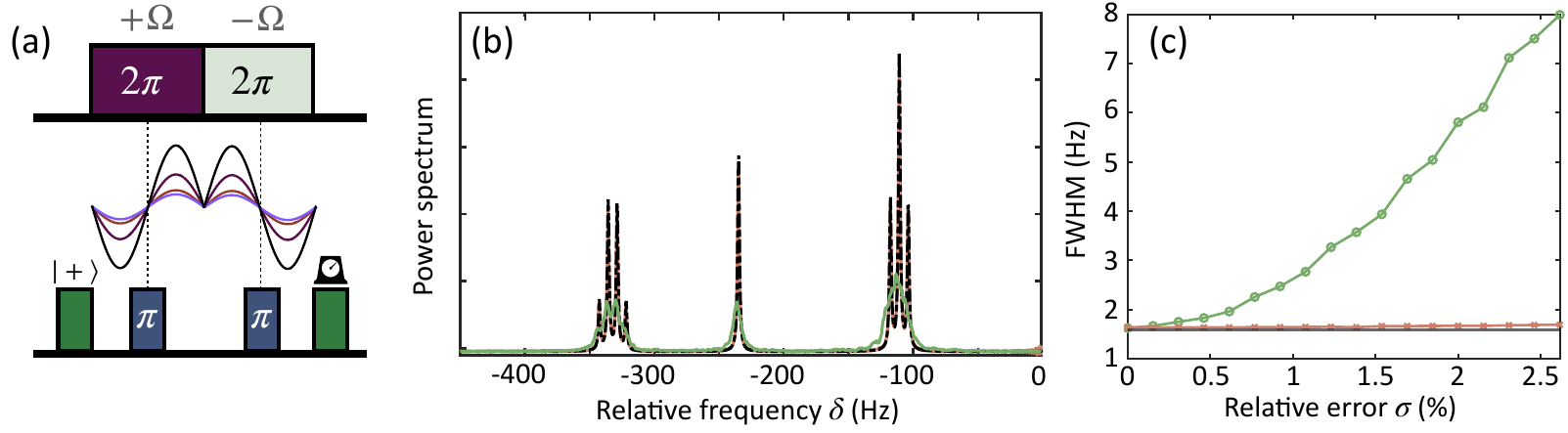}
\caption{\label{fig:appRobust2}  (a) Schematics of the modified AERIS sequence. Two rotations are performed in opposite directions with the RF control, giving raise to a detectable magnetic signal with a $\pi$ phase change in the middle which is measured with two concatenated spin echoes. (b) Spectra comparison for the AERIS sequence with perfect controls (black dotted line) and with $\sigma = 2\%$, $\tau = 0.5$ms and an amplitude shift of 1$\%$ for AERIS (green solid line) and the modified version (orange solid line). (c) FWHM with respect of the relative OU error with $\tau = 1$ms and no constant amplitude shift for AERIS (green line) and the modified version (orange line). The minimum FWHM possible given the nuclear $T_2^*$ is represented as a grey line.}
\end{figure*}  

However, in the presence of more severe noise and  Rabi amplitude shifts (e.g. due to misscalibration) AERIS gives raise to distorted spectra as can be seen in Fig.~\ref{fig:appRobust1} (b). A direct modification of the default sequence leads to a significant improvement on robustness. The alternative sequence is equivalent to the original one but changes the irradiation and NV measurement stages with the scheme represented in Fig.~\ref{fig:appRobust2} (a). The modified version employs a change of sign in the middle of the RF irradiation such that the error accumulated in the first half is the opposite to the one accumulated in the second half leading to cancellation. The XY-4 sequence over the NV is substituted with two $\pi$ pulses in order to accumulate phase from the new magnetic signal. The new version recovers the ideal spectrum in the severe noise example, see Fig.~\ref{fig:appRobust2} (b).

Finally, in Fig.~\ref{fig:appRobust2} (c), we show a comparison of the expected Full Width at Half Maximum (FWHM) of the central spectral peak for AERIS and the modified version with respect to the error amplitude. The modified version recovers a FWHM close to the minimum possible given the nuclear $T_2^*$ for the considered error range. AERIS has motivated further developments on its robustness, see Daly et al. \cite{Daly24}.

In summary, we have devised an NMR signal detection protocol that attains chemical shift level resolution from micron-sized samples while being suitable for large magnetic fields. Our approach relies on the production of a custom field that resonates with dynamically decoupled NV sensors used to extract spectral information from the sample. Actual experiments may require several repetitions to average out the impact of shot noise or inaccurate control sequences. Nevertheless the demand for higher spectral resolution is less stringent at large fields, as chemical shifts increase and J-couplings become clearer. Besides, polarization rates increase, leading to stronger signals that provide measurements with higher contrast.  Both effects contribute to decreasing the required number of repetitions, or, conversely, making small concentration samples adequate to our protocol, which sets the utility of NV sensors for realistic chemical analysis.


\section[High-field NMR in Dipolarly-Coupled Samples]{High-field NMR in\\Dipolarly-Coupled Samples}
\label{sec6}
\label{chapter6}

%
%

\thiswatermark{\put(1,-302){\color{l-grey}\rule{70pt}{60pt}}
\put(70,-302){\color{grey}\rule{297pt}{60pt}}}

%
%

\vfill

\lettrine[lines=2, findent=3pt,nindent=0pt]{A}{n} important limitation of state-of-the-art NV-NMR spectroscopy is caused by the strong dipolar coupling among target nuclei, which results in intricate spectra that challenge data interpretation. This becomes especially critical in solid-state material research, where homonuclear dipole-dipole interactions hinder subtler couplings (heteronuclear interactions are less challenging as they can be eliminated by driving only one of the species). The NMR community has dedicated significant efforts to mitigate the impact of dipolar interactions \cite{Madhu16}, and today solid-state NMR spectroscopy is extensively used in distinct research areas: In pharmaceutics it characterizes active pharmaceutical ingredients (APIs) and their interaction with excipients (inactive substances added to a drug that serve various purposes such as binding or preserving the API) \cite{Pisklak14,Pisklak16,Nie16}, among many other applications (see \cite{Marchetti21} for an extensive review on the topic). In epidemiology, it provides key insights of the structure of molecules related with diseases as present in our societies as Alzeimer \cite{Tycko11}. In energy storage research is used to characterize the local structure of solid materials used in batteries and fuel cells \cite{Pecher17}.

These areas, along with many others, would largely benefit form sensors able to produce narrower spectral lines from smaller solid-state samples. NV-NMR spectroscopy emerges as the prominent technique to access samples in the microscale regime, however, it remains limited to the liquid state scenario where the dipole-dipole interactions get naturally averaged out due to fast molecular motion, leaving a plethora of relevant applications out of its range of action. 

\begin{figure*}[t]
\centering
\includegraphics[width= 1 \linewidth]{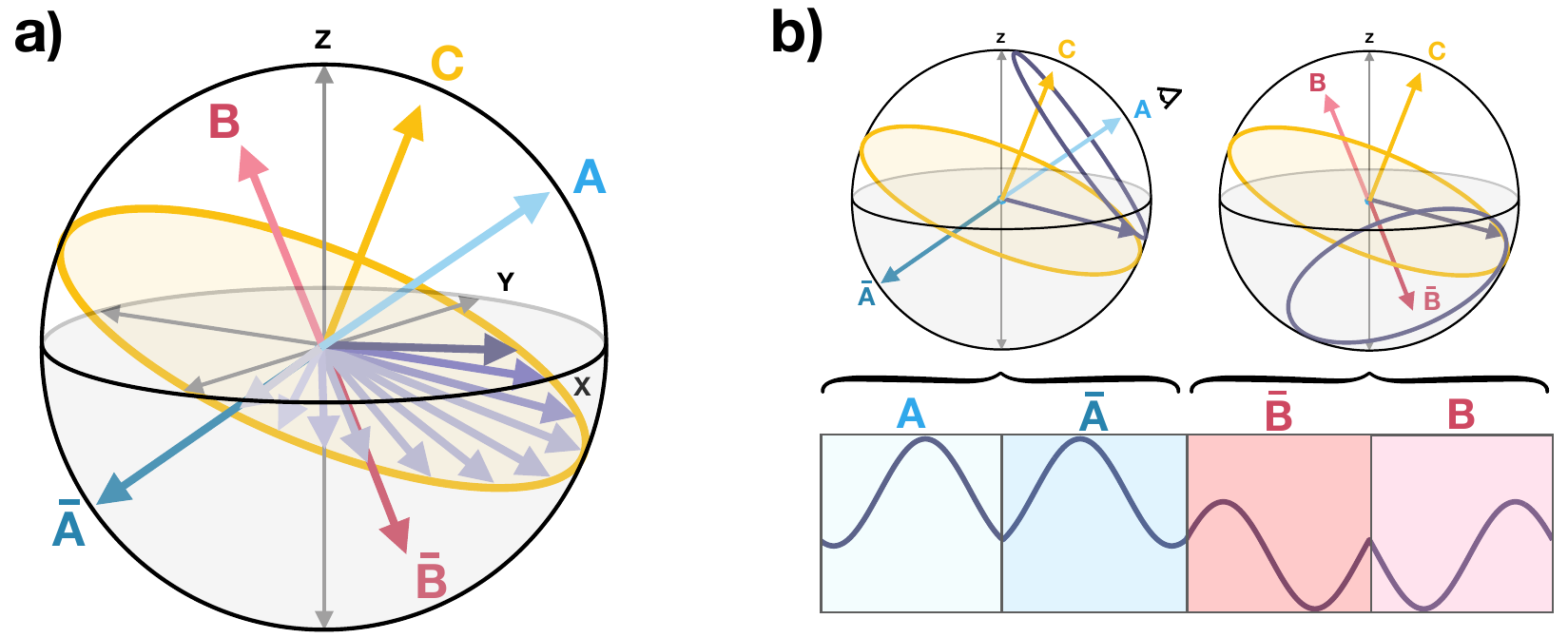}
\caption{\label{rotations} (a) Relative positions of axis $A$ (clear-blue), $\bar{A}$ (dark-blue), $B$ (clear-magenta), $\bar{B}$ (dark-magenta). 
The motion of the magnetization inbetween LG4 blocks ($A\bar A\bar B B$) is shown in purple. This evolution is described as a rotation in the plane (in yellow) perpendicular to the $C$ axis in accordance with the effective Hamiltonian in Eq.~\eqref{eq: Heff} for a single $\delta_i^*$. (b) (Top) Magnetization rotations during each of the four RF drivings. (Bottom) Projection of the magnetization onto $\hat{z}$ as it rotates during a full LG4 block. This projection determines the magnetic signal for the NV ensemble sensor.}
\end{figure*}

In this chapter, we develop a protocol that overcomes these limitations and enables high resolution NV-NMR spectroscopy of ordered samples with strong homonuclear dipolar coupling at elevated external magnetic fields, allowing to take advantage from the higher polarization rates and larger chemical shifts.  Our protocol delivers RF and MW synchronized with measurements on an NV ensemble magnetometer. The RF channel drives the sample with a twofold purpose. On the one hand it effectively decouples the target nuclear spins, diminishing the effect of strong homonuclear couplings in the recorded spectra, and enabling the obtention of nuclear energy shifts. Remarkably, the decoupling benefits from increasing RF intensities could be specially effective in the small volume regime, where current RF antennas have reported nuclear spin rotation velocities within the range of tens to hundreds of kHz \cite{Herb20, Yudilevich23}. On the other hand, the RF bridges the interaction among NV sensors and fast rotating nuclear spin by generating a slow-frequency NMR signal trackable by the NV sensor. Simultaneously, the MW channel delivers a tailored pulse sequence to the NV ensemble enabling the detection of the magnetic field emitted by the driven sample. This sequence is interspersed with measurements of the sensor's state to construct the spectra in a heterodyne frame~\cite{Glenn18, Schmitt17, Boss17} leading to a spectrum only limited by the nuclear sample coherence. 

The content of this chapter is based on the on the work in Ref.~\cite{Munuera24}. This chapter is organized as follows: In Section~\ref{sec:rf_mod}, we analyze the effective dynamics of a nuclear sample being driven by RF fields triggering an LG4 sequence. In Section~\ref{sec:harvest}, we develop a geometric interpretation of the measurement process with the NV and use it to find the optimal pulse timings for maximum resolution. Finally, in Section~\ref{sec:results} we simulate the sequence on a five-spin system, showcasing its effectiveness.

\subsection{RF modulation of nuclear spins}\label{sec:rf_mod}
Lee and Goldburg (LG) showed in a seminal paper~\cite{Lee65} that an off-resonant continuous RF field cancels, up to first order, the contribution to the nuclear spin dynamics of homonuclear dipole-dipole interactions if the {\it LG condition} $\Delta=\pm \Omega/\sqrt{2}$ holds. Here, $\Delta = \omega_L-\omega_d$ is the detuning between the carrier frequency of the RF driving field ($\omega_d$) and the Larmor precession of the spins ($\omega_L$), and $\Omega$ is the Rabi frequency of the RF driving. Subjecting a spin ensemble to an off-resonant RF field leads to collective nuclear spin rotations along an axis tilted with respect to $\hat{z}$ (the direction of the static magnetic field). More specifically, {\it the tilted axis} --in the following $P$-- has a component $\frac{\Delta}{\sqrt{ \Omega^2 + \Delta^2}}$ along $\hat{z}$, while its projection on the orthogonal $xy$ plane is $\frac{\Omega}{\sqrt{ \Omega^2 + \Delta^2}}$.

Further developments have built upon the original LG sequence demonstrating the ability to remove higher order contributions of the dipole-dipole interaction, thus leading to even narrower spectral lines. Prominent examples are the frequency-switched (FSLG)~\cite{Waugh72} and phase-modulated (PMLG)~\cite{Vinogradov99} versions of the original LG sequence. Our protocol incorporates the advanced LG4 sequence~\cite{Halse13} over nuclei, which exhibits remarkable decoupling rates and enhanced robustness against RF control errors. The LG4 consists on concatenated blocks of four consecutive off-resonant RF drivings, all complying with the LG condition, leading to rotations along four different axes. This is, the rotation axis $P$ alternates among $A,\bar A,\bar B$ and $B$, whose relative positions are illustrated in Fig.~\ref{rotations}~(a). Note that, at each block, nuclear spins undergo two sets of complementary rotations along axis pointing in opposite directions ($A, \bar A$ and $B, \bar B$).

The nuclear spin Hamiltonian during each individual rotation of the LG4 sequence reads (see Appendix~\ref{app1}) 
\begin{equation}\label{OnerotationH}
H = \sum_{i=1}^N\left(\frac{\pm\delta_i}{\sqrt{3}} + \bar{\Omega}\right) I^i_P,
\end{equation}
where $\delta_i$ is the {\it target nuclear shift} of the $i$th spin (its sign depends on the direction of the rotation, positive value $``+\delta_i"$ is assigned to rotations along $A$ and $B$, and the negative value $``-\delta_i"$ to rotation along $\bar A$ and $\bar B$), $\bar{\Omega}=\sqrt{\Delta^2+\Omega^2}$ is the effective Rabi frequency, and the spin operator $I^i_P$ takes one of the following forms
\begin{eqnarray}\label{rotaxes}
I^i_A  &=&  \left(\Omega I^i_x\sin{\alpha}+\Omega I^i_y\cos{\alpha}+ \Delta I^i_z\right)/\bar{\Omega}, \nonumber\\
I^i_{\bar{A}} &=&   - I^i_A,\nonumber\\
I^i_B  &=&  \left(-\Omega I^i_x\sin{\alpha}+\Omega I^i_y\cos{\alpha}+ \Delta I^i_z \right)/\bar{\Omega},\nonumber\\
I^i_{\bar{B}} &=& - I^i_B.
\end{eqnarray} 
According to the LG4 scheme~\cite{Halse13}, the phase of the driving is set to $\alpha=55^\circ$ to minimize the line-width of the resonances. 

Note that in Eq.~(\ref{OnerotationH}) we assume that the internuclear interaction Hamiltonian $H_{\rm nn}=\sum_{i>j}^N \frac{\mu_0\gamma^2_n \hbar}{4\pi r_{i,j}^3} \bigg[\vec{I}_i \cdot \vec{I}_j - 3 (\vec{I}_j  \cdot \hat{r}_{i,k})  (\vec{I}_j \cdot \hat{r}_{i,j})\bigg]$ can be neglected due to the introduced decoupling sequence. This assumption simplifies the subsequent analysis. However, $H_{\rm nn}$ will be taken into account in the numerical model in Sec. \ref{sec:results}.

In the remainder of this section, we analyze the signal emitted by the sample subjected to the RF decoupling fields and develop analytical expressions for the target energy shifts.

The magnetic field that originates from the sample during the nuclear spin rotation produced by each RF field of the LG4 follows the general form
\begin{equation}\label{eq:Signal}
s(t) = \Gamma\cos{(\bar\Omega t+\phi)}+b.
\end{equation}
Hereafter, we often refer to $s(t)$ as the signal, as it constitutes the target field for the NV ensemble sensor. In fact, its amplitude $\Gamma$, phase $\phi$, and static bias $b$ depend on the configuration of the nuclear spin ensemble and thereby on the $\delta_i$ energy shifts (see Appendix \ref{app: analytical}), so detecting and properly reading $s(t)$ enables to unravel the desired information.

Consequently, the LG4 meets a twofold goal. Namely: (i) It results in a nuclear spin dynamics with minimal effect from the dipole-dipole interaction (see Appendix~\ref{app1} for the full derivation of the Hamiltonian in Eq. \eqref{OnerotationH}), enabling the identification of the weaker but interesting $\delta_i$ shifts. (ii) It induces a tunable rotation speed in the sample ($\propto \bar{\Omega}$, see Eq.~(\ref{eq:Signal})), facilitating the interaction between nuclear spins and the NV ensemble sensor even at high external magnetic fields. Regarding point (ii), it is important to note that without using RF drivings on the sample, standard techniques based on imprinting in the NVs a rotation speed comparable to the nuclear Larmor frequency would necessitate the application of unrealistic MW fields. For context, in a magnetic field of approximately 2.35 Tesla, hydrogen spins rotate at a speed of $(2\pi)\times 100$ MHz, producing a signal hardly trackable by an NV ensemble sensor operating with conventional methods.~\cite{Glenn18, Bucher20, Arunkumar21}.

Now, we examine the effects of RF decoupling fields in greater detail. Each RF driving (leading to the rotations along $A, \bar A$, $B, \bar B$) is applied for an interval $T = 1/\bar\Omega$. Consequently, the total signal emitted by the sample is a composite of distinct sinusoidal functions, condensed in Eq.~\eqref{eq:Signal}, each persisting for a duration $T$. Figure~\ref{rotations} (b) presents an illustrative example of $s(t)$ by showing the rotation of a single magnetization vector (associated with a specific $\delta_i$) around axes $A$, $\bar{A}$, $B$  and $\bar B$. 

Interestingly, with this RF control, the nuclear spins governed by Eq.~\eqref{OnerotationH} would perform a complete turn at each RF driving, constantly returning to their initial configuration if it were not for the $\delta_i$ shifts. These shifts slightly alter the nuclear spin state (i.e., the sample magnetization), thus imprinting a slower motion within the sample.
More specifically, the sample magnetization at the end of each LG4 block is determined by a set of energy shifts $\delta^*_j$ (distinct from $\delta_i$) according to the effective Hamiltonian (for details see Appendix~\ref{app1}):
\begin{equation}\label{eq: Heff}
H_{\rm eff} = \sum_i \delta^*_i I^i_C,
\end{equation}
where $I^i_C $ is a spin operator along an axis $C$ that bisects $A$ and $B$, see  Fig.~\ref{rotations}~(a), while 
\begin{equation}\label{eq:newshifts}
    \delta_i^* = \delta_i \frac{\sqrt{1 + 2 \cos^2{\alpha}}}{3}.
\end{equation}

In summary, this section demonstrates that each LG4 block alters the sample magnetization $\vec{M}$ through rotations along the $C$ axis, as depicted in Fig.~\ref{rotations} (a). Moreover, we elucidate the mechanism governing the evolution of $\vec{M}$ through the effective Hamiltonian outlined in Eq.$\eqref{eq: Heff}$, while Eq.~\eqref{eq:newshifts} establishes   analytical expressions connecting the rates of the effective rotations, $\delta^*_i$, with the target nuclear shifts $\delta_i$.

In the next section we outline the protocol to monitor this effective precessions with the NV ensemble sensor and extract the desired $\delta_i$ energies from its recordings. 

\subsection{Harvesting nuclear spin parameters with the NV ensemble}\label{sec:harvest}

\subsubsection{Geometrical interpretation of the phase accumulation}

\begin{figure}[t!]
\centering
\includegraphics[width= 0.83 \linewidth]{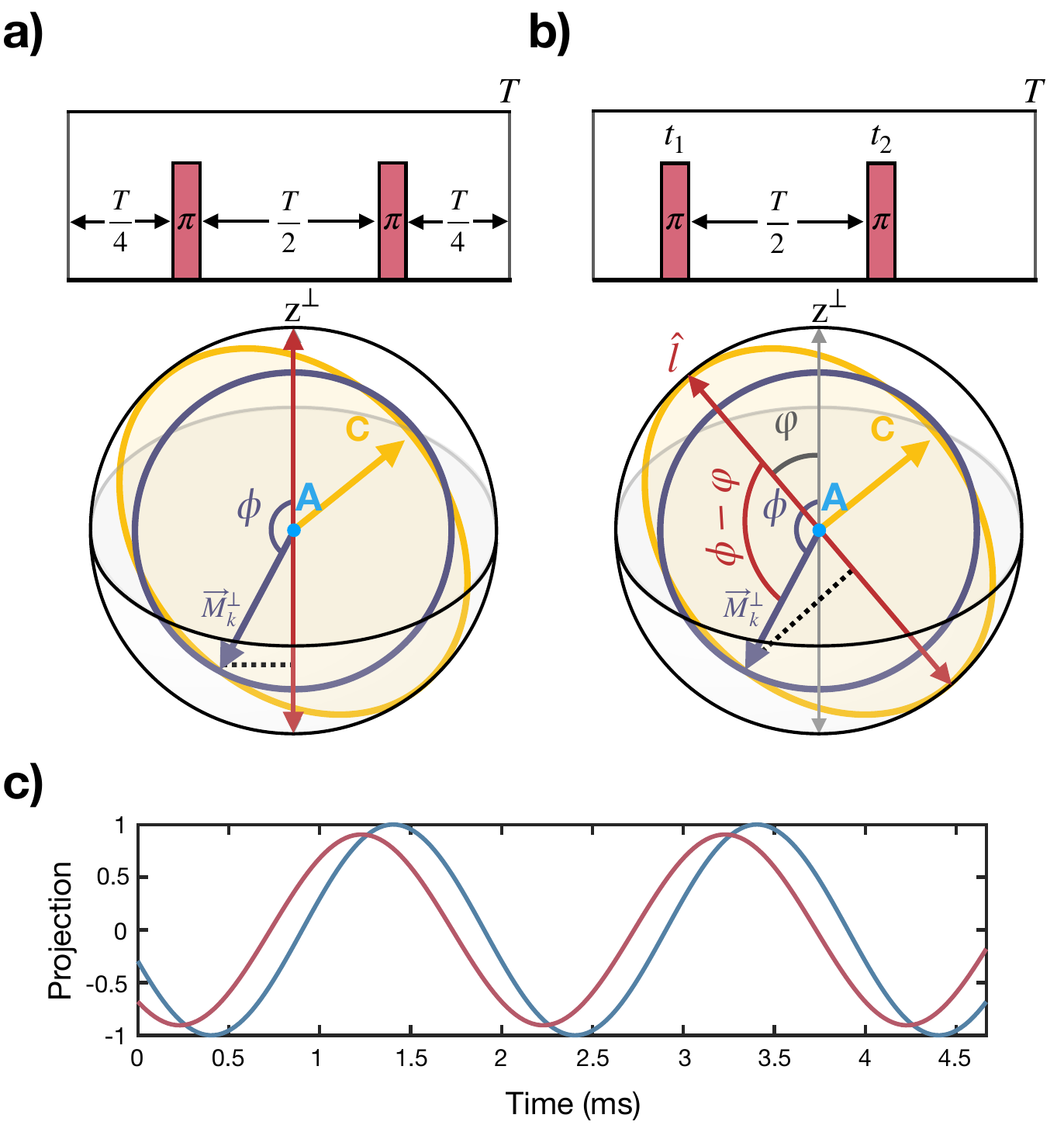}
\caption{\label{control} (a) Illustration of a CPMG pulse block (upper panel) and its geometric interpretation (lower panel). This panel shows a projection of the sphere in Fig.~\ref{rotations} (b)  viewed in a direction parallel to axis $A$. This view facilitates the representation of the projections onto the plane perpendicular to A of (i) The magnetization vector, denoted as $\vec{M}_k^\perp$, and (ii) The $\hat z$ axis, referred to as $\hat z^{\perp}$. In addition, it shows the trajectory followed by a magnetization vector during a rotation around $A$ (purple circle) and after successive LG4 blocks (yellow ellipse). (b) Upper panel, pulse block of our tailored sequence where the initial $\pi$ pulse is delivered at a time $t_1$. With this control, the phase accumulation of the NV is proportional to the projection of  $\vec{M}_k^\perp$ onto $\hat l$ (shown in red), an axis tilted away from $\hat z^\perp$ and aligned with the major axis of the yellow ellipse for optimal contrast. (c) Evolution of the projection of $\vec{M}_k^\perp$ onto axis $\hat z^\perp$ (red) and onto axis $\hat l$ (blue). The amplitude of the projection onto $\hat l$, resulting from the sequence in panel (b), reaches the maximum value of 1. As the phase accumulated by the NV is directly proportional to this projection, the timing of the pulses in (b) ensures the maximum phase accumulation amplitude.}
\end{figure}

The target magnetic field over the NV ensemble sensor is a concatenation of the sinusoidal signals in Eq.~\eqref{eq:Signal} (see lower panel in Fig.~\ref{rotations}(b)). 
A particular RF field at the $k^\text{th}$ LG4 block (note that, the accumulative character of the rotations imposed by Eq.~\eqref{eq: Heff} make it crucial to identify the number of the block from now on), produces a nuclear spin rotation around a certain axis ($A$, $\bar A$, $B$ or $\bar B$) where the amplitude $\Gamma_k$ of the resulting signal $s_k(t) = \Gamma_k\cos{(\bar\Omega t+\phi_k)}+b_k$ is directly proportional to $\vec{M}_k^\perp$ (i.e., to the magnetization component which is orthogonal to the rotation axis --$A$, $\bar A$, $B$ or $\bar B$-- at the start of each RF driving), and the phase $\phi_k$ corresponds to the angle between $\vec{M}_k^\perp$ and $\hat z^\perp$. The latter is the component of $\hat z$ that lies on the plane perpendicular to the rotation axis. See the lower panel in Fig.~\ref{control} (a) and Appendix~\ref{app: analytical} for more details.

We now use this geometric description to analyze the phase accumulated by each NV in the ensemble sensor when subjected to a generic pulse sequence. For this analysis, we choose a standard Carr-Purcell-Meiboom-Gill (CPMG) sequence~\cite{Purcell54, Meiboom58}. In order to keep the discussion accessible, we focus on the signal produced by the nuclear spins rotating around $A$ and limit ourselves to an scenario involving a single effective energy shift, $\delta_i^*$. Note, however, that the following results and the consequent conclusions are valid for signals produced by nuclear spins rotating around axes $B$, $\bar A$ and $\bar B$ and in situations involving multiple shifts.

The phase accumulated by an NV center interacting with the signal $s_k(t)$ and subjected to the CPMG sequence reads (see Appendix~\ref{app: accumulated})
\begin{equation}\label{eq: phase}
\Phi = \frac{4 |\gamma_e|}{\bar \Omega}\Gamma_k \cos{\left(\phi_k\right)}.
\end{equation}
Hence, the phase accumulated by each NV is proportional to the projection of $\vec{M}_k^\perp$ onto  $\hat z^\perp$, or, in other words, to the quantity $\Gamma_k \cos{\left(\phi_k\right)}$. The lower panel of Figure~\ref{control} (a)  provides a clarifying (probably most needed) graphic explanation.

\begin{figure*}[t]
\centering
\includegraphics[width=1\linewidth]{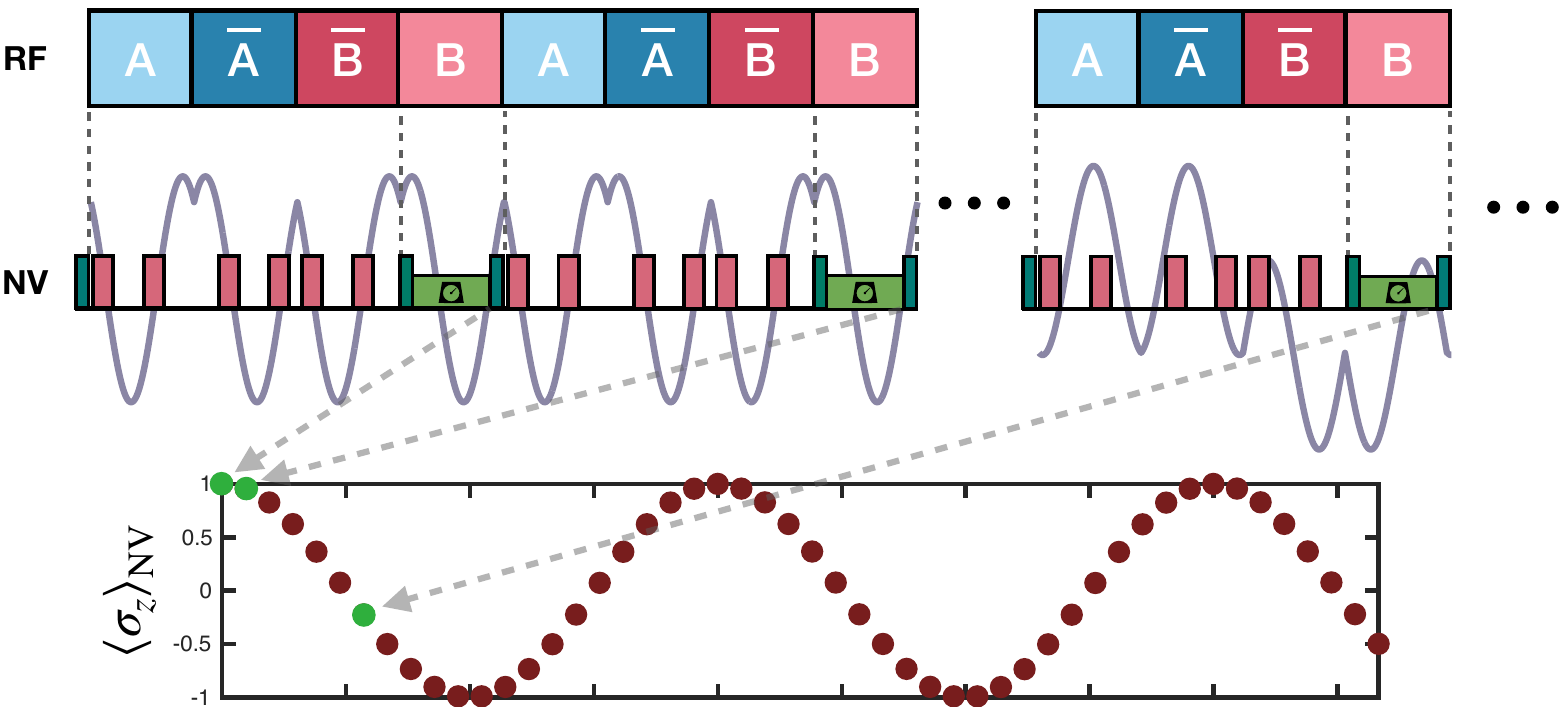}
\caption{\label{scheme} (Top) General layout of our protocol containing the RF control over nuclear spins and the MW pulse sequence on the NV ensemble. The magnetic field emitted by the nuclei as a consequence of the RF rotations is depicted in light purple. As time progresses, this field changes its shape, which changes the phase accumulated by the NV, while MW pulses keep always the same structure. (Bottom) Evolution of the expected results for the measurement over the NVs as the experiments progresses, showing a sinusoidal pattern with frequency $\delta^*$. In the presence of additional shifts, the measurement outcomes evolve as a sum of sinusoidal components with corresponding frequencies $\delta^*_i$, which can be extracted through Fourier transform analysis.}
\end{figure*}

With this description in mind we can summarize the phase acquisition stage as follows: The response of the NV centers to the signal emitted by the sample is determined by the initial sample magnetization. As the protocol advances, the magnetization vector precesses around $C$ with an angular velocity $\delta^*_i$ as described by Eq.~(\ref{eq: Heff}). In the orthogonal plane with respect to $A$, this precession translates into an elliptical motion of the vector $\vec{M}_k^\perp$, shown as a yellow ellipse in Fig.~\ref{control} (a). Thus, the projection of $\vec{M}_k^\perp$ onto the $\hat z^\perp$ axis, and consequently the phase accumulated by the NV in successive blocks of the LG4, follow a sinusoidal function with frequency $\delta^*_i$. The resulting expected value of the $\sigma_z$ operator of each NV in the ensemble at the $k^\text{th}$ LG4 block (after applying a final $\pi/2$ pulse to transform accumulated phase into populations), generalized to every $\delta_i^*$, reads: 
\begin{equation}\label{eq:expectedsigmaz}
\langle \sigma_z\rangle_k \approx 3 D_0 \sum_i \rho_i\cos{\left(\frac{4\delta_i^* k}{\bar{\Omega}} + \nu_0\right)},
\end{equation}
 where $\rho_i$ is the spin density of the $i$th nucleus, $D_0$ is detemined by the pulse sequence and $\nu_0$ depends on both the pulse sequence and the initial nuclear state. A formal derivation of \eqref{eq:expectedsigmaz} as well as further details can be found in Appendix~\ref{app: analytical}. The $0$ subindex in Eq.~(\ref{eq:expectedsigmaz}) indicate that all parameters correspond to the reference CPMG sequence (note that, in the next section we derive an improved sequence). Thus, the NV response  enclosed in Eq.~(\ref{eq:expectedsigmaz}) consists on a sum of sinusoidal functions that encode the different $\delta_i^*$ which can be then extracted via standard Fourier transform. Finally, $\delta_i$ targets can be obtained via a direct application of Eq.~(\ref{eq:newshifts}).

\subsubsection{Sensing MW pulse sequence}

With the geometrical understating developed in the previous section, now we present a tailored MW sequence   to optimally detect the target $\delta_i$ shifts. This sequence retains a CPMG-like structure composed by two $\pi$ pulses spaced by $T/2$ to mitigates noise effects, $T$ being the CPMG block length. Nonetheless, we adjust the timing of the pulses, see  Fig.~\ref{control} (b), in particular the time at which the first pulse is applied ($t_1$ hereafter). Consequently, the phase accumulated by each NV in the ensemble sensor (recall that we are focusing on the signal produced by the nuclear spins rotating around $A$) reads  
\begin{equation}\label{eq: optimal}
\Phi = \frac{4\Gamma_k|\gamma_e|}{\bar{\Omega}}\cos{\left(\phi_k-\varphi\right)}.
\end{equation}
In our geometrical framework, adjusting the timing of the pulses results in an accumulated phase proportional to the projection of $\vec{M}_k^\perp$ onto an axis $\hat l$, which is tilted at an angle $\varphi = \frac{\pi}{2} - \bar{\Omega} t_1$ relative to $\hat{z}^\perp$ (see Fig.~\ref{control} (b)).

The ability to pivot the axis in which $\vec M_k^\perp$ gets projected (note this can be done by selecting distinct values for $t_1$ since $\varphi = \frac{\pi}{2} - \bar{\Omega} t_1$) allows us to design a pulse sequence that maximizes contrast in the recorded spectra. From block to block, $\vec{M}_k^\perp$ evolves following an ellipse, therefore, we design the pulse sequence so that the phase accumulated by each NV is proportional to the projection of $\vec{M}_k^\perp$ into the major axis of the elipse. By doing so, the projecting axis and the direction that contains the extreme points of the elliptic path of $\vec{M}_k^\perp$ match, thereby yielding the maximum amplitude in the oscillation of $\Phi$ in successive blocks, see Fig.~\ref{control} (b). In particular, this is achieved by setting $\varphi = \arccos{\frac{\sqrt{3}\cos{\alpha}}{\sqrt{2 + \cos{2\alpha}}}}$, which determines the timing of the pulses as $t_{1,A}\approx 0.14\, T$ for optimal detection of the signal produced by the nuclear spins rotating around $A$. Repeating the same analysis for the signals produced by nuclear spin rotations around $\bar A$ and $\bar B$ we find $t_{1, {\bar A}} = \frac{T}{2}-t_{1, A}$ and $t_{1, {\bar B}} = t_{1, A}$. 

Summing up, our tailored MW pulse sequence is separated in blocks. Each block contains two $\pi$ pulses specifically timed to optimally detect the signal produced by the corresponding RF field. To maintain synchrony between the two control channels (MW and RF) and to avoid turning off the nuclear decoupling field, the sensor is measured and reinitialized while the RF field is on. Figure~\ref{scheme} shows the general layout of our protocol, including the drivings over the sample and the sensor, and showing the evolution of the expected outcomes in successive measurements, which read 
\begin{equation}
\langle \sigma_z\rangle_k \approx 3 D_{\rm opt} \sum_i \rho_i\cos{\left(\frac{4\delta_i^* k}{\bar{\Omega}} + \nu_{\rm opt}\right)},
\end{equation}
where $D_{\rm opt}\approx 1.1 D_0$ (i.e., with the tailored MW sequence the contrast increases a $10\%$) and $\nu_{\rm opt} = 0$ which corresponds to a initial sample magnetization oriented along the axis perpendicular to the $A$ and $B$ axes, achieved by a RF pulse that triggers the protocol. Finally, we access the effective frequencies $\delta_i^*$ by Fourier transforming the recorded data and obtain the target $\delta_i$ shifts from Eq. \eqref{eq:newshifts}.

\subsection{Results}\label{sec:results}
\begin{figure*}[t]
\centering
\includegraphics[width= 1 \linewidth]{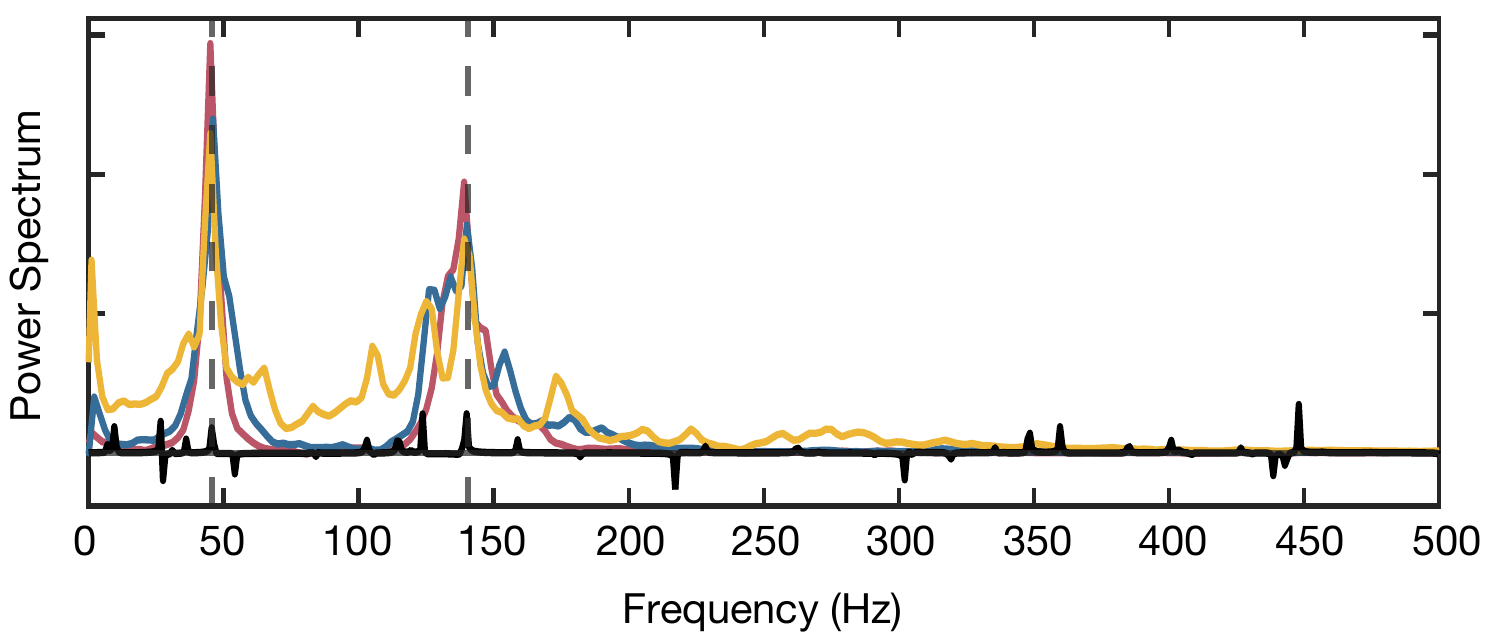}
\caption{\label{ResultsSec}Spectra obtained from three simulations with RF Rabi frequencies of $(2\pi)\times$100 kHz (yellow line), $(2\pi)\times$150 kHz (blue line), and $(2\pi)\times$200 kHz (red line). Vertical dashed gray lines indicate the expected resonances, i.e. the two $\delta^*_i$, of approximately 140 Hz and 45 Hz. For comparison, a spectrum obtained with AERIS is depicted (black line), using an RF Rabi frequency of $(2\pi)\times$150 KHz. In all simulations, the nuclear sample starts in a thermal state corresponding to a temperature of T=300 K in an external magnetic field of 2.1 T. The sample evolves for a total time of 0.5 s.}
\end{figure*}

We test our protocol by simulating its implementation to detect the chemical shifts of hydrogen nuclear spins in ethanol molecules (${\rm C_2H_6O}$) at high external magnetic field. Although ethanol typically exists as a liquid, we employ its solid configuration as an example of an ordered sample with strong homonuclear dipole-dipole couplings. In particular, ethanol molecules exhibit dipolar couplings of up to 17 kHz (the proton attached to the oxygen shows limited dipole interaction with the rest of the system so we exclude it from the simulations). High external fields improve nuclear magnetic resonance procedures, not only because it yields higher polarization rates, but also because it enhances the weaker, and thus harder to detect, energy shifts. Here we consider $B_0=2.1$ T and chemical shifts of 3.66 ppm and 1.19 ppm, resulting in shifts of approximately 327 Hz and 106 Hz, respectively. 

Our numerical simulation unfolds in two phases. First we find the target magnetic signal by simulating the evolution of the nuclear spin sample subjected to the LG4  decoupling sequence. The approximate Hamiltonian in 
Eq.~(\ref{OnerotationH}) facilitates a deeper understanding of the nuclear dynamics and in particular the development of the geometrical interpretation that has set the ground for the design of our protocol. Our simulations, however, make use of the exact nuclear spin Hamiltonian, which reads
\begin{equation}
\begin{aligned}
H(t) &= \sum_{i = 1}^{N = 5} \left\{-\gamma_h \delta_i B_0 I_z^i  + \left[\Omega+\eta(t)\right] I_\phi^i + \Delta I_z^i \right\}+\\
 &+ \sum_{i > {j = 1}}^{N = 5}-\frac{\mu_0 \hbar \gamma_n^2}{8 \pi |r^{ij}|^3}\left(3 r^{ij}_z-1\right)\left[3 I_z^i I_z^j - \vec{I}_i\cdot\vec{I}_j\right],
\end{aligned}
\end{equation}
where $\delta_i$ is the target nuclear shift of the $i$th hydrogen atom and $\eta(t)$ is the driving noise modeled with an Ornstein-Uhlenbeck process with a $1$ ms correlation time and 0.24\% amplitude. 

We assume the sample starting in a completely mixed state. A triggering RF pulse sets the desired initial state $\rho(0)$, a thermal state oriented along the axis perpendicular to $A$ and $B$. From there, the evolution of the nuclear density matrix is simulated using a master equation that includes $T_2^*=0.2$ s relaxation,
\begin{equation}
\dot{\rho} = -i\left[H, \rho\right] + \frac{1}{2 T^*_2}\sum_{j = 1}^{N = 5}\left(4 {I}^j_z \rho {I}^j_z-\rho\right),
\end{equation}
leading to a signal computed as
\begin{equation}
s(t) = \frac{\gamma_h \hbar \mu_0 \sigma_h \mathfrak{f}}{4\pi} \operatorname{Tr}\left[\rho(t)\bar{I_z}\right],
\end{equation}
where $\sigma_h = 5.2\times 10^{28} \mathrm{m^{-3}}$ is the number density of hydrogen spins for ethanol, and $\mathfrak{f} \approx~4.1$ is a geometric factor that relates the sample magnetization and the magnetic field in the NV location, see Refs. \cite{Glenn18, Munuera-Javaloy23} and Section~\ref{seq:signal_produced} for further details.

In the second phase, we simulate the evolution of the NV ensemble interacting with $s(t)$ and subjected to the sensing MW pulse sequence of our protocol. The Hamiltonian that governs the dynamics of each NV centers reads
\begin{equation}
H = \gamma_e s(t)\frac{\sigma_z}{2}+C(t)\frac{\sigma_z}{2}+\frac{\Omega_{\rm NV}(t) \sigma_\phi}{2},
\end{equation}
with $\Omega_{\text{NV}}(t)$  the control Rabi frequency, and $C(t)\frac{\sigma_z}{2}$ describing potential RF-induced crosstalk on the NVs. Following the protocol devised in this work, they interact with the signals produced by the nuclear spin rotations around axis $A$, $\bar A$ and $\bar B$ and accumulate a phase determined by the state of the sample magnetization. During the time that corresponds to the delivery of the RF field over the sample, we transform the accumulated phase into a population difference with a $\pi/2$ pulse and simulate the measurement, after which the sensor is reinitialized and we simulate the detection of the signal produced by the next LG4 block. Finally, a Fourier transform of the measurements provides the spectra displayed in Fig. \ref{ResultsSec}, which proves the ability of our protocol to access the chemical shifts of the molecule.

Figure~\ref{ResultsSec} shows three different experiments with increasing RF intensities. As expected, stronger RF drivings lead to clearer spectra, less distorted by spurious peaks. In particular we simulate RF drivings with Rabi frequencies of $(2\pi)\times$100 kHz, $(2\pi)\times$150 kHz, and $(2\pi)\times$200 kHz, all attainable values by state of the art antennas \cite{Herb20, Yudilevich23}, and set the Rabi frequency of the MW control at 20 MHz in all cases. As intended, our method leads to resonance peaks centered in $\delta_i^*$ from which one can extract the target nuclear shifts $\delta_i$ using Eq. \eqref{eq:newshifts}. For comparison, we include the results of a fourth simulation using AERIS~\cite{Munuera-Javaloy23}, which does not incorporate any dipolar coupling suppression technique. In this case, the obtained spectra (black-curve) is distorted as a consequence of the strong nuclear dipolar couplings. 

In summary, we have designed a protocol that combines LG4 sequences and a tailored NV pulse train to identify chemical shifts in the presence of dipole-dipole interactions. The RF field introduced serves two key purposes: (i) decoupling nuclear spins and (ii) generating a nuclear signal oscillating at a moderate frequency that can be measured by the NVs, allowing the protocol to work at high magnetic fields. By incorporating a tailored MW sequence on the NV for signal detection, we achieve effective retrieval of chemical shifts. Finally, the accuracy of our method is ultimately limited by the nuclear sample decoherence, thus surpassing the limitations imposed by NVs dephasing and thermalization. Our findings pave the way for the advancement of microscale NMR techniques and broaden their application in diverse fields, such as materials science, chemistry, and biology.


\section[Conclusions and perspectives]{Conclusions and perspectives}\label{chapter7}
\fancyhead[LE]{\rightmark}
\fancyhead[RO]{\leftmark}


\thiswatermark{\put(1,-280){\color{l-grey}\rule{70pt}{42pt}}
\put(70,-280){\color{grey}\rule{297pt}{42pt}}}

%
%

\vfill
\lettrine[lines=2, findent=3pt,nindent=0pt]{I}{n} this thesis, we present the quantum control schemes  we have developed to overcome the high-frequency problem in distinct scenarios. Specifically, the high-frequency problem refers to the maximum frequency of an AC signal that the NV center can detect. This limitation arises from driving power constraints, which result in finite-width pulses that create a non-ideal modulation function and, consequently, a significant loss of coupling which seriously affects the sensitivity of NV-based sensors. In extreme cases, pulses can overlap, rendering the protocol completely ineffective.

According to Daly et al. \cite{Daly24}, measuring the proton signal in a magnetic field of 1 T would require pulses of approximately 500 MHz to remain in the instantaneous pulse regime, which is extremely challenging from a technical point of view. Moreover, such high-power driving could damage biological samples, limiting the technique's utility.

Our first approach is to apply Shortcuts to Adiabaticity (STA) to improve the robustness of the technique proposed by Casanova et al. in Ref.~\cite{Casanova18}. In this approach, shaped pulses are designed to modify the modulation function to recover the coupling factor of the instantaneous pulse case. Specifically, STA allows us to reparameterize the problem using the angles traced by the trajectories of the NV state on the Bloch sphere. This flexible parameterization accommodates different conditions with ease. We propose an ansatz for the pulse shape inspired by the Blackman function and optimize the parameters through numerical methods. The application of STA has proven to be highly effective, significantly increasing the robustness of the technique.

This first technique is promising for applications where the signal cannot be manipulated,such as in detection of electrical currents in integrated circuits. However, it necessitates the use of high harmonics to accommodate long shaped pulses, which inevitably leads to a loss of coupling. 

In magnetic resonance related applications, we have control over the system that generates the target signal. If that is the case, we can manipulate the target system to address the high-frequency problem. Specifically, we can couple to target systems through the ZZ interaction, which commutes with the Larmor terms, thus eliminating the need to compensate for high frequencies.

We use this ZZ interaction to couple the NV center to electron spin labels, which exhibit high Larmor frequencies even at moderate magnetic fields due to their large gyromagnetic ratio. Specifically, we propose utilizing a single shallow NV center to measure the coupling between two nitroxide-based electron labels attached to a single molecule. After developing an analytical model of the system, we find that around 30~mT, the nitroxide Hamiltonian becomes especially insensitive to orientation changes. In addition, we show that using labels with different nitrogen isotopes, $^{14}N$ and $^{15}N$, proved beneficial as it avoids spectral peak overlap.

We perform a detailed numerical analysis in the $30$ mT regime and find that the coupling could be inferred by applying a DEER sequence. Our simulations provide numerical evidence that the coupling between nitroxides could be detected under realistic conditions. This finding has potential applications in detecting conformational changes in biological molecules.

Subsequently, we develop a simplified model of the system dynamics that can be computed fast, enabling Bayesian analysis of the simulated experiments. This analysis successfully identifies the coupling between nitroxides and reveals that residual tumbling effects in the system facilitates the extraction of distance information between the nitroxides.

The next work presented in this thesis shifts focus from nitroxide labels to electron spins on the surface of diamonds, exploring their potential to improve polarization transfer. Despite the high frequency associated with these electron spins, we effectively tailor the existing ZZ interaction into a flip-flop Hamiltonian using rotations over both the NV center and the surface electron. We then incorporate $\pi$ pulses to refocus noise, leading to the development of a double PulsePol structure, renowned for its robust properties. This enables us to simultaneously transfer polarization from the surface electron spin to external nuclei by executing a PulsePol sequence between them while simultaneously transferring polarization from the NV center to the surface electron. This work has sparked further developments, see for instance Ref.~\cite{Biteri23}. Utilizing surface electron spins increases the couplings within the system, accelerating polarization transfer. Moreover, we are able to coherently manipulate these electron spins, transforming them from noise sources into active elements in the polarization transfer process.

In the final part of this thesis, we address the high-frequency problem in NV-based NMR spectroscopy. This field is one of the most promising applications for NV centers due to the potential gains in sensitivity and spatial resolution. However, NV-based NMR spectroscopy is significantly impacted by the high-frequency problem, as large magnetic fields are necessary to achieve sufficient chemical shift resolution, increase polarization, and obtain clearer spectra.

To overcome this challenge, we develop a sequence named Amplitude-Encoded Radio Induced Signal (AERIS). By using RF controls, AERIS maps the relevant shifts into a slow signal in the longitudinal magnetization of the sample. The frequency of this induced slow signal depends only on the Rabi frequency of the applied RF, which can be chosen within an optimal range for NV measurement. This approach offers a solution to the high-frequency problem in NV-based NMR spectroscopy, potentially facilitating the development of a new generation of NMR devices. This work has sparked interest in the NV community and has been expanded upon in other works \cite{Alsina23, Daly24} while the experimental group of Prof. Dominik Bucher is planning to  adapt AERIS in their setup. Furthermore, it is part of the roadmap of the QUENCH Research and Innovation Action European project.

The approach of encoding the desired shifts in a crafted signal in the longitudinal magnetization can be extended to other encodings in the, e.g., phase, frequency, amplitude, or in combinations of these parameters. Various protocols can be developed to address the high-frequency problem, each with different strengths and trade-offs. In the final part of this thesis, we develop a protocol that encodes information in both the phase and amplitude of the signal. We use a detuned RF driving, selected to match the Lee-Goldburg condition, which decouples the dipolar interactions within the sample while generating a signal that can be measured by the NV at arbitrarily high fields. Specifically, we apply a Lee-Goldburg 4 (LG4) sequence to the sample and develop a geometric framework to understand and optimize the pulse sequence on the NV center. This protocol provides a way to perform NV-based solid-state NMR at large magnetic fields.

The protocols presented in this thesis are flexible and open to various improvements and modifications. For instance, in the two protocols for NV-based NMR spectroscopy, phase adjustments could be made to enhance robustness. Another interesting path for future research is entangling the NV centers to increase sensitivity, both at the nanoscale and the microscale.

Throughout this thesis, we have significantly deepened our understanding of the high-frequency problem in NV centers. We have explored and developed various quantum control sequences to confront and circumvent this limitation. It  is our believe that the sequences presented here represent a significant step forward in expanding the applications of NV-based quantum sensors.


\titleformat{\section}[display]
{\vspace*{160pt}
\bf\sffamily \Huge}
{{\textcolor{black}{\thesection}}. #1}
{0pt}
{#1}
[]
\titlespacing*{\section}{75pt}{10pt}{40pt}[40pt]

\section*{Appendices}
\phantomsection
\addcontentsline{toc}{section}{Appendices}

\fancyhead[RO]{APPENDICES}

\cleardoublepage

\appendix

\fancyhead[RO]{\leftmark}

\vfill
\titleformat{\section}[display]
{\vspace*{150pt}
\bfseries\sffamily \LARGE}
{\begin{picture}(0,0)\put(-64,-31){\textcolor{black}{\thesection}}\end{picture}}
{0pt}
{\textcolor{black}{#1}}
[]
\titlespacing*{\section}{80pt}{10pt}{50pt}[0pt]
\vfill

\section{Additional details for STA-based shaped pulses}

\fancyhead[RO]{APP. A\quad ADDITIONAL DETAILS REGARDING STA PULSES}

\subsection{Error cancelation condition}\label{app:sta_err}
Here we show the derivation of the approximate transition probability in the presence of errors. We start from the control Hamiltonian including errors. This is 
\begin{equation}
H_c + H_\epsilon = \frac{\Omega(t)(1+\xi_\Omega)}{2}\sigma_{\phi}+\frac{\delta(t)+\xi_\delta}{2}\sigma_z.
\end{equation} 
Now we move to a rotating frame w.r.t. the control. This leads to 
\begin{equation}
H_I = U_0^{\dag}(t)H_\epsilon U_0(t),\label{Interaction Picture}
\end{equation}
where $U_0(t) = \hat T \exp{\left[-i\int_{t_0}^t  \frac{\Omega(s)}{2}\sigma_{\phi}+\sigma_z\frac{\delta(s)}{2} ds\right]}$ is the control Hamiltonian propagator. We can expand now the interaction picture propagator using Dyson series
\begin{equation}
U_I(t_\pi, 0) = \mathbb{I}-i\int_{0}^{t_\pi}dt H_I(t)-\int_{0}^{t_\pi}dt \int_{0}^{t}dt' H_I(t)H_I(t')+...
\end{equation} t
If we now write $H_I(t)$ as in Eq.~(\ref{Interaction Picture}), and multiply the previous expression by $|\phi(0)\rangle$ we get  (up to the second order)
\begin{equation}
\begin{split}
|\phi(t_\pi)\rangle_I \approx \ &|\phi_0(0)\rangle-i\int_{0}^{t_\pi}dt H_I (t) |\phi_0(0)\rangle_I-\\
&-\int_{0}^{t_\pi}dt \int_{0}^{t}dt' H_I(t)H_I(t') |\phi_0(0)\rangle_I,
\end{split}
\end{equation}
where $|\cdot\rangle_I$ represents the state in the interaction picture, while $|\phi_0(t)\rangle$ is the state evolved without errors.  

Now we apply $U_0(t_\pi, 0)$ to $|\phi(t_\pi)\rangle_I$ and find
\begin{equation}
\begin{split}
|\phi(t_\pi)\rangle \approx \ & |\phi_0(t_\pi)\rangle-i\int_{0}^{t_\pi}dt U_0 (t_\pi, t)H_\epsilon |\phi_0(t)\rangle-\\
&-\int_{0}^{t_\pi}dt \int_{0}^{t}dt' U_0(t_\pi, t) H_\epsilon U_0 (t, t')H_\epsilon |\phi_0(t')\rangle.
\end{split}
\end{equation}

At this point we make use of the relation $U_0(t_f, 0) = |\phi_0(t_f)\rangle\langle\phi_0(0)|+|\phi_0^\perp(t_f)\rangle\langle\phi_0^\perp(0)|$, where $|\phi_0^\perp(t)\rangle  = \left[ \sin \left(\frac{\theta}{2}\right)e^{i\frac{\beta}{2}} |1\rangle -\cos\left(\frac{\theta}{2}\right)  e^{-i\frac{\beta}{2}} |0\rangle \right] e^{-i \gamma}$ is the orthogonal state to $|\phi (t) \rangle = \left[\cos\left(\frac{\theta}{2}\right) e^{i\frac{\beta}{2}} |1\rangle + \sin \left(\frac{\theta}{2}\right) e^{-i\frac{\beta}{2}} |0\rangle \right] e^{i \gamma}$. Writing the full expression of $H_\epsilon$ and using the identity $\int_{a}^{b}dx\int_{a}^x dy f(x, y) = \frac{1}{2}\int_{a}^{b}dx\int_{a}^{b} dy f(x, y)$ if $f(x, y) = f(y, x)$ in the integration range, we can obtain that the probability to find $|\phi_0(t_\pi)\rangle$ at the end of the pulse up to order two, i.e. $P(t_\pi) = |\langle\phi_0(t_\pi)|\phi(t_\pi)\rangle|^2$, is
\begin{equation}
P(t_\pi) \approx 1 - \left|\int_{0}^{t_\pi}dt\langle\phi_0^\perp(t)|\left(\frac{\xi_\delta}{2}\sigma_z+\frac{\Omega\ \xi_\Omega}{2}\sigma_\phi\right)|\phi_0(t)\rangle\right|^2.
\end{equation}
Finally, using expression (4) we get the {\it error cancelation condition}
\begin{equation}
P(t_\pi) \approx 1-\left|\int_{0}^{t_\pi}dt \frac{e^{i2 \gamma(t)}}{2}\left(\xi_\delta\sin(\theta) - i2\xi_\Omega\dot{\theta}\sin^2(\theta)\right)\right|^2\ .
\end{equation}

\section{Additional details for nitroxide coupling sensing}
\fancyhead[RO]{APP. B\quad ADDITIONAL DETAILS REGARDING NITROXIDE LABELS}

\subsection{Details about the system Hamiltonian}\label{app:nitro_ham}
In the following, we derive Eq.~(1) of the main text. We start with the Hamiltonian for a driven NV center interacting with two driven nitroxide electron-spin labels. The full Hamiltonian is
\begin{align}
\begin{aligned}
H = &D \left(S^z\right)^2 +B^z|\gamma_e| S^z + H_{n_1}+H_{n_2} +\sum_{i=1,2} H_{{\rm NV}-n_i}\\ + &H_{ee} + \sqrt{2}\Omega_{\rm MW}S^x\cos(\omega_{\rm MW}t) + 2 \Omega_{\rm RF} \left(J_1^x+J_2^x\right) \cos(\omega_{\rm RF} t). \label{eq:SfullHamiltonian}
\end{aligned}
\end{align}
Here, $D = 2\pi \times 2.87$ GHz is the zero-field splitting of the NV center, $|\gamma_e| = 2\pi\times 28$\ MHz/mT is the electron gyromagnetic ratio, $H_{{\rm NV}-n_i} = \frac{\mu_0 \gamma_e^2 \hbar}{4 \pi d_i^3}\left[\vec{S}\cdot\vec{J}_i-\frac{3\left(\vec{S}\cdot\vec{r}_i\right)\left(\vec{J}_i\cdot\vec{r}_i\right)}{d_i^2}\right]$ is the dipolar interaction between the NV and the $i$th label electron, and $H_{ee} = \frac{\mu_0 \gamma_e^2 \hbar}{4 \pi d_{12}^3}\left[\vec{J}_1\cdot\vec{J}_2-\frac{3\left(\vec{J}_1\cdot\vec{r}_{12}\right)\left(\vec{J}_2\cdot\vec{r}_{12}\right)}{d_{12}^2}\right]$ is the dipolar interaction between label electrons. In these equations, $\vec{r}_i$ is the vector joining the NV with the $i$th label, $\vec{r}_{12} = \vec{r}_1 - \vec{r}_2$ is the relative vector that connects the two labels, $d_i = |\vec{r}_i|$, and $d_{12} = |\vec{r}_{12}|$. The last two terms on the right hand side of Eq.~(\ref{eq:SfullHamiltonian}) are the MW and RF driving terms, respectively.
We move to the interaction picture with respect to $H_0 = D \left(S^z\right)^2+|\gamma_e| B^z S^z = \omega_+ |1\rangle\langle 1 | + \omega_- |-1\rangle\langle -1|$, with $\omega_\pm = D\pm|\gamma_e|B^z$. In addition, we set the MW field on resonance with the $0\leftrightarrow 1$ NV transition ($\omega_{\rm MW} = \omega_+$). In the interaction picture, the Hamiltonian takes the form
\begin{align}
\begin{aligned}
H_I = &H_{n_1} + H_{n_2} +S^z\left(\vec{A}_1 \cdot \vec{J}_1+\vec{A}_2 \cdot \vec{J}_2\right)+ H_{ee}\\ + &\frac{\Omega_{\rm MW}}{2}\left(|0\rangle\langle1|+|1\rangle\langle0|\right) + 2 \Omega_{\rm RF} \left(J_1^x+J_2^x\right) \cos(\omega_{\rm RF} t).
\end{aligned}
\end{align}
Note that we substituted $H_{{\rm NV}-n_i} \rightarrow S^z \vec{A}_i \cdot \vec{J}_i$, with $\vec{A}_i = \frac{\mu_0 \gamma_e^2 \hbar}{4 \pi d_i^3}\left[\hat{z}-\frac{3 r_i^z \vec{r}_i}{d_i^2}\right]$. This is justifiable since the strength of the interaction between the NV and labels is much smaller than the transition energies of the NV. Therefore, the terms proportional to $S^{x,y}$ can be neglected in the RWA.

We initialize the NV in the $m_s = 1,0$ manifold. Moreover, the Hamiltonian does not contain terms that can significantly populate the $m_s = -1$ subspace. Consequently, we can project the Hamiltonian on the $m_s = 1,0$ manifold and treat the NV as a two-level system. The resulting Hamiltonian is
\begin{align}
\begin{aligned}
H_I = &H_{n_1} + H_{n_2} +\frac{\mathbb{I}+\sigma^z}{2}\left(\vec{A}_1 \cdot \vec{J}_1+\vec{A}_2 \cdot \vec{J}_2\right)\\ +&H_{ee} +\frac{\Omega_{\rm MW}}{2}\sigma^x + 2 \Omega_{\rm RF} \left(J_1^x+J_2^x\right) \cos(\omega_{\rm RF} t),
\end{aligned}
\end{align}
where $\mathbb{I} = |1\rangle\langle 1| + |0\rangle\langle 0|$, $\sigma^z = |1\rangle\langle 1| - |0\rangle\langle 0|$, and $\sigma^x = |1\rangle\langle 0| + |0\rangle\langle 1|$.

The next approximation consists in removing the counter-rotating terms in $H_{ee}$. These are terms of the form $J_1^+J_2^+$ and $J_1^-J_2^-$ that precess under the externally applied magnetic field $B^z$. They can be neglected in the RWA since the strength of the coupling between label electrons is much smaller than their Zeeman energy. Under this assumption, $H_{ee}$ simplifies to
\begin{align}
\begin{aligned}
H_{ee}\approx \frac{\mu_0 \gamma_e^2 \hbar}{4 \pi d_{12}^3}\left[1-3\left(\frac{r_{12}^z}{d_{12}}\right)^2\right]\left[J_1^z J_2^z-\frac{1}{4}\left(J_1^ + J_2^- + J_1^-J_2^+\right)\right],
\end{aligned}
\end{align}
with $J_i^\pm = J_i^x\pm i J_i^y$ and $g_{12} = \frac{\mu_0 \gamma_e^2 \hbar}{4 \pi d_{12}^3}\left[1-3\left(\frac{r_{12}^z}{d_{12}}\right)^2\right]$. Combining the above approximations leads to Eq.~(1) of the main text.

\subsection{Details about the nitroxide simplified model}\label{app:nitro_simple}
In this section, we develop a simplified model of electron-nucleus dynamics within each nitroxide label. Our approach is based on a perturbative treatment of the nitroxide nuclear degrees of freedom that are orthogonal to the $z$ direction, i.e., the direction of the external magnetic field. We first introduce the rotation matrices used to relate the laboratory axes and the nitroxide principal axes, namely,
\begin{align}
R^y(\theta) =
\begin{pmatrix}
\cos\theta & 0 & \sin\theta\\
0 & 1 & 0\\
-\sin\theta & 0 & \cos\theta
\end{pmatrix}, \quad
R^z(\varphi) =
\begin{pmatrix}
\cos\varphi & -\sin\varphi & 0\\
\sin\varphi & \cos\varphi & 0\\
0 & 0 & 1
\end{pmatrix}.
\end{align}
The two frames are related via the combined rotation $R(\theta,\varphi) = R^z(\varphi)R^y(\theta)$. In particular, a tensor $\mathbb{O}^{(P)}$ defined with respect to the nitroxide principal axes takes the form $\mathbb{O} = R(\theta,\varphi)\mathbb{O}^{(P)} R(\theta,\varphi)^{\intercal}$ in the laboratory frame. For instance, consider the nitroxide Hamiltonians [Eq.~(2) of the main text]. In the present analysis, we ignore the quadrupolar and nuclear Larmor terms since they are small and do not contribute significantly to the dynamics. The validity of this last assumption was confirmed by our detailed numerical simulations. With this assumption, the nitroxide Hamiltonians expressed in the laboratory frame become
\begin{align}
\begin{aligned}
H_{n_i} \approx &\mu_B B^z \hat{z} \cdot \mathbb{C}_i \cdot \vec{J}_i+\vec{J}_i \cdot \mathbb{G}_i \cdot \vec{I}_i =\mu_B B^z \hat{z} \cdot R(\theta_i,\varphi_i) \mathbb{C}_i^{(P)} R(\theta_i,\varphi_i)^{\intercal} \cdot \vec{J}_i \\&+ \vec{J}_i \cdot R(\theta_i,\varphi_i)\mathbb{G}_i^{(P)} R(\theta_i,\varphi_i)^{\intercal} \cdot \vec{I}_i.
\end{aligned}
\end{align}
Due to the symmetry of the $\mathbb{C}_i^{(P)}$ and $\mathbb{G}_i^{(P)}$ tensors ($C^x\approx C^y\approx 2.007$ and $G^x\approx G^y\approx 2\pi \times 14 \ {\rm MHz}$) we can set $\varphi_i = 0$ without loss of generality. We find
\begin{align}
\begin{aligned}
H_{n_i} = &\frac{\mu_B B^z}{2} (C^\parallel - C^\perp) \sin (2 \theta_i) J_i^x + \frac{\mu_B B^z}{2} \left[( C^\parallel- C^\perp) \cos (2 \theta_i)+ C^\perp+ C^\parallel\right]J_i^z \\
&+\frac{G^{\parallel }- G^{\perp }}{2} \sin(2\theta_i)J_i^z I_i^x +\left(\frac{G^{\parallel } + G^{\perp}}{2} + \frac{G^{\parallel} - G^{\perp}}{2} \cos(2\theta_i)\right)J_i^z I_i^z \\
&+G^{\perp} J_i^y I_i^y + \left(\frac{G^{\parallel } + G^{\perp}}{2} - \frac{G^{\parallel} - G^{\perp}}{2} \cos(2\theta_i) \right)J_i^x I_i^x \\&+ \frac{G^{\parallel}- G^{\perp}}{2} \sin(2\theta_i)J_i^x I_i^z. \label{eq:Sroot}
\end{aligned}
\end{align}

\subsubsection{Simplified dynamics of a nitroxide label hosting $^{14}$N}\label{app:nitro_14n}
\begin{figure*}[h]
\centering
\hspace{-0. cm}\includegraphics[width=0.9\linewidth]{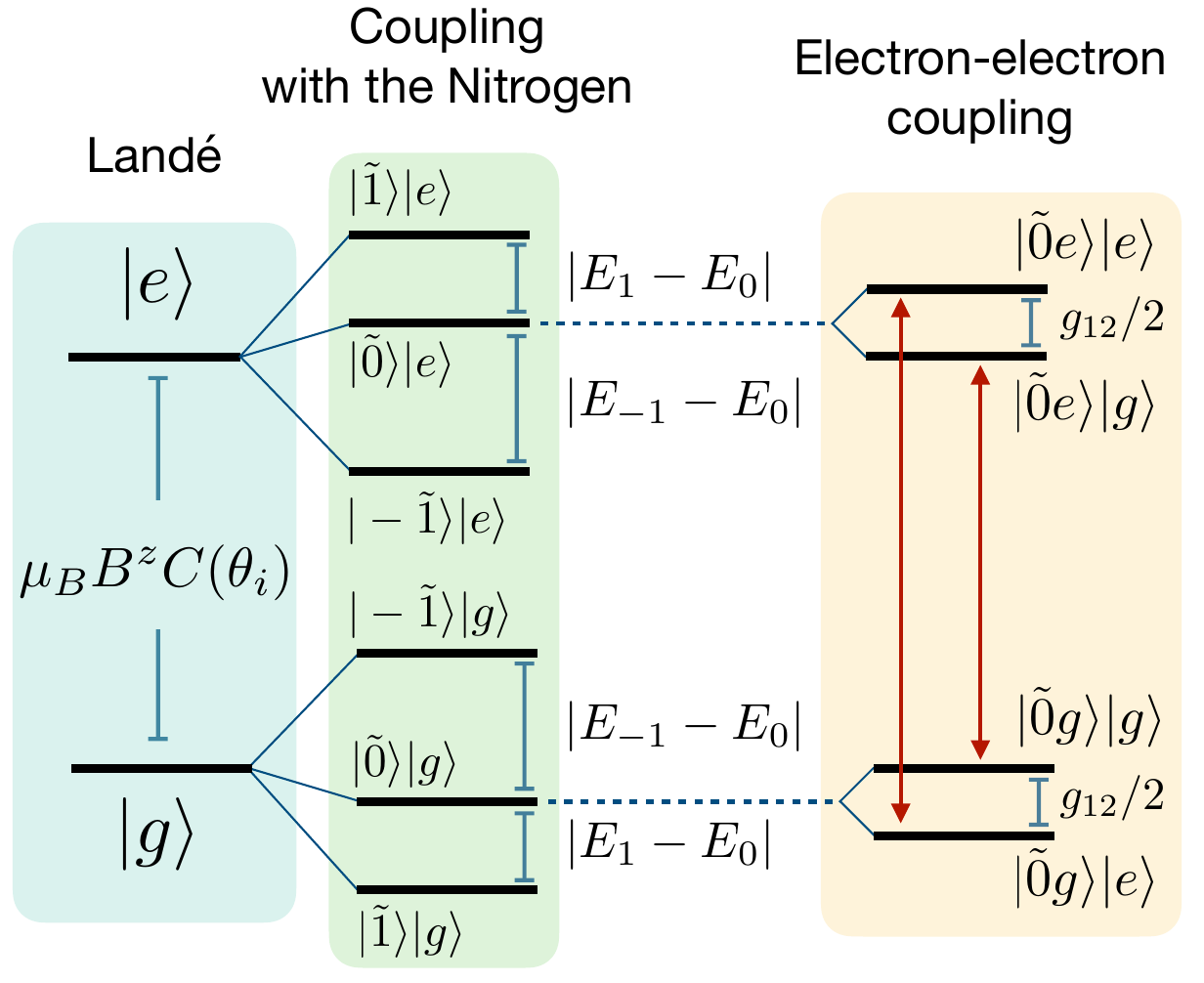}
\caption{Energy diagram of a nitroxide label hosting a $^{14}$N. A magnetic field $B^z$ is applied along the laboratory $z$ axis, leading to a Land\'e splitting of the electronic states (blue panel). The hyperfine coupling between the label electron and the nitrogen then splits the energy levels (green panel). Finally, each level is further split in two by the coupling between label electrons (yellow panel). The red arrows in the yellow panel indicate the transitions targeted by our protocol. \label{fig:Ssplittings}}
\end{figure*}

\begin{figure*}[t]
\centering
\hspace{-0. cm}\includegraphics[width=0.9\linewidth]{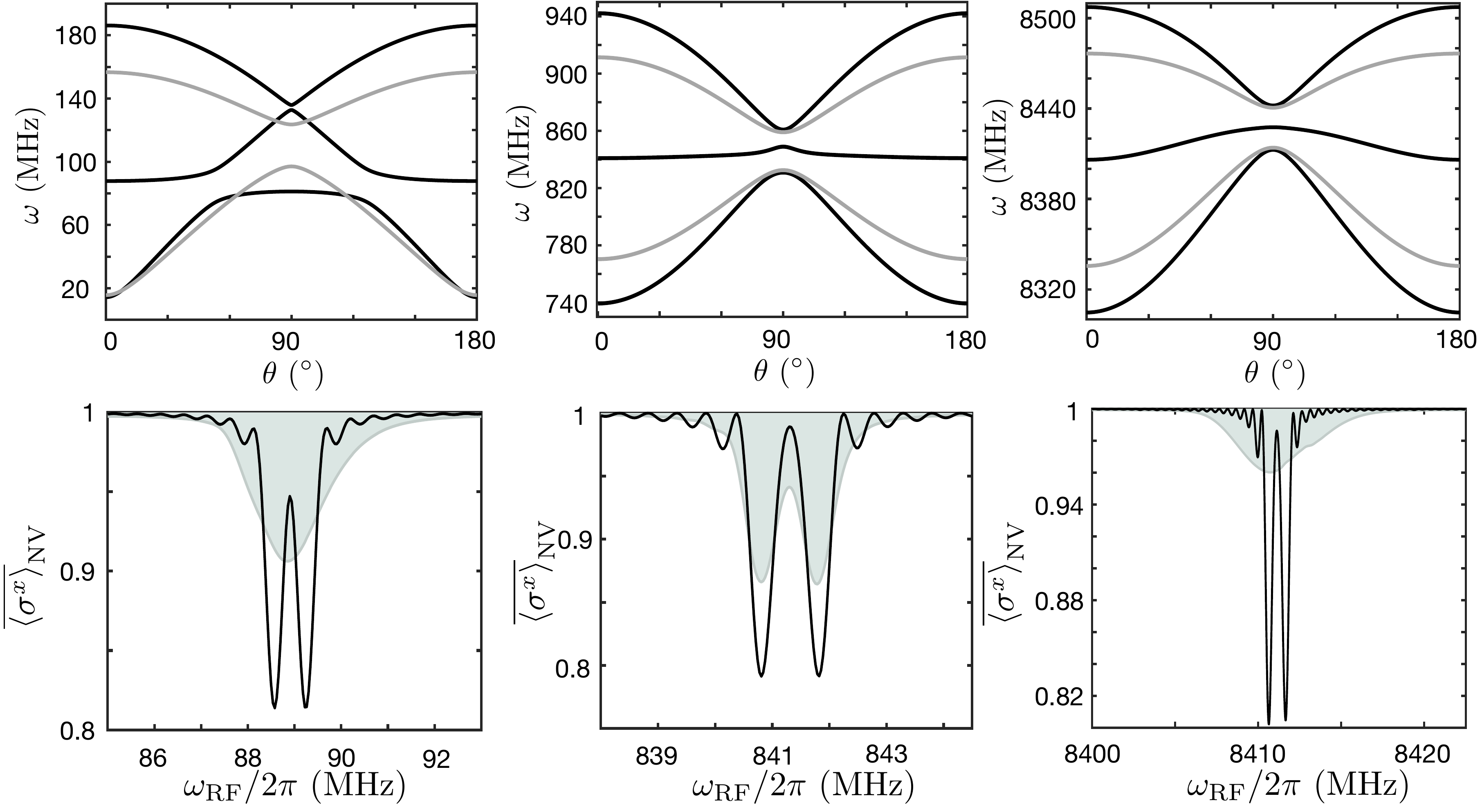}
\caption{(Top panels) Energy-transition branches of a nitroxide as a function of its azimuth angle $\theta$ for $B =$ $3$, $30$ and $300$\ mT (left to right). Solid black (grey) lines correspond to a nitroxide hosting $^{14}$N ($^{15}$N). All branches are obtained by diagonalizing Eq.~(2) of the main text. (Bottom panels) Numerical simulation of the average NV spectrum $\overline{\langle\sigma^x\rangle}_{\rm NV}$ for the same magnetic fields as in the top panels. All simulations are performed by unitary propagation of Eq.~(1) of the main text and using the full nitroxide Hamiltonian, Eq.~(2). We show the spectrum for an equilibrium azimuth $\theta_{\rm eq} = 30 ^\circ$ (black lines) and the spectra averaged over a Gaussian distribution of angles centered at $\theta_{\rm eq}$ and with standard deviation $\sigma_\theta = 6.25 ^\circ$ (shaded areas). At intermediate magnetic fields, the spectrum is resilient to tumbling due to a weak dependence of the central branch $E_0$ on $\theta$. \label{fig:SOrientation}}
\end{figure*}

As a first approximation, we keep only the terms proportional to $J_i^z$ in Eq.~(\ref{eq:Sroot}) . This is a reasonable first approximation since the hyperfine interaction is typically smaller than the electronic Zeeman energy. The resulting Hamiltonian is
\begin{align}
\begin{aligned}
H_{\rm diag} &= \frac{\mu_B B^z}{2} \left[( C^{\parallel}- C^{\perp}) \cos (2 \theta_i )+ C^{\perp}+ C^{\parallel}\right]J_i^z \\&+\left(\frac{G^{\parallel } + G^{\perp}}{2}+ \frac{G^{\parallel} - G^{\perp}}{2} \cos(2\theta_i)\right) J_i^z I_i^z + \frac{G^{\parallel }- G^{\perp }}{2} \sin(2\theta_i)J_i^z I_i^x \\
&= \mu_B B^z C(\theta_i) J_i^z \\&+ \left[ \left(\frac{G^{\perp} + G^{\parallel}}{2} + \frac{G^{\parallel} - G^{\perp}}{2}\cos(2\theta_i) \right) I_i^z + \frac{G^{\parallel }- G^{\perp }}{2} \sin(2\theta_i)I_i^x \right]J_i^z, \label{eq:Sdiagonal}
\end{aligned}
\end{align}
where $C(\theta) =\frac{1}{2} \left[( C^\parallel- C^\perp) \cos (2 \theta )+ C^\perp+ C^\parallel\right]$. To find the shifts on $J_i^z$ associated to specific nuclear spin states, we diagonalize the term
\begin{align}
\begin{aligned}
\left(\frac{G^{\perp} + G^{\parallel}}{2} + \frac{G^{\parallel} - G^{\perp}}{2}\cos(2\theta_i) \right) I_i^z + \frac{G^{\parallel }- G^{\perp }}{2} \sin(2\theta_i)I_i^x. \label{eq:Stodiagonalise}
\end{aligned}
\end{align}
Since $^{14}$N is a spin-1 particle, diagonalization leads to three nuclear eigenstates $|\widetilde{0}\rangle$, $|\widetilde{1}\rangle$, and $|-\widetilde{1}\rangle$ . The states have eigenenergies
\begin{align}
\begin{aligned}
&\omega_0^i=0,\\
&\omega_1^i=\frac{1}{\sqrt{2}} \sqrt{\left[(G^{\parallel})^2- (G^{\perp})^2\right] \cos (2 \theta_i )+(G^{\perp})^2+(G^{\parallel})^2}, \\
&\omega_{-1}^i=-\frac{1}{\sqrt{2}} \sqrt{\left[(G^{\parallel})^2- (G^{\perp})^2\right] \cos (2 \theta_i )+(G^{\perp})^2+(G^{\parallel})^2}.
\end{aligned}
\end{align}
Thus, our leading approximation for the nitroxide Hamiltonian is
\begin{align}
\begin{aligned}
H_{n_i} \approx \left[E_1^i |\widetilde{1}\rangle\langle \widetilde{1}| + E_0^i |\widetilde{0}\rangle\langle \widetilde{0}| + E_{-1}^i |-\widetilde 1\rangle\langle -\widetilde{1}|\right] J_i^z, \label{eq:firstOrder}
\end{aligned}
\end{align}
where
\begin{align}
\begin{aligned}
&E_0^i= \mu_B B^z C(\theta_i), \\
&E_1^i= \mu_B B^z C(\theta_i) + \frac{1}{\sqrt{2}} \sqrt{\left[(G^{\parallel})^2- (G^{\perp})^2\right] \cos (2 \theta_i )+(G^{\perp})^2+(G^{\parallel})^2}, \\
&E_{-1}^i=\mu_B B^z C(\theta_i)-\frac{1}{\sqrt{2}} \sqrt{\left[(G^{\parallel})^2- (G^{\perp})^2\right] \cos (2 \theta_i )+(G^{\perp})^2+(G^{\parallel})^2}.
\end{aligned}
\end{align}
The above expressions reveal the existence of three energy-transition branches. There is one central branch with energy $\mu_B B^z C(\theta_i)$. The other two branches are shifted from the central branch by $\pm \left|\frac{1}{\sqrt{2}} \sqrt{\left[(G^{\parallel})^2- (G^{\perp})^2\right] \cos (2 \theta_i )+(G^{\perp})^2+(G^{\parallel})^2}\right|$. A full energy diagram is given in Fig.~\ref{fig:Ssplittings}.
Note that for moderate magnetic fields, $E_0^i$ depends much more weakly than $E_{1,-1}^i$ on the nitroxide azimuth $\theta_i$ since the Land{\'e} tensor is almost isotropic, $C^\parallel \simeq C^\perp$, while the hyperfine tensor is strongly anisotropic, $G^\parallel \neq G^\perp$. Thus, the central branch is particularly important since it can be robust against molecular tumbling (i.e., robust against changes in $\theta_i$). Note, however, that the dependence of $E_0^i$ on $\theta_i$ due to the small anistropy of the Land{\'e} tensor can become significant if the magnetic field becomes too large. At low magnetic fields, the above approximation starts to break down. We now investigate the leading correction to the central branch $E_0^i$ due to the hyperfine interaction in that regime. This correction arises from the nondiagonal terms in Eq.~\eqref{eq:Sroot}, i.e., the terms proportional to $J_i^{x,y}$. We first rewrite the nondiagonal terms in Eq.~\eqref{eq:Sroot} as
\begin{align}
\begin{aligned}
V = \alpha_i J_i^x+\beta_i J_i^y,
\end{aligned}
\end{align}
where
\begin{align}
\begin{aligned}
&\alpha_i = \left(G^\perp\cos^2(\theta_i)+G^\parallel\sin^2(\theta_i)\right)I^x+\left(G^\parallel-G^\perp\right)I^z\cos(\theta_i)\sin(\theta_i), \\ &\beta_i = G^\perp I^y.
\end{aligned}
\end{align}
We find the second-order correction due to $V$ by performing operator perturbation theory. More precisely, we move to a rotating frame with respect to $\mu_B B^z C(\theta_i) J_i^z$ and find the effective contribution of $V$ by keeping time-independent terms in its Dyson series (expanded up to second order). This leads to an effective interaction $\frac{(\alpha_i-i \beta_i)(\alpha_i+i\beta_i)}{2\mu_B B^z C(\theta_i)} J_i^z$. Projecting this expression onto the $|\widetilde{0}\rangle$ subspace gives the leading correction to $E_0^i$, 
\begin{align}
\begin{aligned}
\frac{1}{2 \mu_B B^z C(\theta_i)}\frac{2 (G^{\perp} G^{\parallel})^2 +\left[(G^{\perp})^4-(G^{\perp} G^{\parallel})^2\right] \sin^2(\theta_i)}{(G^{\perp})^2 \sin^2(\theta_i)+(G^{\parallel})^2 \cos^2(\theta_i)}.
\end{aligned}
\end{align}
This expression shows that the anisotropy of the hyperfine tensor can lead to a strong dependence of $E_0^i$ on $\theta_i$ when the magnetic field becomes small enough to be comparable to the hyperfine coupling. From the above arguments, we therefore expect that there exists an optimum magnetic field strength where the central branch is most robust to variations in the azimuth $\theta_i$. This is illustrated in Fig.~\ref{fig:SOrientation}.

In summary, we write the nitroxide Hamiltonian as
\begin{align}
\begin{aligned}
H_{n_i} \approx \left[ E_1^i |\widetilde{1}\rangle\langle \widetilde{1}|_i + E_0^i |\widetilde{0}\rangle\langle \widetilde{0}|_i + E_{-1}^i |-\widetilde 1\rangle\langle -\widetilde{1}|_i \right] J_i^z,
\end{aligned}
\end{align}
with
\begin{align}
\begin{aligned}
&E_0^i= \mu_B B^z C(\theta_i) + \frac{1}{2 \mu_B B^z C(\theta_i) }\frac{2 (G^{\perp} G^{\parallel})^2 +\left[(G^{\perp})^4-(G^{\perp} G^{\parallel})^2\right] \sin^2(\theta_i)}{(G^{\perp})^2 \sin^2(\theta_i)+(G^{\parallel})^2 \cos^2(\theta_i)}
, \\
&E_1^i= \mu_B B^z C(\theta_i) + \frac{1}{\sqrt{2}} \sqrt{\left[(G^{\parallel})^2- (G^{\perp})^2\right] \cos (2 \theta_i )+(G^{\perp})^2+(G^{\parallel})^2}, \nonumber\\
&E_{-1}^i=\mu_B B^z C(\theta_i)-\frac{1}{\sqrt{2}} \sqrt{\left[(G^{\parallel})^2- (G^{\perp})^2\right] \cos (2 \theta_i )+(G^{\perp})^2+(G^{\parallel})^2}.
\end{aligned}
\end{align}

\subsubsection{Simplified dynamics of a nitroxide label hosting $^{15}$N}\label{app:nitro_15n}
For a $^{15}$N, (i.e., a spin-1/2 particle) the diagonalization of Eq.~\eqref{eq:Stodiagonalise} yields two nitrogen eigenstates $|\widetilde{1/2}\rangle$ and $|-\widetilde{1/2}\rangle$ with eigenergies
\begin{align}
\begin{aligned}
E_{1/2,-1/2}^i=\mu_B B^z C(\theta_i) \pm \frac{\sqrt{(G^\perp)^2+(G^\parallel)^2+\left[(G^\parallel)^2-(G^\perp)^2\right]\cos(2\theta_i)}}{2 \sqrt{2}}.
\end{aligned}
\end{align}

\subsection{Dissipative model}\label{app:nitro_dissi}
Our detailed numerical simulations account for decoherence in ambient conditions using a Lindblad master equation of the form
\begin{align}
\begin{aligned}
\dot{\rho} =& -i\left[H, \rho\right] + \frac{1}{2T_2}\left(\sigma^z \rho \sigma^z - \rho\right) \\
&+ \sum_{i=1,2}\bigg[\Gamma (\bar{n} + 1) \left(J_i^- \rho J_i^+ - \frac{1}{2}J_i^+J_i^-\rho-\frac{1}{2}\rho J_i^+J_i^-\right) \\&+ \Gamma \bar{n} \left(J_i^+ \rho J_i^- - \frac{1}{2}J_i^-J_i^+\rho-\frac{1}{2}\rho J_i^-J_i^+\right)\bigg].
\end{aligned}
\end{align}
Here, $T_2 = 20\ \mu$s is the NV coherence time for an NV depth of $4$\ nm~\cite{Shi15}, $\Gamma \approx 2\pi \times 2.68$ Hz is the relaxation rate of the label electrons, and $\bar{n} = 1/\left[\exp\left(\hbar \gamma_e B^z/k_B T\right) - 1\right]$ is the thermal occupation of a bosonic thermal bath of temperature $T$ that generates electronic transitions. For the magnetic field $B^z = 30$\ mT and the temperature $T = 300$\ K used in our simulations, this corresponds to an electronic relaxation time $T_1 = 1/\left[(2\bar{n}+1)\Gamma\right] \approx 4\ \mu$s~\cite{Shi15}. The coherent part of the evolution in the presence of driving is described by the Hamiltonian
\begin{align}
\begin{aligned}
H = &\frac{\Omega_{\rm MW}}{2} \sigma^x + g_{12}\left[J_1^z J_2^z-\frac{1}{4}\left(J_1^+ J_2^- + J_1^- J_2^+ \right)\right] \\
+& \sum_{i=1,2}\bigg[\frac{1}{2}\left(\mathbb{I} + \sigma^z \right) \vec{A}_i \cdot \vec{J}_i + \mu_B B^z \hat{z}\cdot \mathbb{C}_i \cdot\vec{J}_i+ \gamma_N B^z I_i^z\\+&\vec{I}_i\cdot\mathbb{Q}_i\cdot\vec{I}_i+\vec{J}_i\cdot\mathbb{G}_i\cdot \vec{I}_i + 2 \Omega_{\rm RF} J_i^x \cos(\omega_{\rm RF} t)\bigg]. \label{eq:Slindblad}
\end{aligned}
\end{align}
The above model was previously used to accurately describe the experimental results of Ref.~\cite{Shi15}.

\subsection{Molecular tumbling}\label{app:nitro_tumbl}
\begin{figure*}[h]
\centering
\hspace{-0. cm}\includegraphics[width=1\linewidth]{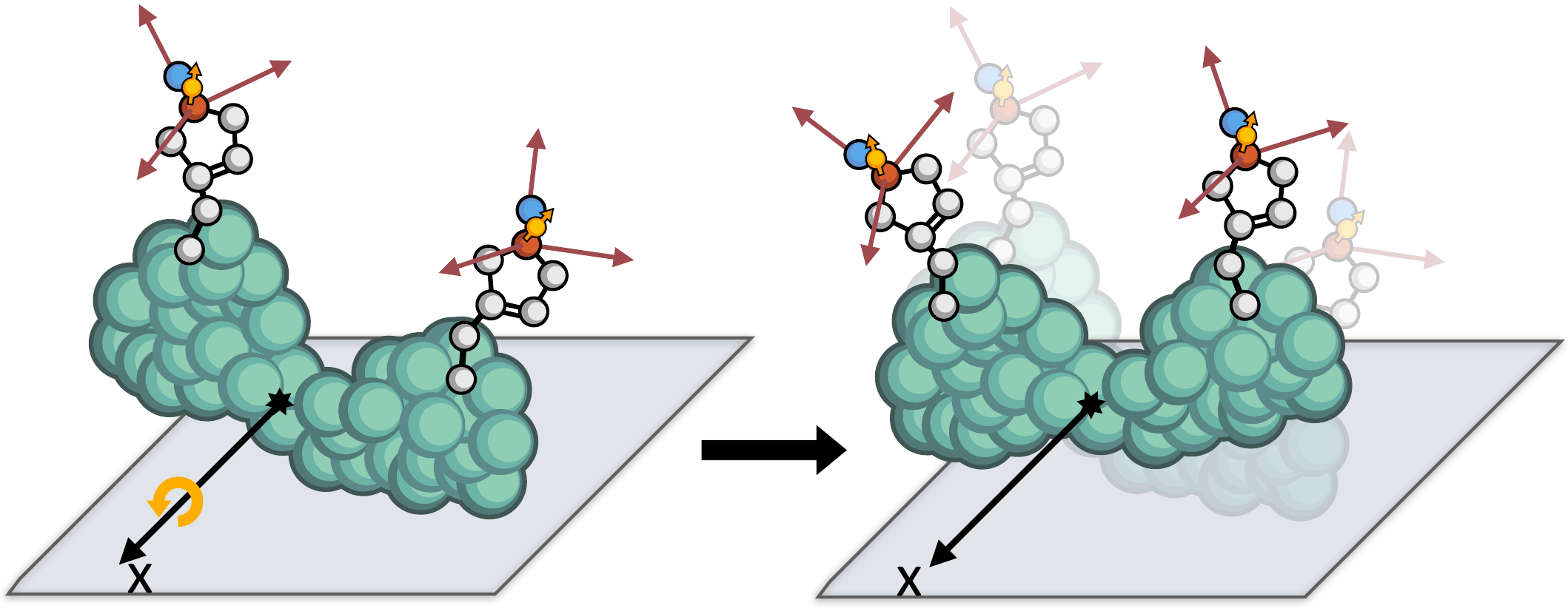}
\caption{Starting from an initial configuration (left), the molecule is slightly rotated around an axis parallel to the laboratory $x$ axis (right). In the right panel, the transparent image shows the initial configuration for better comparison. \label{fig:STumbling}}
\end{figure*}

Figure~\ref{fig:STumbling} depicts the effect of molecular tumbling on the orientations and positions of the nitroxides before rotation and after rotation. For all simulations, the molecule rotates around an axis that is parallel to the laboratory $x$ axis and that contains the point $\vec{r}_0 = (3,0,6)$\ nm. For the simulations of Fig.~3(a,d), the equilibrium positions of the two nitroxides are $\vec{r}_{1,{\rm eq}} = (-2.10,2.17,6.24)$\ nm and $\vec{r}_{2,{\rm eq}} = (0.4,0.3,7.3)$\ nm and their equilibrium orientations are $(\theta_{1,{\rm eq}},\varphi_{1,{\rm eq}}) = (11.46,-91.67)^\circ$ and $(\theta_{2,{\rm eq}},\varphi_{2,{\rm eq}}) = (91.67,154.70)^\circ$. For the simulation of Fig.~3(b), the equilibrium positions are $\vec{r}_{1,{\rm eq}} = (-2.10,2.17,6.24)$\ nm and $\vec{r}_{2,{\rm eq}} = (0.89,0.39,8.27)$\ nm and the equilibrium orientations are $(\theta_{1,{\rm eq}},\varphi_{1,{\rm eq}}) = (11.46,-91.67)^\circ$ and $(\theta_{2,{\rm eq}},\varphi_{2,{\rm eq}}) = (91.67,154.70)^\circ$. For the simulation of Fig.~3(c), the equilibrium positions are $\vec{r}_{1,{\rm eq}} = (-2.10,2.17,6.24)$\ nm and $\vec{r}_{2,{\rm eq}} = (0.4,0.3,7.3)$\ nm and the equilibrium orientations are $(\theta_{1,{\rm eq}},\varphi_{1,{\rm eq}}) = (63.03,-91.67)^\circ$ and $(\theta_{2,{\rm eq}},\varphi_{2,{\rm eq}}) = (51.57,154.70)^\circ$. All positions are measured with respect to an origin located at the NV center. The average spectrum $\overline{\langle\sigma^x\rangle}_{\rm NV}$ is obtained by averaging spectra assuming that the rotation angle $\delta$ follows a Gaussian distribution with zero mean and standard deviation $6.25^\circ$. Finally, note that the molecular tumbling has a correlation time of $\sim$1 ms for typical proteins at room temperature~\cite{Shi15}. Since the NV signal must be averaged over several seconds, all tumbling configurations are averaged over in the course of the experiment

\subsection{Inference model \label{app:nitro_infer}}
In this section, we motivate the expression used for Bayesian parameter inference in the main text. We make the following simplifying assumptions:
\begin{enumerate}
\item{The system is simplified to a single NV interacting with a single target nitroxide ($^{14}$N) with orientation $(\theta,\varphi)$. The only role of the other nitroxide is to introduce an energy shift $\pm g_{12}(\beta)/2$ on the target nitroxide depending on its state. Here, $g_{12}(\beta)\propto d_{12}^{-3}(1-3\cos^2\beta)$ is the dipole-dipole coupling and $\beta$ is the angle between the magnetic field and the vector $\vec{r}_{12}$ joining the nitroxides.}
\item{The longitudinal coupling $a^z$ between the NV and target nitroxide stays approximately constant during molecular tumbling. This is a reasonable approximation for small molecular tumbling angles.}
\item{The driving frequency $\omega_{\rm RF}$ is close to resonance with the $E_0(\theta)$ energy-transition branch of the target nitroxide and far off resonance with the other branches. Therefore, non-trivial evolution of the target nitroxide occurs only for nuclear eigenstate $|\widetilde{0}\rangle$.}
\item{The $\pi$-pulse on the NV is simply described by the action of the operator $-i \sigma^x$, i.e., the interaction between the NV and label electrons during irradiation can be neglected.}
\end{enumerate}
With these assumptions, the Hamiltonian of the target nitroxide during free evolution is $H_{{\rm free},\pm} = \left\{\left[E_0(\theta) - \omega_{\rm RF} \pm \frac{g_{12}(\beta)}{2}\right] J^z + \frac{a^z}{2} \sigma^z J^z\right\}\otimes|\widetilde{0}\rangle\langle\widetilde{0}|$ . Similarly, the Hamiltonian of the target nitroxide during irradiation is \begin{equation}
H_{{\rm pulse},\pm} = \left\{\left[E_0(\theta) - \omega_{\rm RF} \pm \frac{g_{12}(\beta)}{2}\right]J^z + \Omega_{\rm RF} J^x\right\}\otimes|\widetilde{0}\rangle\langle\widetilde{0}|.
\end{equation}
In these expressions, the index $\pm$ labels the states of the other nitroxide electron. For coherent dynamics, these Hamiltonians yield the propagator 
\begin{equation}
U_\pm = e^{-i H_{{\rm free},\pm} \tau_{\rm free}} e^{-i H_{{\rm pulse},\pm} \frac{\pi}{\Omega_{\rm RF}}} (-i \sigma^x) e^{-i H_{{\rm free},\pm} \tau_{\rm free}}.
\end{equation}

We assume an initial state $\rho = | + \rangle\langle +| \otimes \frac{\mathbb{I}}{2} \otimes \frac{\mathbb{I}}{3}$, i.e., the target nitroxide is in a fully mixed state. We also assume that the other nitroxide is in a fully mixed state so that the evolutions $U_\pm$ have equal probability. With these initial conditions, we find that the spectrum $\mathcal{S}(\omega_{\rm RF},\theta,\beta) = \langle\sigma^x\rangle_{\rm NV}$ has the form
\begin{align}
\begin{aligned}
\mathcal{S}(\omega_{\rm RF}, \theta, \beta) &\approx \sum_{s=+,-} \frac{1}{2}{\rm Tr}\left(U_s \rho U^\dag_s \sigma^x\right) \\&= \mathcal{S}_0 - \sum_{s=+,-} \mathcal{C}_s \left[\frac{\Omega_{\rm RF}}{\Omega_s(\omega_{\rm RF},\theta,\beta)}\right]^2\sin^2\left[\frac{\pi}{2} \frac{\Omega_s(\omega_{\rm RF},\theta,\beta)}{\Omega_{\rm RF}}\right], \label{eq:ScoherentSpectrum}
\end{aligned}
\end{align}
where $\Omega_\pm^2(\omega_{\rm RF},\theta,\beta) = \Omega_{\rm RF}^2 + \left[\omega_{\rm RF}-E_0(\theta) \pm g_{12}(\beta)/2 \right]^2$, $\mathcal{S}_0 = 1$, and \\$\mathcal{C}_s = \frac{1}{6}\sin^2\left(\frac{a^z \tau_{\rm free}}{2}\right)$.

For a fixed value of $\tau_{\rm free}$, we expect the main effect of dissipation to be two-fold. First, NV decoherence can modify the baseline NV response $\mathcal{S}_0$ when the drive is off-resonant. Second, NV decoherence and relaxation of the nitroxide electrons can reduce the contrast $\mathcal{C}_s$ in a way that may depend on $s = \pm$. Therefore, we leave $\mathcal{S}_0$, $\mathcal{C}_+$, and $\mathcal{C}_-$ as adjustable parameters for the dissipative model. In the presence of molecular tumbling, the average spectrum $\overline{\mathcal{S}}(\omega_{\rm RF})$ is obtained by averaging Eq.~\eqref{eq:ScoherentSpectrum} over a suitable distribution of the tumbling angle $\delta$, with the angles $\theta$ and $\beta$ both depending on $\delta$. For tumbling by an angle $\delta$ around an axis parallel to the laboratory $x$ axis (see Fig.~\ref{fig:STumbling}), the azimuth angle $\theta$ of the target nitroxide is related to $\delta$ by
\begin{align}
\begin{aligned}
\theta(\delta) = \arccos\left[\cos(\delta)\cos(\theta_{\rm eq})+\sin(\delta)\sin(\theta_{\rm eq})\sin(\varphi_{\rm eq})\right],
\end{aligned}
\end{align}
where $(\theta_{\rm eq},\varphi_{\rm eq})$ is the equilibrium orientation of the target nitroxide. Moreover, the dependence of the angle $\beta$ on the tumbling angle $\delta$ is described by the expression $\cos^2[\beta(\delta)]=A_\beta^2\cos^2(\delta + \phi_\beta)$, where $A_\beta$ and $\phi_\beta$ are parameters capturing the orientation of $\vec{r}_{12}$ in the $x-y$ plane. Using these relations, the average spectrum $\overline{\mathcal{S}}(\omega_{\rm RF})$ is obtained by numerically averaging $\mathcal{S}[\omega_{\rm RF},\theta(\delta),\beta(\delta)]$ over $\delta$ assuming that $\delta$ is Gaussian-distributed with unknown standard deviation $\sigma_\delta$. When trying to infer $d_{12}$, there are nine adjustable parameters in total. We assume that $\mathcal{S}_0$ is calibrated and known. The remaining eight adjustable parameters are ${\bf V}=\{\mathcal{C}_+,\mathcal{C}_-,\theta_{\rm eq},\varphi_{\rm eq},A_\beta,\phi_\beta,d_{12},\sigma_\delta\}$.

\subsection{Inference of the distance between nitroxide electron-spin labels}\label{app:nitro_dis}
\begin{figure*}[h]
\centering
\hspace{-0. cm}\includegraphics[width=0.95\linewidth]{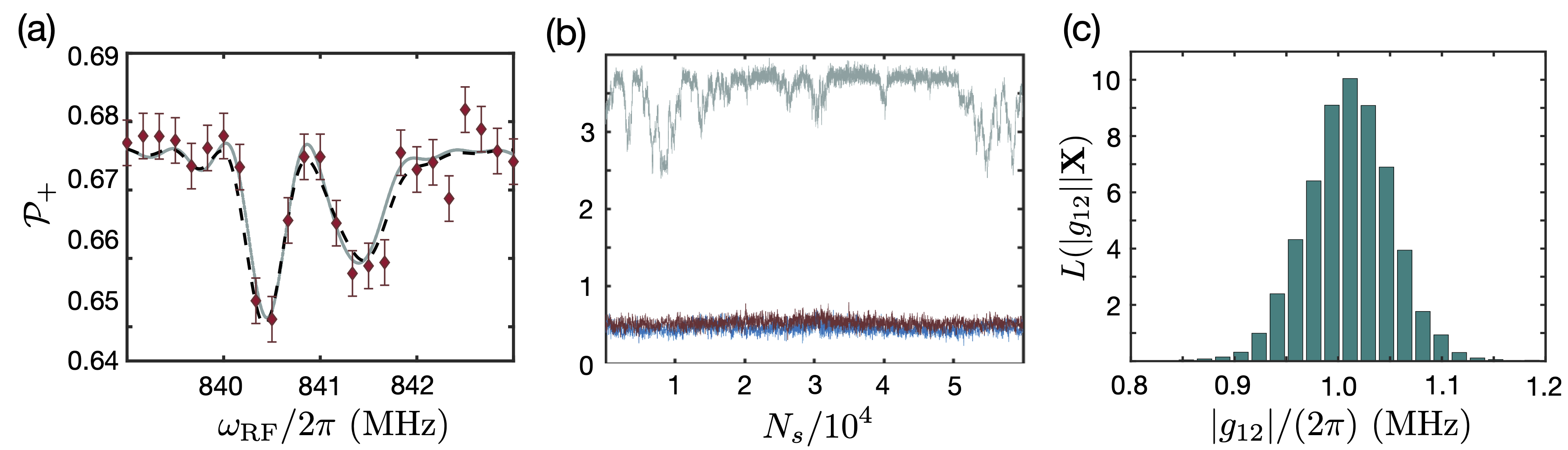}
\caption{(a) Simulated experimental dataset ${\bf X}$ (red points). The dataset contains $M=25$ frequencies $\omega_{{\rm RF},j}$ swept across the target nitroxide resonance, with $N_m = 2\times 10^4$ ideal measurements for each frequency. The error bars are given by $\sigma_m = \sqrt{\left(1-\overline{S}^2\right)/4N_m} \approx \sqrt{0.22/N_m}$. The ``true'' spectrum $\left(\overline{\langle\sigma^x\rangle}_{\rm NV}+1\right)/2$ (solid line) and the simplified model $\left(\overline{\mathcal{S}}(\omega_{\rm RF})+1\right)/2$ (dashed line) are also shown (see text for details about the used parameters). (b) Metropolis Markov chain for the parameters $\mathcal{C}_\pm$ (red and blue) and $d_{12}$ (grey). The contrasts $\mathcal{C}_\pm$ are dimensionless and are multiplied by $10$ for easier viewing. The distance $d_{12}$ is measured in nm. (c) Marginal posterior $L(|g_{12}||{\bf X})$ obtained from the Markov chain by combining the parameters $d_{12}$, $A_\beta$, and $\phi_\beta$ according to $g_{12}=\left(\mu_0 \gamma_e^2 \hbar/4\pi d_{12}^3\right) \left(1-3A_\beta^2\cos^2 \phi_\beta \right)$. All samples before $N_s = 10^4$ were discarded. The expectation value is $|g_{12}|/2\pi=1.010(41)$\ MHz, in good agreement with the ideal value $|g_{12}|/2\pi = 1$\ MHz. The error is given by the standard deviation of the marginal posterior. \label{fig:SInference}}
\end{figure*}

We now describe how to infer system parameters from the simplified model discussed in App.~\ref{app:nitro_infer}. We first simulate the acquisition of experimental data from a spectrum obtained numerically using Eq.~\eqref{eq:Slindblad}. Here, we use the spectrum of Fig.~3(a) in the main text. The data has the form ${\bf X}=\{X_1,\ldots,X_M\}$, where $X_j$ is an estimate of the probability $\mathcal{P}_+ = \left(1+\overline{\langle \sigma^x\rangle}_{\rm NV}\right)/2$ of the NV occupying the $+1$ eigenstate of $\sigma^x$ when the target nitroxide is driven at frequency $\omega_{{\rm RF},j}$. Assuming ideal NV measurements, the simulated values of $X_j$ are drawn independently from a Binomial distribution, $X_j \sim B(N_m,\mathcal{P}_+)/N_m$. Fig.~\ref{fig:SInference}(a) shows an example simulated dataset ${\bf X}$. The ``true'' spectrum $\mathcal{P}_+ = \left(1+\overline{\langle \sigma^x\rangle}_{\rm NV}\right)/2$ and the approximate spectrum $\mathcal{P}_+ = \left(1+\overline{\mathcal{S}}(\omega_{\rm RF})\right)/2$ are also shown for comparison.

The values of $\theta_{\rm eq},\varphi_{\rm eq}, A_\beta, \phi_\beta, d_{12}$ and $\sigma_\delta$ used to plot the approximate spectrum are the same as for the ``true'' spectrum, while the contrasts $C_{\pm}$ and baseline $\mathcal{S}_0$ are adjusted to fit the ``true'' spectrum. The two spectra are in good agreement, suggesting that our simplified model can be safely used to infer the system parameters ${\bf V}$. To infer the parameters, we assume that $X_j$ is Gaussian distributed with mean $\left(1+\overline{\mathcal{S}}(\omega_{{\rm RF},j})\right)/2$ and variance $\sigma_m^2 = \left(1-\overline{\mathcal{S}}(\omega_{{\rm RF},j})^2\right)/4N_m$. The Gaussian approximation is justifiable when the number $N_m$ of measurements per frequency is large. Since $\overline{\mathcal{S}}\approx 0.34$ at all frequencies for the spectrum of Fig.~3(a), we fix $\sigma_m^2$ to a constant value $\sigma_m^2 \approx 0.22/N_m$. In addition, we assume a uniform prior for ${\bf V}$. The posterior distribution $L({\bf V}|{\bf X})$ is then sampled using a standard Metropolis algorithm~\cite{Gilks96}. Fig.~\ref{fig:SInference}(b) shows an example Markov chain for the parameters $d_{12}$, $\mathcal{C}_+$, and $\mathcal{C}_-$ (the contrasts $\mathcal{C}_\pm$ are multiplied by a factor $10$ for easier viewing). The resulting marginal posterior for $d_{12}$ is shown in Fig.~4(d) of the main text. For reference, the marginal posterior of $|g_{12}|$ is also shown in Fig.~\ref{fig:SInference}(c). The posteriors were obtained by discarding the first $10^4$ steps as a burn-in phase. The Metropolis acceptance rate was $\sim 30\%$.

When the measurements are not ideal, preparation of the NV in the $m_S = 0$ state (mapped to $\sigma^x = +1$) leads to a photon detection with probability $p \ll 1$, while preparation of the NV in the $m_S = \pm 1$ state (mapped to $\sigma^x = -1$) leads to no photon detection. Under these assumptions, we estimate that the measurement noise variance scales as $\sigma_m^2 \approx (1+\overline{\mathcal{S}})/2pN_m$~\cite{Danjou14}. For $\overline{\mathcal{S}}\approx 0.34$ and for an experimentally achievable ground state detection efficiency $p \approx 0.12$~\cite{Wan18}, this gives $\sigma_m^2 \approx 5.6/N_m$. Comparing with the ideal case, we find that the same variance as for ideal measurements is achieved with approximately $25$ times more measurements. In the present case, this would require $N_m \approx 5\times 10^5$ non-ideal measurements per RF frequency. The execution time of the sequence ($4.6\ \mu$s), plus measurement and reinitialization ($\sim3\ \mu$s) and some additional time to assure reliable thermalization of the labels ($\sim3 T_1$) leads to a total time of $\sim20\ \mu$s per shot. Assuming $M = 25$ inspected frequencies and $N_m \approx 5\times 10^5$, the experiment would require 250 s to be completed.

\section{Additional details for AERIS}
\fancyhead[RO]{APP. C\quad ADDITIONAL DETAILS REGARDING AERIS}

\subsection{Radio field intensity estimation\label{app2}}

In this section, we estimate the radio signal amplitude for the example in the main text. We numerically compute the geometrical factor $\mathfrak{f} =  \int  f(r) dV$ for different hemispheres while we consider the NV axis perpendicular to the diamond surface. This leads to an asymptotical value of $\mathfrak{f} \sim4.1$. Note that half of the asymptotic value is reached for integration hemispheres with a radius of 2-3 times the depth of the NV, which leads to detectable signals even for picoliter volume samples. Considering a pure ethanol sample with a density of 789 kg m$^{-3}$ and a molar mass of 46 g mol$^{-1}$, we obtain a proton density of $\rho =$ 6.2 $\times\ 10^{28}$ m$^{-3}$.  With this into consideration, the total amplitude obtained in a 2.1 T external field at room temperature is $b\sim 2.56$ nT. Finally, we can distribute this amplitude throughout the ethanol spectral peaks according to the following rules: $b/3$ (signal produced by 2 out of 6 hydrogens of the molecule) distributed in four peaks with ratios 1:3:3:1, a single peak of $b/6$, and $b/2$ (signal produced by 3 out of 6 hydrogens of the molecule) distributed in three peaks with ratio 1:2:1, to obtain
\beq
b_k \in \{106 ,320, 320, 106, 426, 320, 640, 320\}\ {\rm pT}.
\eeq

\section{Additional details for dipolarly-coupled samples}
\fancyhead[RO]{APP. D\quad ADDITIONAL DETAILS REGARDING LG4}

\subsection{Nuclear spin dynamics under RF drivings~\label{app1}}

Under an RF field (that will rotate nuclei around the $A , \bar{A}, B,$ or $\bar{B}$ axes) the nuclear spin Hamiltonian including dipole-dipole terms among nuclear spins reads
\begin{equation}\label{HS}
\begin{aligned}
H =&  \sum_i \left[ \gamma_n B_z I^i_z +  \delta_j I^i_z + 2 \Omega I^i_x \sin{\left(\omega_d t - \alpha\right)} \right] 
\\&+\sum_{i>j} \frac{\mu_0\gamma^2_n \hbar}{4\pi r_{i,j}^3} \bigg[\vec{I}_i \cdot \vec{I}_j - 3 (\vec{I}_i  \cdot \hat{r}_{i,j})  (\vec{I}_j \cdot \hat{r}_{i,j})\bigg],
\end{aligned}
\end{equation}
where $r_{i,j}$ is the distance between each pair of nuclei ($\hat{r}_{i,j}$ its a unitary vector such that $\vec{r}_{i,j} = r_{i,j}\hat{r}_{i,j}$), $\gamma_N$ is the nuclear gyromagnetic ratio, $\mu_0$ is the vacuum permeability, $\vec I_i ={\left(I^i_x,I^i_y,I^i_z\right)}$ are the nuclear spin operators for the $i$th spin, $\Omega$ and $\omega_d$ are the Rabi and carrier frequencies of the radio-frequency (RF), $\alpha$ is a tunable phase of the RF, and $\delta_i$ represents the deviation (i.e. the energy shift) of the $i$th nuclear spin from the Larmor precession rate $\omega_L = \gamma_n B_z $ owing to its particular magnetic environment. The accurate determination of $\delta_j$ is the target of the sensing protocol introduced here. 

In a rotating frame w.r.t. $ (\omega_L - \Delta) \sum_i I_z^i$, and under the secular approximation of the dipole-dipole term that eliminates fast rotating terms by invoking the rotating wave approximation, Eq.~(\ref{HS}) simplifies to

\begin{equation}\label{RFin}
\begin{aligned}
H =& \sum_i \left[ (\Delta + \delta_i) I_z^i + \Omega I^i_\alpha \right]
\\&+ \sum_{i>j} \frac{\mu_0\gamma^2_n \hbar}{4\pi r_{i,j}^3} \left[1-3\left(r_z^{i,j}\right)^2\right] \left[I^i_z I^j_z -\frac{1}{2} (I^i_\alpha I^j_\alpha + I^i_{\alpha^\perp} I^j_{\alpha^\perp})\right].
\end{aligned}
\end{equation}
where $I^i_\alpha = \left(I^i_x \sin\alpha + I^i_y \cos\alpha\right)$, and ${I^i_{\alpha}}^\perp = \left(I^i_x \cos\alpha - I^i_y \sin\alpha\right)$.

Introducing a new spin basis, defined by rotating the original axes around $\alpha^\perp$, 
\begin{eqnarray}\label{change}
&&I_P^j = \cos{(\theta)}I_z^j + \sin{(\theta)}I_\alpha^j,\nonumber\\
&&I_Q^j = \cos{(\theta)}I_\alpha^j - \sin{(\theta)}I_z^j,\nonumber\\
&&I_{Q^{\perp}}^j = I_{\alpha^{\perp}}^j,
\end{eqnarray}
and defining the rotation angle through  $\cos{\theta}=\frac{\Delta}{\sqrt{\Omega^2 + \Delta^2}}$, and $\sin{\theta} = \frac{\Omega}{\sqrt{\Omega^2 + \Delta^2}}$, allows to rewrite the sample Hamiltonian as

\begin{equation}
\begin{aligned}\label{hlg}
&H = \bar \Omega \sum_i I_P^i + \sum_i \delta_i[\cos{(\theta)}I_P^i -  \sin{(\theta)}I_Q^i]
\\&+\sum_{i>j} \frac{\mu_0\gamma^2_n \hbar}{4\pi r_{i,j}^3} \left[1-3(r_{i,j}^z)^2\right] \Bigg\{\cos^2{(\theta)}I_P^i I_P^j  + \sin^2{(\theta)}I_Q^i I_Q^j- \cos(\theta) \sin(\theta)(I_P^i I_Q^j + I_Q^i I_P^j )\\&-\frac{1}{2}\bigg[\cos^2{(\theta)}I_Q^i I_Q^j
 + \sin^2{(\theta)}I_P^i I_P^j + \cos(\theta) \sin(\theta)(I_P^i I_Q^j + I_Q^i I_P^j )+ I_{Q^{\perp}}^i I_{Q^{\perp}}^j\bigg]\Bigg\},
\end{aligned}
\end{equation}
where the effective rotation rate around $I_P$ reads $\bar \Omega = \sqrt{\Delta^2 + \Omega^2}$. Finally, many terms can be neglected by a secular approximation with respect to $\bar \Omega \sum_i I_P^i$. The remaining terms in the dipolar interaction disappear {\it magically} when the angle that defines the change of basis in Eq. \eqref{change} satisfies $\cos(\theta) = \pm 1/\sqrt 3$, or, equivalently, when the Lee-Goldburg condition $\Delta = \pm \Omega/\sqrt{2}$ is met, leading to 

\begin{equation}
H = \sum_{i=1}^N\left(\frac{\pm\delta_i}{\sqrt{3}} + \bar{\Omega}\right) I^i_P,
\end{equation}

where $\pm\delta_j/\sqrt{3}$ are the parallel components of the shifts with respect to the effective rotation axis {\it P}, and its sign is the same as the sign of $\Delta$. Note that any combination of $\Omega$ and $\Delta$ that complies with the Lee-Goldburg condition produces the described decoupling effect. In particular, for a given intensity of the RF field, this can be detuned from the top and from the bottom with respect to the Larmor. Moreover, the previous derivation is valid for any phase $\alpha$ of the RF field. This freedom has been exploited to develop more elaborated control schemes that concatenate various RF fields, such as the LG4 sequence implemented in our protocol.

In the LG4 sequence, each driving axis is applied during a time $T=2\pi/{\bar\Omega}$ following the order $A$, $\bar A$, $\bar B$, and $B$ (see main text). In order to obtain the effective dynamics of a full LG4 block, we write the explicit propagator
\begin{equation}
\begin{aligned}
U_{\rm LG4} =& U_{B}U_{\bar B}U_{\bar A}U_A =
e^{-i\sum_{i=1}^N\left(\frac{\delta_i}{\sqrt{3}} + \bar{\Omega}\right) I^i_B T}e^{-i\sum_{i=1}^N\left(-\frac{\delta_i}{\sqrt{3}} + \bar{\Omega}\right) I^i_{\bar B}T}\\&\cdot  e^{-i\sum_{i=1}^N\left(-\frac{\delta_i}{\sqrt{3}} + \bar{\Omega}\right) I^i_{\bar A}T}e^{-i\sum_{i=1}^N\left(\frac{\delta_i}{\sqrt{3}} + \bar{\Omega}\right) I^i_A T}.
\end{aligned}
\end{equation}
In every propagator, we can do the following change
\begin{equation}
e^{-i\sum_{i=1}^N\left(\pm\frac{\delta_i}{\sqrt{3}} + \bar{\Omega}\right) I^i_P T}=e^{-i\sum_{i=1}^N\pm\frac{\delta_i}{\sqrt{3}} I^i_P T}e^{-i\sum_{i=1}^N\bar \Omega I^i_P T}=e^{-i\sum_{i=1}^N\pm\frac{\delta_i}{\sqrt{3}} I^i_P T},
\end{equation}
where we used that $e^{-i\sum_{i=1}^N\bar \Omega I^i_P T}=e^{-i\sum_{i=1}^N 2\pi I^i_P}=\mathbb{I}$. Assuming that $\bar\Omega>>\pm\frac{\delta_i}{\sqrt{3}}$, we can Trotterize the LG4 propagator to obtain
\begin{equation}
e^{-i\sum_{i=1}^N\left(\frac{\delta_i}{\sqrt{3}} I^i_B-\frac{\delta_i}{\sqrt{3}} I^i_{\bar B}-\frac{\delta_i}{\sqrt{3}} I^i_{\bar A}+\frac{\delta_i}{\sqrt{3}} I^i_A\right) T}.
\end{equation}
Finally, substituting the expression for each axis operator of Eq.~\eqref{rotaxes}, obtain the propagator
\begin{equation}
e^{-i\sum_{i=1}^N\left[\frac{\delta_i}{\sqrt{3}\bar \Omega} \left(\Omega I^i_y\cos\alpha+\Delta I^i_z\right)\right]4T}.
\end{equation}
From this expression, we reach the final effective Hamiltonian \eqref{eq: Heff} after rearranging the terms
\begin{equation}
H_{\rm eff} = \sum_i \delta^*_i I^i_C,
\end{equation}
where $I^i_C =  \frac{\sqrt{2}I^i_y\cos{\alpha} + I^i_z}{\sqrt{2\cos^2{\alpha} + 1}}$ and $\delta_i^* = \delta_i \frac{\sqrt{1 + 2 \cos^2{\alpha}}}{3}$.

\subsection{Accumulated Phase\label{app: accumulated}}
As stated in the main text, the signals received by the NV adhere to a general form Eq.\eqref{eq:Signal}. When a two pulse CPMG sequence is applied on the NV sensor (see Fig.\eqref{rotations}) with $\pi$ pulses applied at times $t_1$ and $t_2$, the phase accumulated by the NV at stage $k$ is:

\begin{equation}
\begin{aligned}
\Phi_k=&\int_0^{t_1}\left[|\gamma_e|\Gamma_k\cos{\left(\bar\Omega t + \phi_k\right)}+b_k\right]dt-\int_{t_1}^{t_2}\left[|\gamma_e|\Gamma_k\cos{\left(\bar\Omega t + \phi_k\right)}+b_k\right]dt\\&+\int_{t_2}^{T}\left[|\gamma_e|\Gamma_k\cos{\left(\bar\Omega t + \phi_k\right)}+b_k\right]dt,
\end{aligned}
\end{equation}

where we choose the separation of both pulses to be $\frac{T}{2}$, which ensures the cancellation of the static $b_0$ term

\begin{equation}\label{NVPhase}
\Phi_k = \frac{2|\gamma_e|\Gamma_k}{\bar \Omega}\left[\sin{(\bar \Omega t_1 + \phi_k)} - \sin{(\bar \Omega t_2 + \phi_k)}\right] = \frac{4|\gamma_e|\Gamma_k}{\bar\Omega}\cos{\left(\phi_k-\varphi\right)},
\end{equation}
with $\varphi =  \frac{\pi}{2}-\bar\Omega t_1$.

As the sample evolves under the LG4 sequence, the amplitude $\Gamma_k$ and phase $\phi_k$ of the NMR signal evolve, see Fig.~\eqref{control} (a, b). If the signal gets projected about some axis, e.g. $\Gamma_k\cos{\phi_k}$, the variation of this projection is a simple sinusoidal function (see Fig.~\eqref{control} (c)) which is exactly what we need in order to extract the information using a discrete Fourier transform. This result can be understood geometrically, see main text.

Once we choose a projection angle axis, we can compute the adequate timing for the CPMG sequence as
\begin{equation}\label{Atiming}
\varphi = \varphi_{\rm opt} \rightarrow \varphi_{\rm opt} = \frac{\pi}{2}-\bar\Omega t_1 \rightarrow t_1 = \frac{\pi}{2\bar\Omega} - \frac{\varphi_{\rm opt}}{\bar\Omega}.
\end{equation}

For optimal pulse positions, we select the angle matching the major axis of the ellipse. This axis is orthogonal to both $\hat A$ and $\hat C$, i.e., $\left(0, -\frac{1}{\sqrt{2 + \cos{2\alpha}}}, \frac{\sqrt{2}\cos{\alpha}}{\sqrt{2 + \cos{2\alpha}}}\right)$. Then, the angle $\theta_A$ is measured with respect to the orthogonal component of $\hat z$ concerning $\hat A$. This angle is:

\begin{equation}
\varphi_{\rm opt} = \arccos{\frac{\sqrt{3}\cos{\alpha}}{\sqrt{2 + \cos{2\alpha}}}}.
\end{equation}

\subsection{Analytical expression}\label{app: analytical}
Here we provide details of the derivation of the analytical expression for the expected value of the measurements performed with the NV. Our starting point is the fact that the NV will couple to a signal proportional to the $\hat z$ component of the sample magnetization. 

Focusing on the $k$th driving stage around $\hat A$, we can describe the expected signal as:
\begin{equation}\label{eq: signal_origin}
s \propto \hat M(t)\cdot \hat z = \hat M(t)\left(\hat z^\perp\sin{\theta_{\rm LG}} + \hat A\cos{\theta_{\rm LG}}\right),
\end{equation}
where $\hat M(t)$ is the magnetization vector and the $\hat z$ axis was split in the parallel and perpendicular components with respect to axis $\hat A$, and $\theta_{\rm LG} = \arccos{\frac{1}{\sqrt{3}}}$ is the magic angle. We can describe the time dependency of the magnetization during the driving stage $A$ by employing the Rodrigues' rotation formula as 
\begin{equation}
    \hat{M}(t) = \hat{M}_k\cos{\bar \Omega t}+\left(\hat A\times \hat{M}_k\right)\sin{\bar \Omega t}+\hat A\left(\hat A\cdot \hat{M}_k\right)\left(1-\cos{\bar \Omega t}\right),
\end{equation}
where $\hat M_k$ is the magnetization vector at the beginning of the $k$th sequence. Substituting in Eq. \eqref{eq: signal_origin}, we get 
\begin{equation}
s\propto \left[\hat M_k\cdot \hat z^\perp\cos{\bar \Omega t}+\left(\hat A\times \hat M_k\right)\cdot \hat z^\perp \sin{\bar \Omega t}\right]\sin{\theta_{\rm LG}}+\hat A \cdot \hat M_k \cos{\theta_{\rm LG}}.
\end{equation}
We can now split the magnetization vector into its parallel and perpendicular components with respect to $\hat A$ as $\hat M_k = \left(\vec M_k^{\parallel} + \vec M_k^{\perp}\right)$. With this we reach expression
\begin{equation}
s\propto|\vec M_k^{\perp}|\sin{\theta_{\rm LG}}\cos{\left(\bar \Omega t+\phi\right)}+|\vec M_k^{\parallel}|\cos{\theta_{\rm LG}},
\end{equation}
where $\phi$ is the angle between $\vec M_k^{\perp}$ and $\hat z^\perp$. Notice how this expression exactly matches the shape of Eq.\eqref{eq:Signal} in the main text.

Substituting in Eq. \eqref{eq: optimal}, we obtain
\begin{equation}
    \Phi \propto -\frac{4\gamma_e\sin{\theta_{\rm LG}}}{\bar \Omega}|\vec M_k^{\perp}|\cos{\left(\phi-\varphi\right)} = -\frac{4\gamma_e\sin{\theta_{\rm LG}}}{\bar \Omega}\hat M_k\cdot\hat l
\end{equation}
with $\hat l$ a vector perpendicular to $\hat A$ and tilted $\varphi$ with respect to $\hat z^\perp$. 

We can now generalize to all the driving stages by describing the precession motion of the initial magnetization vectors employing Rodrigues' formula once again
\begin{equation}
\begin{aligned}
\hat{M}_k =& \hat{M}_0\cos{\left(\frac{4\delta_i^* k}{\bar{\Omega}}\right)}+\left(\hat C\times  \hat{M}_0\right)\sin{\left(\frac{4\delta_i^* k}{\bar{\Omega}}\right)}\\&+\hat C\left(\hat C\cdot  \hat{M}_0\right)\left[1-\cos{\left(\frac{4\delta_i^* k}{\bar{\Omega}}\right)}\right].
\end{aligned}
\end{equation}
Starting with an initial magnetization $\hat{M}_0$ in the orthogonal plane with respect to $\hat C$ and an angle $\mu$ with respect to $\hat x$ (which resides in this plane), and including factors for the signal amplitude, we obtain the formula for the accumulated phase
\begin{equation}
    \Phi_k =D_\varphi\rho_i\cos{\left(\frac{4\delta_i^* k}{\bar{\Omega}}+\mu-\beta_\varphi\right)},
\end{equation}
where $D_\varphi = \frac{-\gamma_e\hbar^2\gamma_h^2\mu_0 \mathfrak{f} \sin{\left(\theta_{\rm LG}\right)}}{8\bar{\Omega}\pi^2 k_B T}\sqrt{\left(\frac{\cos{\varphi}\sin{\alpha}}{\sqrt{3}}-\cos{\alpha}\sin{\varphi}\right)^2+\frac{\left(\sqrt{3}\cos{\alpha}\cos{\varphi}+\sin{\alpha}\sin{\varphi}\right)^2}{2+\cos{\left(2\alpha\right)}}}$, $\rho_i$ is the spin density of the $i$th nucleus, and \\$\beta_\varphi = \arctan{\frac{3\left(\sqrt{3}\cos{\alpha}\cos{\varphi}+\sin{\alpha}\sin{\varphi}\right)}{\sqrt{2+\cos{(2\alpha)}}\left(\sqrt{3}\cos{\varphi}\sin{\alpha}-3\cos{\alpha}\sin{\varphi}\right)}}$. Here, $\mathfrak{f}$ is a geometric factor that relates the sample geometry with the signal amplitude in the NV site (see Section~\ref{seq:signal_produced}), $k_B$ is the Boltzmann constant, and $T$ is the temperature. See \cite{Glenn18, Munuera-Javaloy23} for further details on the signal amplitude expression. It can be checked that $\varphi_{\rm opt}$ does indeed maximize $D_\varphi$. The total accumulated phase of the three drivings $A$, $\bar A$, $B$ is simply $3\Phi_k$, provided that $\nu_\varphi = \mu-\beta_\varphi$ is the same in the three stages, which in our case we choose to add up to $0$.

Finally, to consider all effective chemical shifts $\delta^*_i$ it suffices to sum all the contributions. Assuming a small angle $\Phi_k$, the final formula for the expected value of $\sigma_z$ is
\begin{equation}
    \langle\sigma_z\rangle_k \approx 3 D_\varphi\sum_i{\left[\rho_i\cos{\left(\frac{4\delta_i^* k}{\bar{\Omega}}+\nu_\varphi\right)}\right]},
\end{equation}
which gives us the desired spectrum upon Fourier transform.


\gdef\thesubsection{}


%

\renewcommand{\refname}{Bibliography}

\bibliographystyle{X}

\let\oldbibliography\bibliography

\renewcommand{\bibliography}[1]{{%
\let\section\subsection
\oldbibliography{#1}}}

\end{document}